\documentclass[reprintamsfonts,amssymb,amsmath,showpacs,floatfix]{revtex4}

 \usepackage[pdftex]{graphicx,color}

\usepackage{mathrsfs}
\usepackage{amsmath,amssymb}
\usepackage{hhline}
\usepackage{dcolumn}
\usepackage{bm}

\def\mrm{\mathrm}

\def\vphi{\varphi}
\def\pbar{\overline{p}}
\def\goto{\rightarrow}

\def\proot{\psi_{\rm rt}}
\def\pav{\psi_{\rm av}}
\def\pavc{\psi_{{\rm av},c}}

\def\smax{s_{\rm max}}
\def\sroot{s_{\rm rt}}
\def\sav{s_{\rm av}}
\def\pl{p_{c1}}
\def\pu{p_{c2}}
\def\dgamma{\gamma_{\rm d}}
\def\tilp{\tilde{p}}
\def\fps{p_*^{\rm stable}}
\def\fpu{p_*^{\rm unstable}}

\begin{document}

\title{Critical Phase in Complex Networks: a Numerical Study}

\author{Takehisa Hasegawa}
 \email{hasegawa@m.tohoku.ac.jp}
\affiliation{%
Graduate School of Information Science, 
Tohoku University, 
6-3-09, Aramaki-Aza-Aoba, Sendai, Miyagi, 980-8579, JAPAN
}
\author{Tomoaki Nogawa}
\email{nogawa@med.toho-u.ac.jp}
\affiliation{
Faculty of Medicine, Toho University, 5-21-16, Omori-nishi, Ota-ku, Tokyo 143-8540, JAPAN
}
\author{Koji Nemoto}
\email{nemoto@statphys.sci.hokudai.ac.jp}
\affiliation{
Department of Physics, Hokkaido University,
Kita 10 Nishi 8, Kita-ku, Sapporo, Hokkaido, 060-0810, JAPAN
}

\begin{abstract}
We compare phase transition and critical phenomena 
of bond percolation on Euclidean lattices, nonamenable graphs, and complex networks. 
On a Euclidean lattice, percolation shows a phase transition 
between the nonpercolating phase and percolating phase at the critical point. 
The critical point is stretched to a finite region, called the critical phase, on nonamenable graphs. 
To investigate the critical phase, we introduce a fractal exponent, which characterizes a subextensive order of the system. 
We perform the Monte Carlo simulations for percolation on two nonamenable graphs --
the binary tree and the enhanced binary tree.
The former shows the nonpercolating phase and the critical phase, whereas the latter shows all three phases. 
We also examine the possibility of critical phase in complex networks. 
Our conjecture is that networks with a growth mechanism have only the critical phase and the percolating phase. 
We study percolation on a stochastically growing network with and without a preferential attachment mechanism, 
and a deterministically growing network, called the decorated flower, to show that the critical phase appears in those models. 
We provide a finite-size scaling by using the fractal exponent, which would be a powerful method for numerical analysis 
of the phase transition involving the critical phase.
\end{abstract}

\pacs{89.75.Hc,87.23.Ge,05.70.Fh,64.60.aq}

\maketitle


\section{Introduction}

\noindent
In our world, we have many complex networks, {\it e.g.}, the WWW, the Internet, social networks, and airlines.
The study of complex networks has been one of the most popular topics for many research fields since the late 1990's 
\cite{albert2002statistical,newman2003structure,boccaletti2006complex,barrat2008dynamical}.
This activity has stemmed from the discoveries of 
the {\it small-world} property \cite{watts1998collective} and the {\it scale free} property \cite{barabasi1999emergence} in many real networks.
The former means that the mean shortest path length $\bar{\ell}$ between nodes (sites, vertices) scales as 
$\bar{\ell} \propto \log N$, where $N$ is the number of nodes in the network (graph, lattice), 
while the latter means that the degree distribution obeys a power law $P(k) \propto k^{-\dgamma}$, 
where the degree $k$ is the number of edges (bonds, links) connected to a node.
The small-world property often means that 
the network has both a logarithmic size dependence of $\bar{\ell}$ and a high clustering coefficient, {\it i.e.}, $C>0$, 
where $C$ is the probability that two randomly-chosen neighbors of a randomly-chosen node are connected to each other.

Among various studies in network science, 
many kinds of dynamics on complex networks, such as percolation, epidemic processes, spin systems, and coupled oscillators, 
have been extensively studied to stimulate our interests in the relationships between network topology and critical phenomena 
they exhibit \cite{dorogovtsev2008critical}. 
On networks with complex connectivity, 
the type of phase transition can be different from that on the Euclidean lattices. 
Indeed, a new phase called the {\it critical phase} appears in some complex networks (but not all). 
The authors have studied phase transitions and critical phenomena involving the critical phase 
in complex networks and nonamenable graphs (explained in Sec.~\ref{sec:NAG}) \cite{nogawa2009monte,nogawa2009reply,hasegawa2010critical,hasegawa2010generating,nogawa2012generalized,nogawa2012criticality,hasegawa2013absence,hasegawa2013profile,nogawa2013meta}. 
In this paper, we mainly focus on bond percolation, which is the simplest model showing a phase transition, 
and review our results in an attempt to unveil the nature of the critical phase. 

The organization of this paper is as follows.
In Sec.~\ref{sec:Euc}, we start with the introduction of percolation on the Euclidean lattices. 
We consider bond percolation on the square lattice to introduce phase transition and critical phenomena. 
In Sec.~\ref{sec:NAG}, we study percolation on nonamenable graphs, which is the first example showing the critical phase. 
We start with the definition of nonamenable graphs and multiple phase transition, which includes the critical phase. 
After reviewing the previous results for infinite nonamenable graphs, 
we express the critical phase in finite nonamenable graphs and corroborate it by using some nonamenable graphs. 
In Sec.~\ref{sec:Nets}, we move on to complex networks. 
In this section, we consider two types of networks, 
namely stochastically growing networks and deterministically growing (hierarchical small-world) networks, 
to show that such networks also have critical phase. 
In the last part of this section, we will introduce a finite-size scaling method for complex networks. 
Section~\ref{sec:Discussion} is devoted to the discussion for future works.
\begin{figure}\begin{center}
\includegraphics[width=5cm]{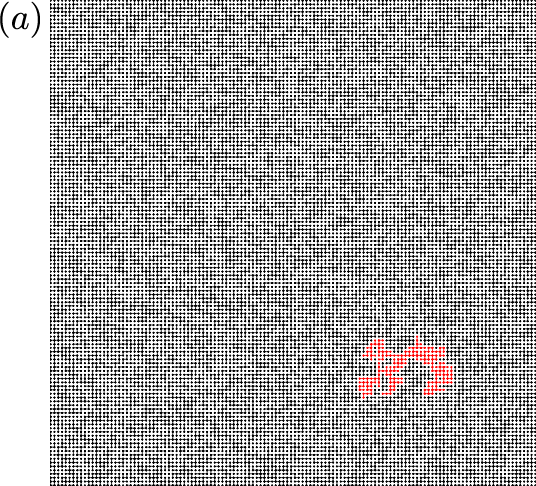}
\includegraphics[width=5cm]{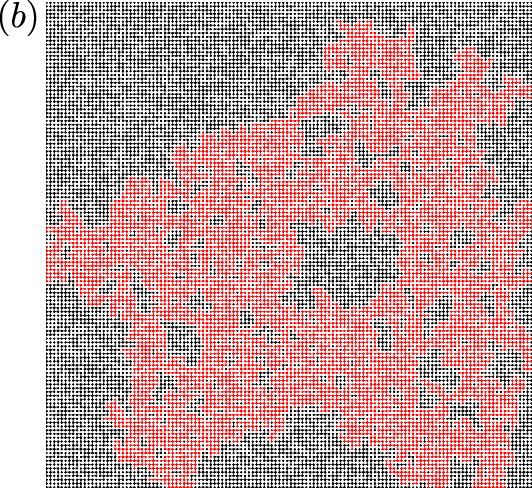}
\includegraphics[width=5cm]{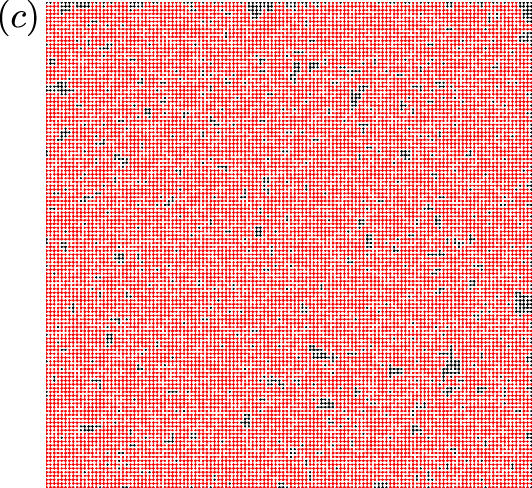}
\end{center}
\caption{
Snapshots of bond percolation on the square lattice for (a) $p=0.4<p_c$, (b) $p=p_c=0.5$, and (c) $p=0.6>p_c$.
In each panel, the red-colored cluster has the largest size.
}
\label{fig:Euc:lattice}
\end{figure}

\begin{figure}
\begin{center}
\includegraphics[width=100mm]{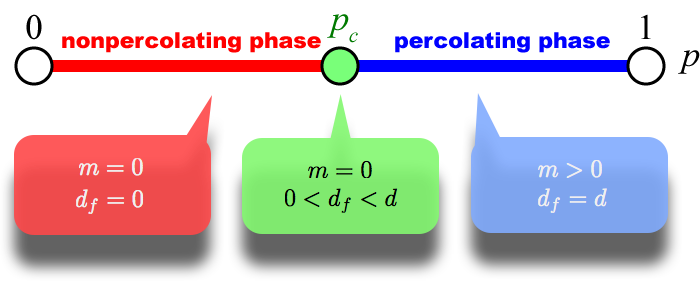}
\end{center}
\caption{Schematic of phase diagram for percolation on the Euclidean lattice.}
\label{fig:Euc:diagram}
\end{figure}

\section{Percolation on the Euclidean lattices \label{sec:Euc}}

\noindent
In this section, we recall some elementary properties of phase transition and critical phenomena 
in bond percolation on the Euclidean lattices \cite{stauffer1994introduction}.
Let us consider a square lattice of linear dimension $L$. The number of nodes is $N=L^2$.
Bond percolation with open bond probability $p$ is a very simple process: 
each bond is independently open (undamaged) with probability $p$, and closed (damaged, to be removed) otherwise 
(Fig.~\ref{fig:Euc:lattice}).
We call a component connected by open bonds a cluster.
The size of a cluster is given by the number of nodes that belong to it.

Once the state of each bond is set to be open or closed at a given value of $p$, 
an important problem is whether a {\it giant component} (percolating cluster), 
which is a cluster that occupies a finite fraction of the whole system in the large size limit $N \to \infty$, exists or not. 
For small $p$, a large number of closed bonds divide the lattice into finite clusters 
and no giant components exist (Fig.~\ref{fig:Euc:lattice}(a)).
A giant component appears when $p$ exceeds a certain value (Fig.~\ref{fig:Euc:lattice}(c)). 
In the language of physics, the system is said to show a second order phase transition 
from the nonpercolating phase, in which only finite clusters exist, 
to the percolating phase, in which a giant component almost surely exists beside finite clusters, 
via a critical state at the {\it critical point} (also known as the percolation threshold) $p_c$ (Fig.~\ref{fig:Euc:lattice}(b)).
The schematic of the phase diagram of percolation on the Euclidean lattices is shown in Fig.~\ref{fig:Euc:diagram}. 
For the square lattice, $p_c=1/2$.

\begin{figure}\begin{center}
\includegraphics[width=7.5cm]{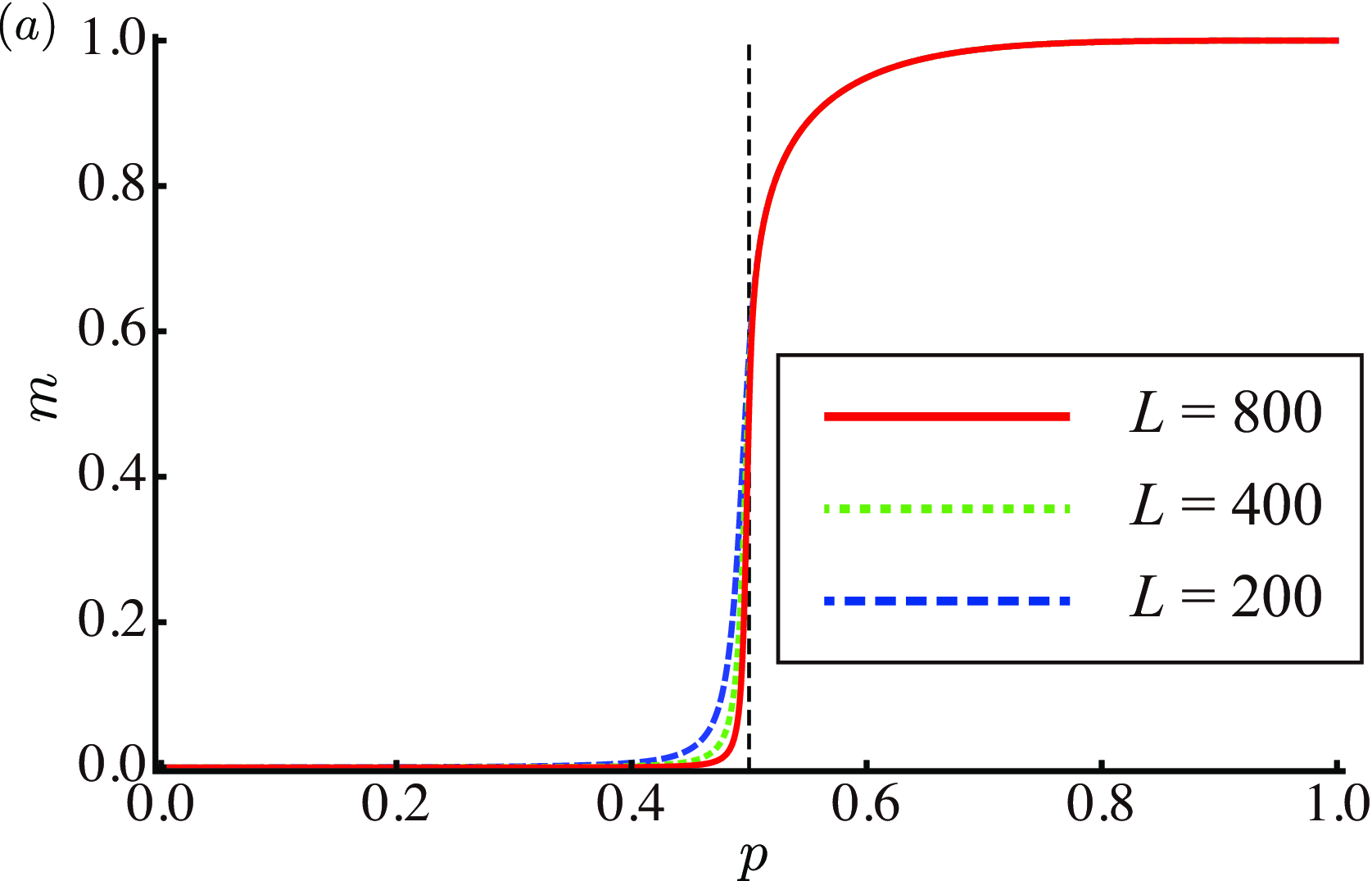}
\includegraphics[width=7.5cm]{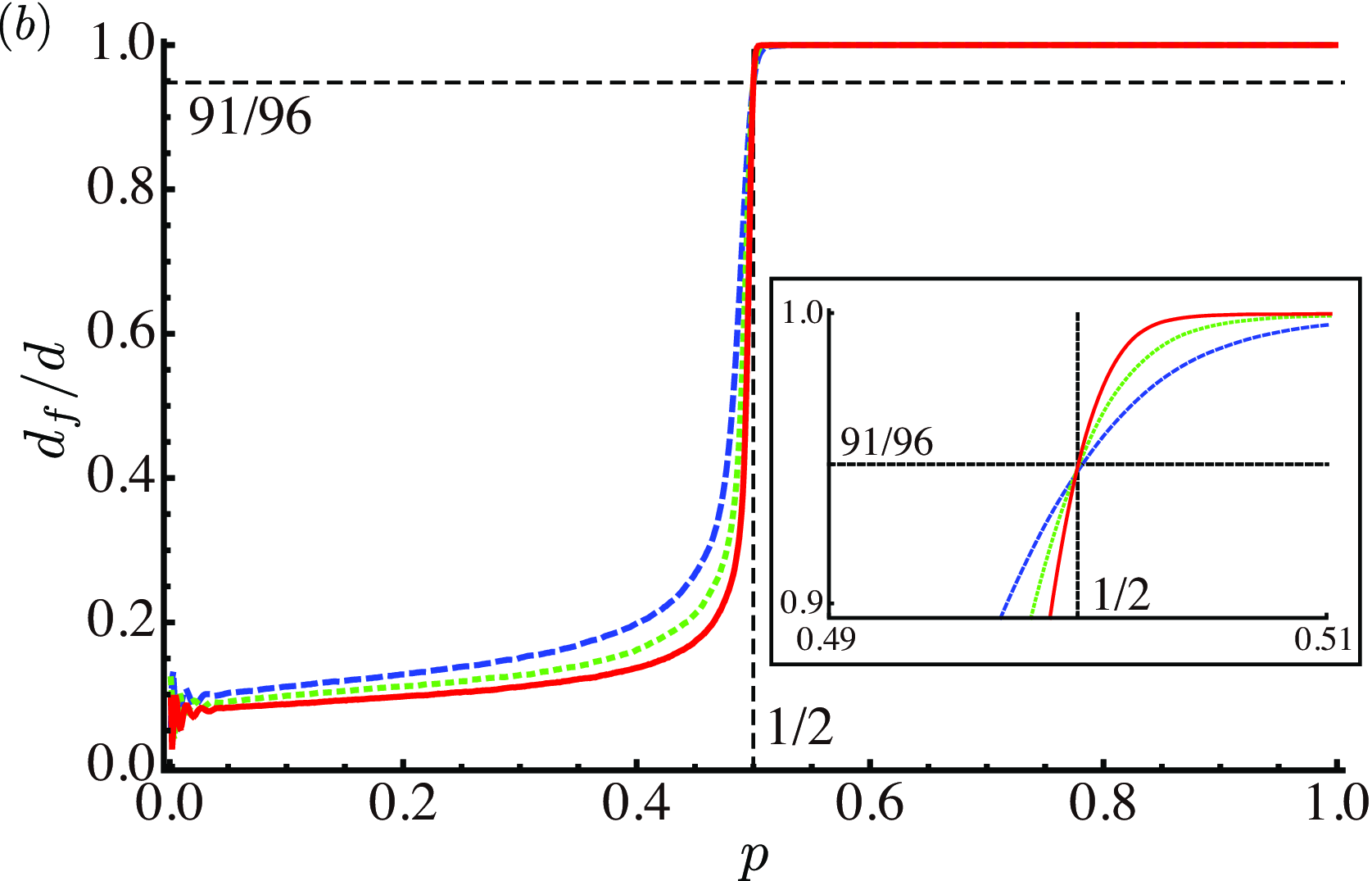}
\includegraphics[width=7.5cm]{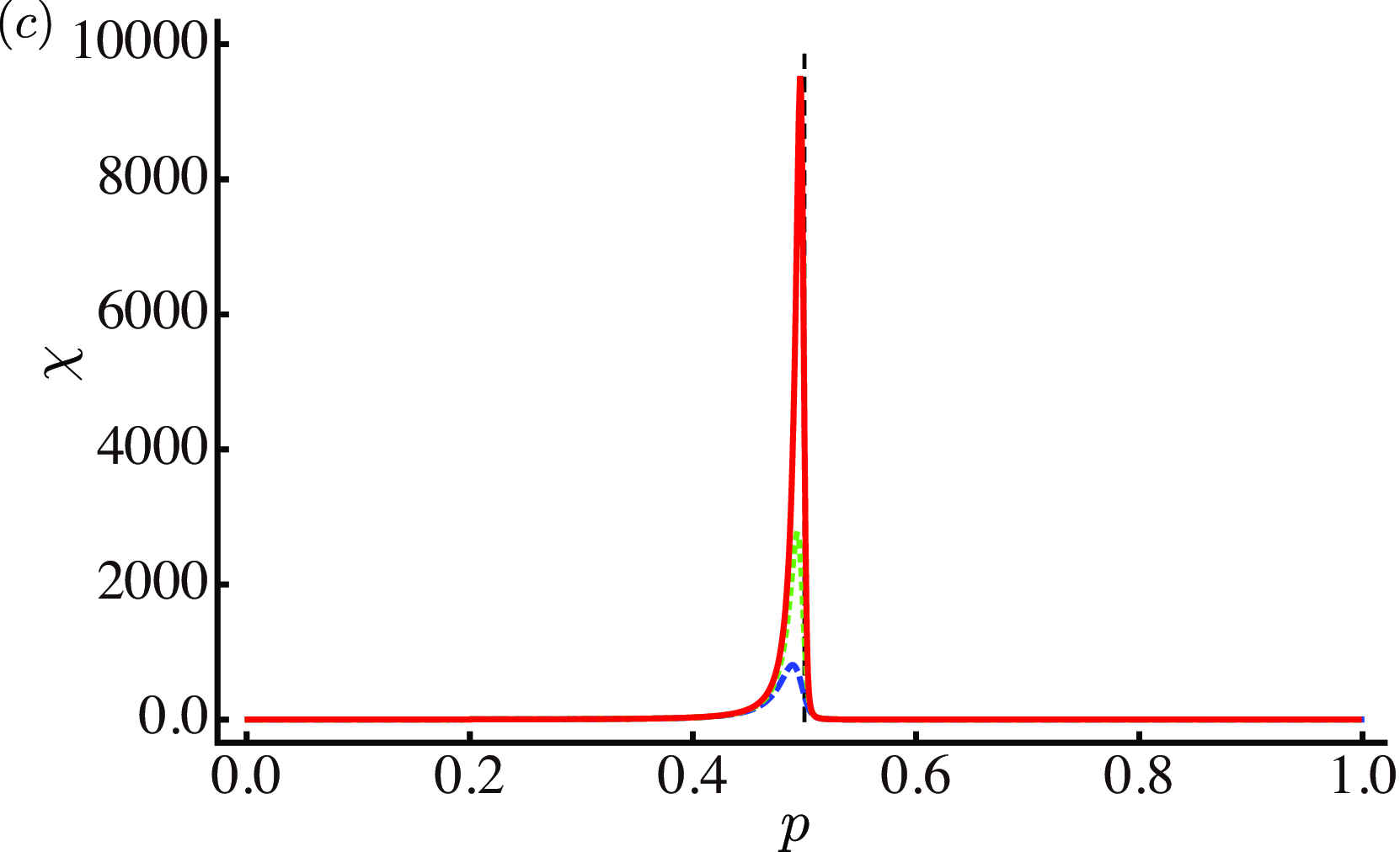}
\includegraphics[width=7.5cm]{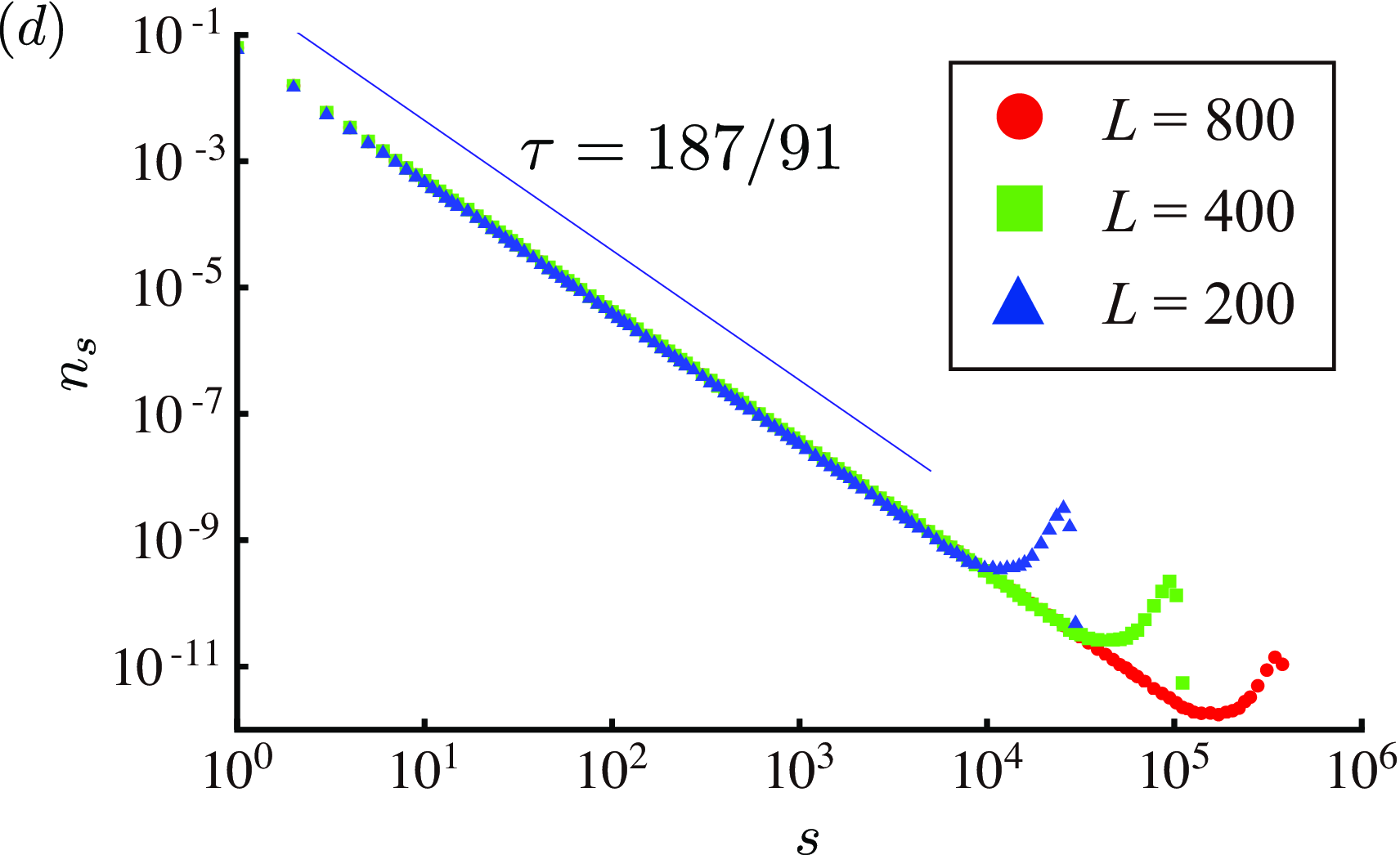}
\end{center}
\caption{
(a) $m(p,N)$, (b) $d_f/d$, (c) $\chi(p,N)$, and (d) $n_s(p_c)$ for the square lattice.
}
\label{fig:Euc:m}
\end{figure}

We usually introduce the order parameter $m(p)$ to characterize the phase transition. 
Let us denote by $\smax(p,N)$ the size of the largest cluster, averaged over many trials, for percolation on the lattice with $N$ nodes. 
The order parameter $m(p)$ in the large size limit is defined by 
\begin{equation}
m(p)=\lim_{N \to \infty} m(p,N), \quad m(p,N)=\smax(p,N)/N.
\end{equation}
In Fig.~\ref{fig:Euc:m}(a), we plot $m(p,N)$, obtained by Monte Carlo simulations, as a function of $p$ with several sizes.
In the large size limit, $m(p)=0$ for $p < p_c$ because only finite clusters exist, 
whereas $m(p)>0$ for $p>p_c$ because a giant component exists. 
At $p_c$ the critical state does not include a giant component, so that $m(p_c)=0$.

We can characterize each phase in another way. 
In Fig.~\ref{fig:Euc:m}(b), we numerically plot $d_f/d$ for the square lattice with several sizes. 
Here the lattice dimension $d$ and the effective dimension of the largest cluster $d_f$ are given as 
\begin{equation} 
N \propto L^d, \quad \smax(p,N) \propto L^{d_f} \propto N^{d_f/d},
\end{equation}
respectively. For the square lattice, $d=2$.
Figure \ref{fig:Euc:m}(b) indicates that in the large size limit, $d_f/d$ becomes a step function of $p$. 
For $p<p_c$, $d_f=0$ because the sizes of all clusters are finite. 
For $p>p_c$, $d_f=d$ because the giant component has a size on the same order as the lattice size.
Just at $p=p_c$, $d_f$ is not zero but slightly smaller than $d$ because the largest cluster is fractal. 
For the square lattice, $d_f/d=91/96$ at $p_c=1/2$ \cite{stauffer1994introduction}.
When we plot $d_f/d$ for several system sizes, 
we have one crossing point at $(p_c, d_f/d)=(1/2,91/96)$, as shown in the inset of Fig.~\ref{fig:Euc:m}(b).

Near the critical point $p_c$, the systems show critical phenomena, {\it i.e.}, 
some physical quantities exhibit power-law behaviors. 
For example, the order parameter $m(p)$ and the mean size of all clusters except the largest one (per node) $\chi(p)$, 
which we call the susceptibility by analogy with magnetic susceptibility in spin systems, behave as 
\begin{eqnarray} 
m(p) &\propto& (p-p_c)^\beta, \quad {\rm for} \quad p>p_c, \\ 
\chi(p) &\propto& |p-p_c|^{-\gamma}, \quad {\rm for} \quad p \neq p_c, 
\end{eqnarray}
respectively. The latter diverges in the large size limit at $p=p_c$ (Fig.~\ref{fig:Euc:m}(c)). 
This behavior is dominated by a divergent length $\xi$, which is called the correlation or connectivity
length. The correlation function $\xi$ is defined as mean distance between two nodes belonging to the same cluster, and behaves as
$\xi \propto |p-p_c|^{-\nu}$.
Here the critical exponents 
$\beta$, $\gamma$ and $\nu$ are universal, {\it i.e.}, they depend only on the dimensionality of the lattice 
(for example, the critical exponents of percolation on the square lattice are the same as those on the triangular lattice). 
The dominance of the correlation length $\xi(p)$ is well understood by considering
the mean number $n_s(p)$ of clusters with size $s$ per node, which we call the cluster size distribution.
At the critical point where $\xi$ diverges, $n_s(p_c)$ obeys a power law (Fig.~\ref{fig:Euc:m}(d)), and for finite $\xi(p)$, 
it is modified such that it decreases rapidly for $s \gg \xi^{d_f}$: 
\begin{equation}
n_s(p_c) \propto s^{-\tau},\quad 
n_s(p)=n_s(p_c)f(s/\xi^{d_f}(p)).
\end{equation}
where $f(x)$ is a scaling function decreasing rapidly with $x$. 
This form reflects a fractal nature of the system.
One can easily show from this distribution that a scaling relation $\beta+\gamma=d_f\nu$ holds.
For the square lattice, $\tau=187/91$, $\beta=5/36$, $\gamma=43/18$, and $\nu=4/3$ \cite{stauffer1994introduction}.

To determine the critical point and the critical exponents from Monte Carlo simulations of finite size systems, 
we often use a finite-size scaling method.
The correlation length $\xi$ determines the relevancy of the finite linear dimension $L$ to the behavior of observables.
For $\xi \ll L$, all observables are governed by $\xi$, while they are governed by $L$ for $\xi \gg L$.
If an observable $X$ is expected to behave as $|p-p_c|^{-\lambda}$ in the large size limit, 
finite-size scaling predicts that it will obey a scaling law
\begin{equation}
X(L,p)=\xi^{\lambda/\nu}\bar{X}(L/\xi)=|p-p_c|^{-\lambda} \tilde{X}((p-p_c)L^{\frac{1}{\nu}}),
\end{equation}
where 
$\bar{X}(x)$ and $\tilde{X}(x)=\bar{X}(x^{\frac{1}{\nu}})$ are scaling functions.

To summarize percolation on the Euclidean lattices, both $m=0$ and $d_f=0$ in the nonpercolating phase 
and $m>0$ and $d_f=d$ in the percolating phase. 
At the critical point, $0<d_f/d<1$, although $m=0$, and $n_s$ obeys a power law.
The phase diagram schematically shown in Fig.~\ref{fig:Euc:diagram} holds for all Euclidean lattices (although $p_c=1$ when $d=1$).
However, this is not always the case for non-Euclidean lattices. 
In the next section, we consider a well-defined example using percolation on nonamenable graphs.
\begin{figure}\begin{center}
 (a) 
\hspace{-0.75cm}
\includegraphics[width=55mm,clip]{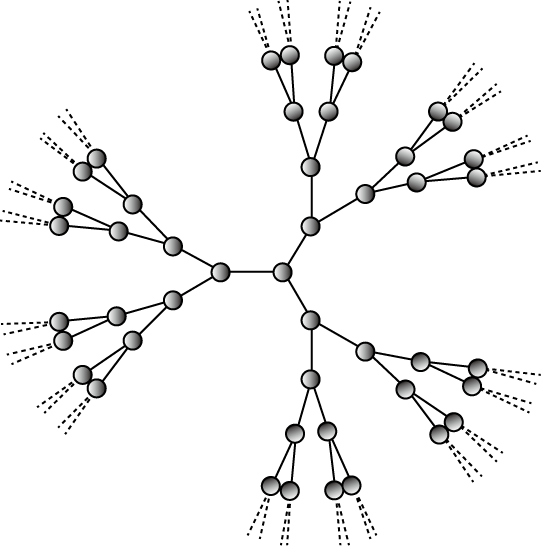}
(b) 
\hspace{-0.75cm}
\includegraphics[width=55mm,clip]{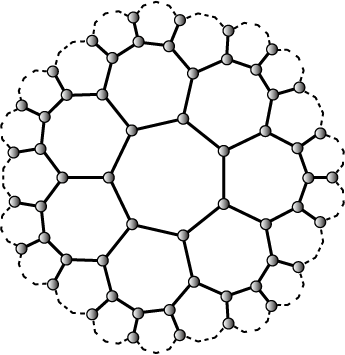}
(c) 
\hspace{-0.75cm}
\includegraphics[width=55mm,clip]{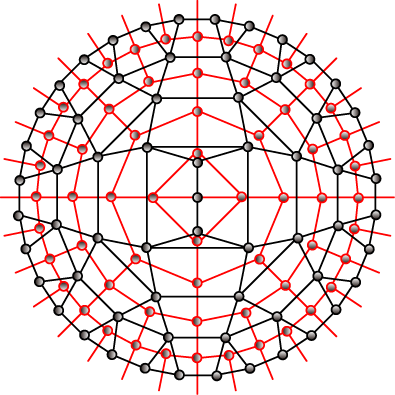}
\end{center}
\caption{
Examples of NAGs; (a) the Bethe lattice, (b) the hyperbolic lattice, and (c) the EBT (black). 
In (c), the red bonds form the dual lattice of the EBT (the dual EBT).}
\label{fig:NAG:example}
\end{figure}

\begin{figure}\begin{center}
\includegraphics[width=100mm]{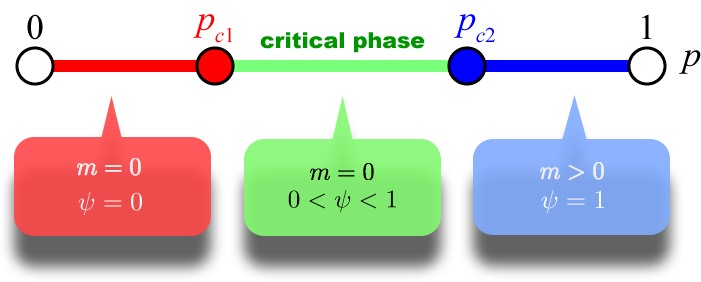}
\end{center}
\caption{
Schematic of phase diagram for percolation on NAGs.}
\label{fig:NAG:diagram}
\end{figure}

\section{Percolation on nonamenable graphs \label{sec:NAG}}

\noindent
Percolation on a nonamenable graph (NAG) along with its novel phase transition, 
called {\it multiple phase transition} (MPT) \cite{benjamini1996percolation}, has been studied in the field of probability theory, 
where an infinite graph is usually assumed. 
In the first part of this section, we briefly review previous results for percolation on infinite NAGs.
Next we provide an expression of the MPT in finite size systems and corroborate it with two NAGs -- 
the binary tree (BT) and the enhanced binary tree (EBT).

\subsection{Theoretical framework}

\subsubsection{Nonamenable graph and multiple phase transition}

\noindent
Let us consider an {\it infinite} (almost-)transitive graph $G$. 
Here ``transitive'' means that all nodes in $G$ play the same role ({\it i.e.}, a transitive graph is a regular lattice).
An infinite graph $G$ is said to be {\it nonamenable}/{\it amenable} if the Cheeger constant $h(G)$ is positive/zero.
The Cheeger constant $h(G)$ is given by 
\begin{equation}
h(G)=\inf_K \frac{|\partial K|}{|K|},
\end{equation} 
where $K$ is an arbitrary nonempty subset of $V(G)$, $V(G)$ being the set of all nodes in $G$, 
and $\partial K$ consists of all nodes in $V(G)-K$ that have a neighbor in $K$.
Figure ~\ref{fig:NAG:example} shows typical examples of NAGs, such as 
the Bethe lattice, hyperbolic lattice, and the enhanced binary tree (EBT).
A fundamental property of NAGs is that the number of reachable nodes from an arbitrary node increases exponentially with the distance from the starting node.
Such an exponential volume growth is never observed in Euclidean lattices (which are amenable). 
In the language of the network science, NAGs are regular lattices having a small-world property: 
the mean shortest path length $\bar{\ell}$ of a finite NAG with $N$ nodes is $\bar{\ell} \propto \log N$. 

As reviewed in \cite{lyons2000phase,schonmann2001multiplicity}, 
percolation on NAGs exhibits a phase transition different from the standard second order transition on the Euclidean lattices, {\it i.e.}, MPT. 
Figure \ref{fig:NAG:diagram} is a schematic of the phase diagram of the percolation on NAGs.
The system on the NAG shows the following three phases, depending on the value of $p$:  
\begin{itemize}
\item{ {\it nonpercolating phase}, where there are only finite clusters, for $0 \le p < \pl$,} 
\item{ {\it critical phase} (also called intermediate phase \cite{lyons2000phase,schonmann2001multiplicity}), 
where infinitely many infinite clusters exist, for $\pl \le p < \pu$, and} 
\item{ {\it percolating phase}, where a unique infinite cluster exists and other clusters are finite, for $\pu \le p \le 1$.}
\end{itemize}
Here an {\it infinite cluster} is defined as a cluster whose size is infinite. 
For transitive NAGs, $\pl<\pu$ \cite{lyons2000phase,schonmann2001multiplicity}, 
while for the Euclidean lattices, $\pl=\pu(=p_c)$, {\it i.e.}, the critical point is unique \cite{burton1989density}.

Whether $\pu=1$ or $\pu<1$ on a given $G$ depends on the number of ${\it ends}$, 
which is a graph property that measures a sort of vulnerability of the graph.
The number of ends of $G$, $e(G)$, is given as the supremum of the number of infinite connected components in $G \backslash S$, 
where $G \backslash S$ is the graph obtained from $G$ by removing an arbitrary finite subset $S$ of nodes or edges. 
The number of ends of an infinite transitive (amenable and nonamenable) graph is either of 1, 2, or $\infty$ \cite{mohar1991some}, 
{\it e.g.}, an infinite tree (with a branching number larger than one) has infinitely many ends, 
whereas the hyperbolic lattices and the EBT have one end \footnote{$e(G)=2$ is realized only for infinite one-dimensional chain.}.
If $G$ is locally-finite ({\it i.e.}, degrees of all nodes are bounded) and transitive, 
$\pu = 1$ when $e(G) = \infty$, and $\pu < 1$ when $e(G)=1$.
Indeed, Benjamini and Schramm proved the existence of the MPT ($0<\pl<\pu<1$) on planar transitive NAGs with one end \cite{benjamini2001percolation}. 

\subsubsection{Critical phase in finite size systems \label{sec:finite_NAG}}

\noindent
The above results apply to infinite NAGs.
Here we should note that infinite NAGs might be different from the asymptotic graphs for a sequence of size-increasing finite NAGs, 
{\it e.g.}, in the language of statistical physics, 
the Bethe lattice is defined as an ginterior regionh of the infinite Cayley tree such that any boundary effect disappears. 
The Bethe lattice is an infinite NAG, and the infinite Cayley tree is the limit of a sequence of size-increasing finite NAGs. 
The latter has non-negligible boundary effects, unlike the Bethe lattice, although  
such an asymptotic graph also shows a critical phase, as we show later.
What we want to qualify here is the asymptotic behavior of the sequence of size-increasing finite graphs.
Below we mention how percolation behaves on a {\it finite} NAG.

Plainly speaking, the critical phase is {\it a finite region in $p$ where the system is in a critical state}. 
In other words, the critical phase is a set of critical points.
A quantity that demonstrates it well is the cluster size distribution $n_s(p)$.
In the critical phase, $n_s(p)$ always obeys a power law, {\it i.e.}, 
\begin{equation}
n_s(p) \propto s^{-\tau(p)}, \label{eq:NAG:ns}
\end{equation}
as $n_s(p)$ of the Euclidean lattices does at the critical point $p_c$.
Moreover, $n_s(p)$ in the critical phase changes its exponent $\tau(p)(>2)$ with $p$.

In \cite{nogawa2009monte}, 
we introduced the {\it fractal exponent} $\psi(p)$ to characterize the critical phase. 
The fractal exponent is defined as 
\begin{equation}
\psi(p) = \lim_{N \to \infty} \psi(p, N), \quad 
\psi(p, N) = \frac{{\rm d} \ln \smax(p, N)}{{\rm d} \ln N}, \label{eq:psi}
\end{equation} 
which is a generalization of $d_f/d$ for the $d$-dimensional Euclidean lattice systems.
Actually, the fractal exponent $\psi(p)$ corresponds to $d_f/d$ for the Euclidean lattices. 
In the critical phase, $\psi(p)$ takes a certain value between zero and one 
as $d_f/d$ of the Euclidean lattices does at $p=p_c$. 
The fractal exponent $\psi(p)$ in the critical phase continuously increases with $p$ and is related to the exponent $\tau(p)$ 
because $\psi(p)$ plays a role of a {\it natural cutoff} exponent 
\footnote{See \cite{dorogovtsev2001size} for a natural cutoff of $P(k)$ in complex networks.}
of $n_s(p)$. 
Assuming that a power law of $n_s$ holds asymptotically and 
the number of clusters with sizes larger than $\smax(p,N)$ should be at most one in a graph with $N$ nodes, 
the largest cluster size $\smax(p,N)$ satisfies
\begin{equation}
N \int_{\smax(p,N)}^{\infty} n_s(p) ds \simeq 1 \to \smax(p,N) \propto N^{\frac{1}{\tau(p)-1}}. \label{eq:cutoff}
\end{equation}
From Eqs.(\ref{eq:psi}) and (\ref{eq:cutoff}), we have 
\begin{equation}
\tau(p)=1+\psi(p)^{-1}. \label{eq:psi-tau}
\end{equation}
Here this equation reduces to a well-known scaling relation $\tau=1+(d_f/d)^{-1}$ for Euclidean lattices. 
By using data of $\psi(p,N)$ from systems of various size, we can characterize each phase as follows: 
\begin{itemize}
\item
$\psi(p, N)$ and $m(p, N) $ go to zero with increasing $N$ in the nonpercolating phase.
\item
$\psi(p, N)$ goes to $0<\psi(p)<1$ and $m(p, N)$ goes to zero with $N$ in the critical phase.
\item
$\psi(p, N)$ goes to unity and $m(p, N)$ goes to $m(p)>0$ when $N$ increases in the percolating phase.
\end{itemize} 
The fractional value of the fractal exponent in a critical phase indicates that 
the largest cluster size diverges in the limit $N \to \infty$, {\it i.e.}, it is an infinite cluster. 
However, it does not occupy a finite fraction of the whole system because $\lim_{N \to \infty} N^{\psi(p)}/N \to 0$ as long as $\psi(p)<1$.
Thus, the order parameter $m(p)$ is zero in the critical phase and becomes nonzero only when $\psi(p)=1$ 
(and then the largest cluster can be regarded as the giant component).
For numerical evaluation, we approximate $\psi(p,N)$ by the difference as, for example, 
\begin{equation}
\psi(p,N) 
\approx 
\frac{\ln \smax(p,2 N)-\ln \smax(p,N/2)}{\ln(2 N)-\ln(N/2)}. 
\end{equation}

Finally we note that a standard finite-size scaling analysis does not work to
determine $\pl$ and $\pu$ by using the data in the critical phase.
Suppose that $\smax(p,N) \propto N^{\psi(p)}$ holds with an increasing analytic function of $\psi(p)$ in the critical phase. 
Then one can expand $\psi(p)$ around any $p^*$ in the critical phase so that
$\psi(p)-\psi(p^*) \propto p-p^*$, which leads us to
\begin{equation}
\smax(p,N) N^{-\psi(p^*)}=g(\ln N(p-p^*)),
\end{equation}
where $g(.)$ is a ``scaling function'' around $p^*$. 
For this reason, this sort of finite-size scaling provides an artificial threshold, depending on the range of the size in the system.
Thus the boundary $p_{c1}$ and $p_{c2}$ is hardly determined from this type of analysis.
In Sec.~\ref{Sec:FSS}, we provide a finite scaling analysis using only data where $p>\pu$.

\begin{figure}[t]\begin{center}
\includegraphics[width=75mm,clip]{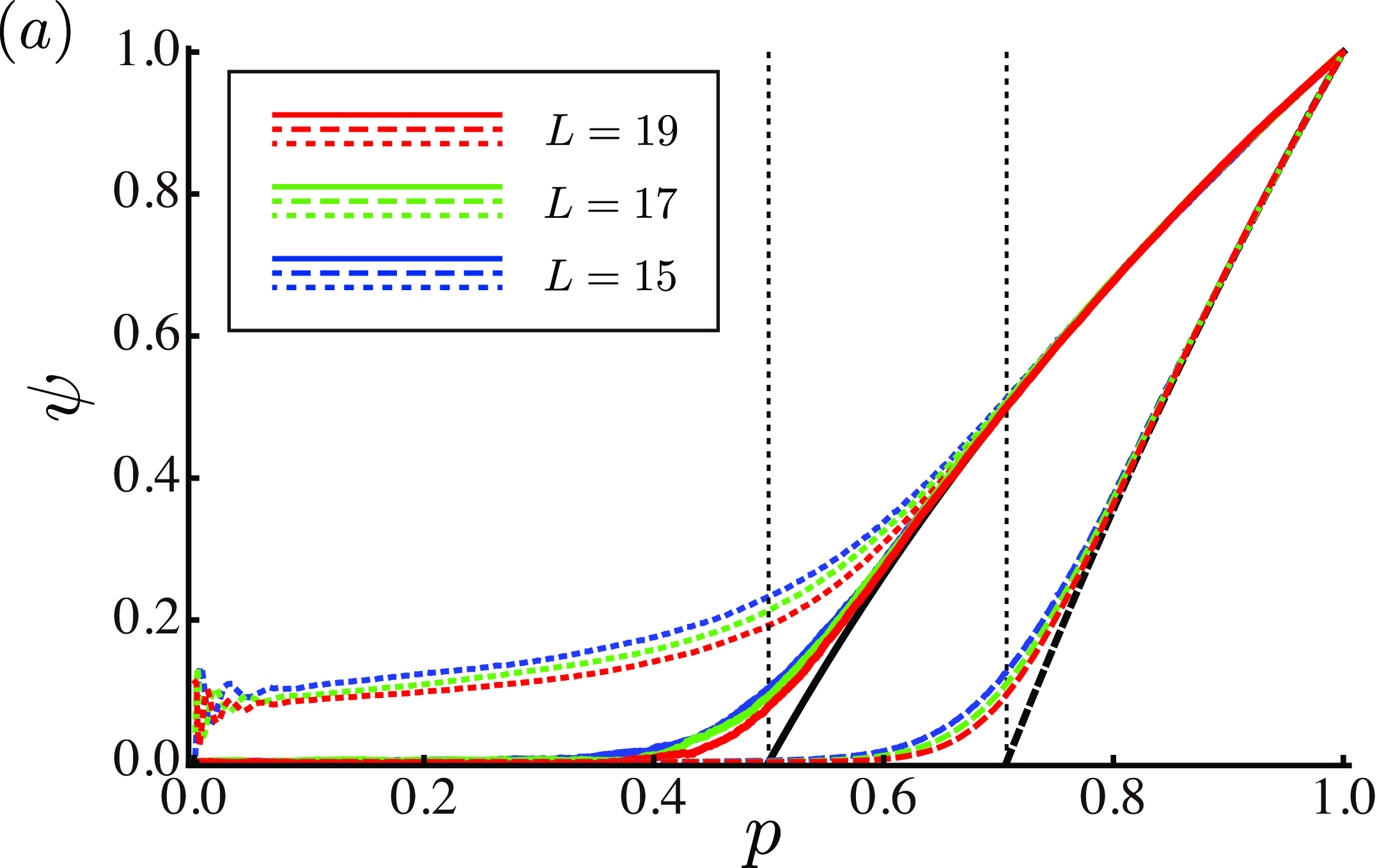}
\includegraphics[width=75mm,clip]{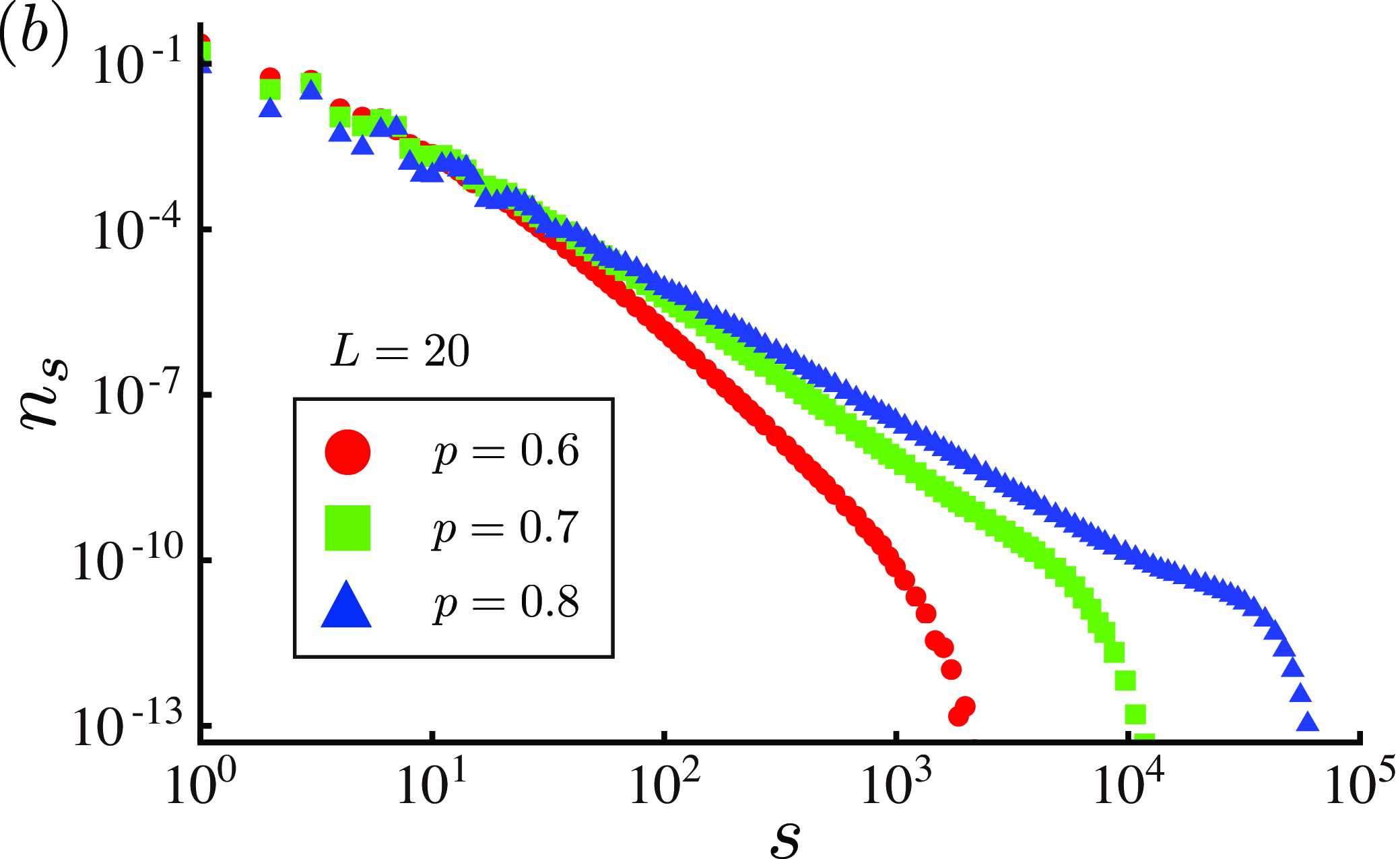}
\end{center}
\caption{
(a) The fractal exponent $\psi(p,N_L)$ and (b) $n_s(p)$ for the BT.
The solid, dotted, and dashed lines in (a) represent the numerical results of $\proot$, $\psi$, and $\pav$, respectively.
The black-solid and black-dashed lines represent analytical predictions.
Two vertical lines indicate $\pl=1/2$ and $p_s=1/\sqrt{2}$.
}
\label{fig:BT:psi}
\end{figure}

\subsection{Percolation on the binary tree \label{SecBT}}

\noindent
As an example of finite NAGs, let us consider the binary tree (BT).
In a BT that has $L$ generations, each node, 
except in the last generation, has two descendants of nodes in the
next generation, so that
the node $v_{n,m}$ has two bonds linked to the nodes $v_{n+1,2m}$ and $v_{n+1,2m+1}$,
where $v_{n,m}$ denotes the $m$-th node in generation $n$ ($n=0, 1, \cdots, L-1$ and $m=0, 1, \cdots, 2^n-1$).
There are $N_L=2^L-1$ nodes in total.
 
Percolation on the BT has $\pl=1/2$ from analysis by a branching process and $\pu=1$ because it is a tree.
We consider the root cluster, {\it i.e.}, the cluster containing the root node $v_{0,0}$. 
In some models discussed in this paper, we calculate the mean size of the root cluster, $\sroot(p,N)$,  
and its fractal exponent, $\proot(p)$, such that $\sroot(p,N) \propto N^{\proot(p)}$, instead of those of the largest cluster. 
Because of the fact that our fractal exponent focuses on the $N$-dependence of a cluster size, 
it is sufficient that the focal node is contained in the largest cluster with a nonzero probability, and then $\proot(p)=\psi(p)$. 
Nodes in the central area of finite NAGs, or hubs of scale-free networks, would be good candidates in most cases.
We easily obtain $\proot(p)$. 
The mean root cluster size of the BT having $L$ generations, $\sroot(p,N_L)$ follows the recursive equation
\begin{equation}
\sroot(p,N_{L+1})=1+2 p \sroot(p,N_{L}), 
\end{equation}
with $\sroot(p,N_{1})=1$.
Then, we have 
\begin{eqnarray}
\sroot(p,N_{L})
\propto \left\{
\begin{array}{ccc}
{\rm const.} & \mathrm{for} & p < 1/2
\\
\ln N_L & \mathrm{for} & p = 1/2
\\
N_L^{\proot(p)} & \mathrm{for} & p > 1/2
\end{array}
\right. ,
\end{eqnarray}
where 
\begin{equation}
\proot(p)=\frac{\ln 2p}{\ln 2}.\label{eq:BT:psi}
\end{equation}
As $p$ increases, $\proot(p)$ increases continuously from $\proot=0$ at $p=\pl=1/2$ to $\proot=1$ at $p=\pu=1$.
In Fig.~\ref{fig:BT:psi}(a), we compare the analytical result (\ref{eq:BT:psi}) and numerical results 
for the fractal exponent of the root cluster and the largest cluster, defined as 
\begin{equation}
\psi(p,N_L) = \frac{d \ln \smax(p,N_L)}{d \ln N_L}, 
\quad
\proot(p,N_L) = \frac{d \ln \sroot(p,N_L)}{d \ln N_L}, 
\end{equation}
respectively. 
The numerical values of the fractal exponents for systems of various sizes 
are almost independent of $N$, and correspond to the analytical line, 
except near above and below $\pl$, where the generation of the BT is not sufficient for convergence of the fractal exponent. 
Also, $\psi(p,N_L)$ deviates from the analytical line when compared with $\proot(p,N_L)$. 
However, both fractal exponents approach the analytical line as $L$ increases.

The root cluster size tells us what happens at $\pl$.
In percolation, the correlation function between two nodes in a graph is defined as 
the probability that the two nodes belong to the same cluster.
Because there is only one path to connect these two nodes in a tree, 
the correlation function $C(p,\ell)$ between $\ell$-distant nodes is given as
\begin{eqnarray}
C(p,\ell) = p^\ell = \exp[-\ell/\xi(p)], \quad \xi(p)^{-1} = \ln(1/p).
\end{eqnarray}
Here the correlation length $\xi(p)$ is finite as long as $p<1$, 
in contrast to that of the Euclidean systems, which diverges at the critical point $p_c$.
At $\pl$, the correlation length itself does not diverge, 
but {\it the sum of the correlation functions} does.
This is due to the exponential volume growth of the NAG. 
The number of nodes such that the distance from the root is $\ell$, $A(\ell)$, is $A(\ell)=2^\ell=e^{y_d \ell}$, where $y_d = \ln 2$.
Since the root cluster size is written in terms of the sum of the correlation function as 
\begin{eqnarray}
\sroot(p,N_L)= \sum_{\ell=0}^{L} A(\ell) C(p, \ell) \propto \sum_{\ell=0}^L e^{ [y_d - \xi(p)^{-1}] L}, 
\end{eqnarray}
it diverges if only $\xi(p)>1/y_d$. 
In the critical phase, some clusters diverge in size (the sum of correlation functions), 
though the correlation length $\xi(p)$ remains finite for $\pl<p<\pu$, and
it is at $\pu$ that $\xi(p)$ (if properly defined) diverges. 
Conversely, the appearance of the critical phase requires an exponential volume growth or a small-world property.

As mentioned in the previous section, the cluster size distribution $n_s$ always obeys a power law in the critical phase. 
In Fig.~\ref{fig:BT:psi}(b), we plot $n_s(p)$ of systems of various size at $p=$ 0.6, 0.7, and 0.8.
We find that a power law of $n_s$ changes its slope $\tau$ with $p$, 
although it is difficult to confirm a clear slope near above $\pl$ 
because $\tau$ is a decreasing function of $p$, becoming infinite at $\pl$.

The continuous change of the power law of $n_s$ with $\tau=\infty$ at $\pl$ and $\tau \approx 2$ at $\pu$ means that 
the $l$-th moment of $n_s$ starts to diverge as $p$ approaches 
the value satisfying $l+1=\tau(p_l)$ from below.
Any order of moments of $n_s$ does not diverge at $\pl$,
but these moments sequentially diverge in descending order as $p$ increases in the critical phase.
In terms of spin systems, this type of phase transition can be understood by defining 
a free energy of percolation $F$ as a generating function of a ghost field $h$: 
\begin{equation}
F(h)={\sum_s} n_s \exp(-h s), 
\end{equation}
so that the $l$-th moment of $n_s$ is obtained by the $l$-th derivative of $F$ with respect to $h$.
Then, the present transition of this model is often 
called the sequence of phase transitions, from the infinite order transition (at $\pl$) to the first order transition (at $\pu$). 
The same type of phase transition has already been reported in the spin system on the Cayley tree 
\cite{eggarter1974cayley,muller1974new}. 
Finally, in this context, we should mention that 
the mean cluster size $\sav(p,N)=\sum_s s^2 n_s$ (as well as the susceptibility $\chi=\sum_{s \neq \smax} s^2 n_s$) 
diverges above $p=p_s<\pu$, at which $\psi=1/2$.
We can also define the fractal exponent of the mean cluster size, $\pav(p)$. 
Since $\sav(p, N) \sim \int^{\smax(p, N)} {\rm d}s s^2 n_s \propto N^{2\psi-1}$, 
$\pav(p)$ is related to $\psi(p)$ as $\pav(p) = 2 \psi(p)-1$ for $p>p_s$. 
As shown in Fig.~\ref{fig:BT:psi}(a), the numerically-obtained $\pav(p)$ satisfies this relation.

\begin{figure}[t]\begin{center}
\includegraphics[width=80mm,clip]{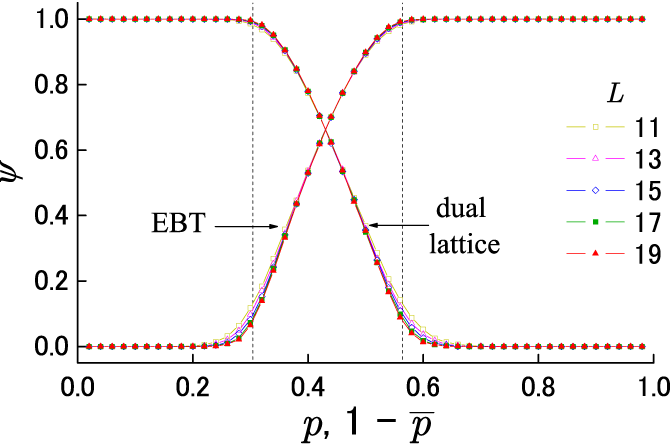}
\end{center}
\caption{
The fractal exponent $\proot(p,N_L)$ for percolation on the EBT and the dual EBT.
For the latter, the horizontal axis is $1-\pbar$ for a check of the duality relations.
The two vertical lines indicate $\pl=0.304$ and $\pu=0.564$, respectively.
}
\label{fig:NAG:psi}
\end{figure}

\begin{figure}[t]\begin{center}
\includegraphics[width=55mm,clip]{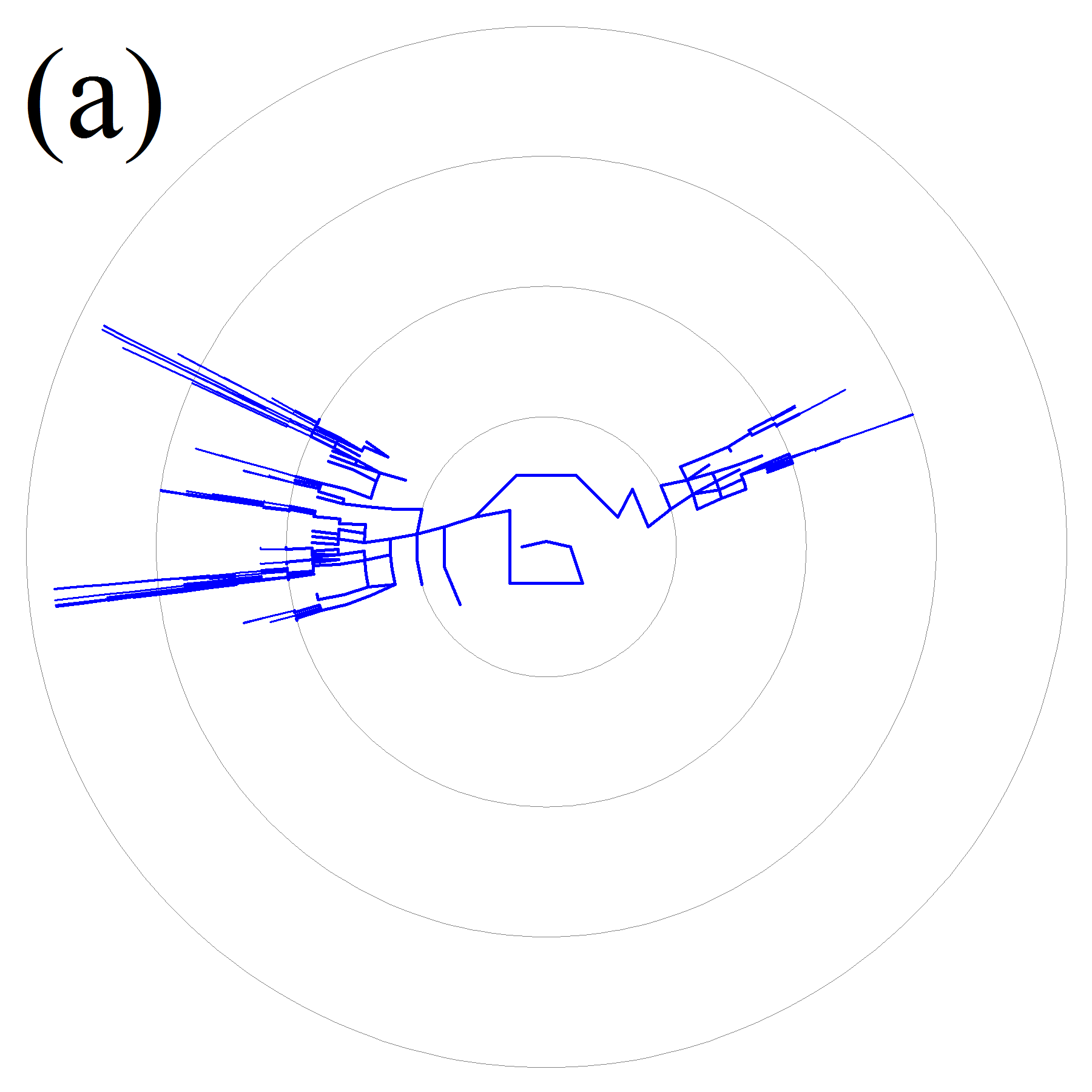}
\includegraphics[width=55mm,clip]{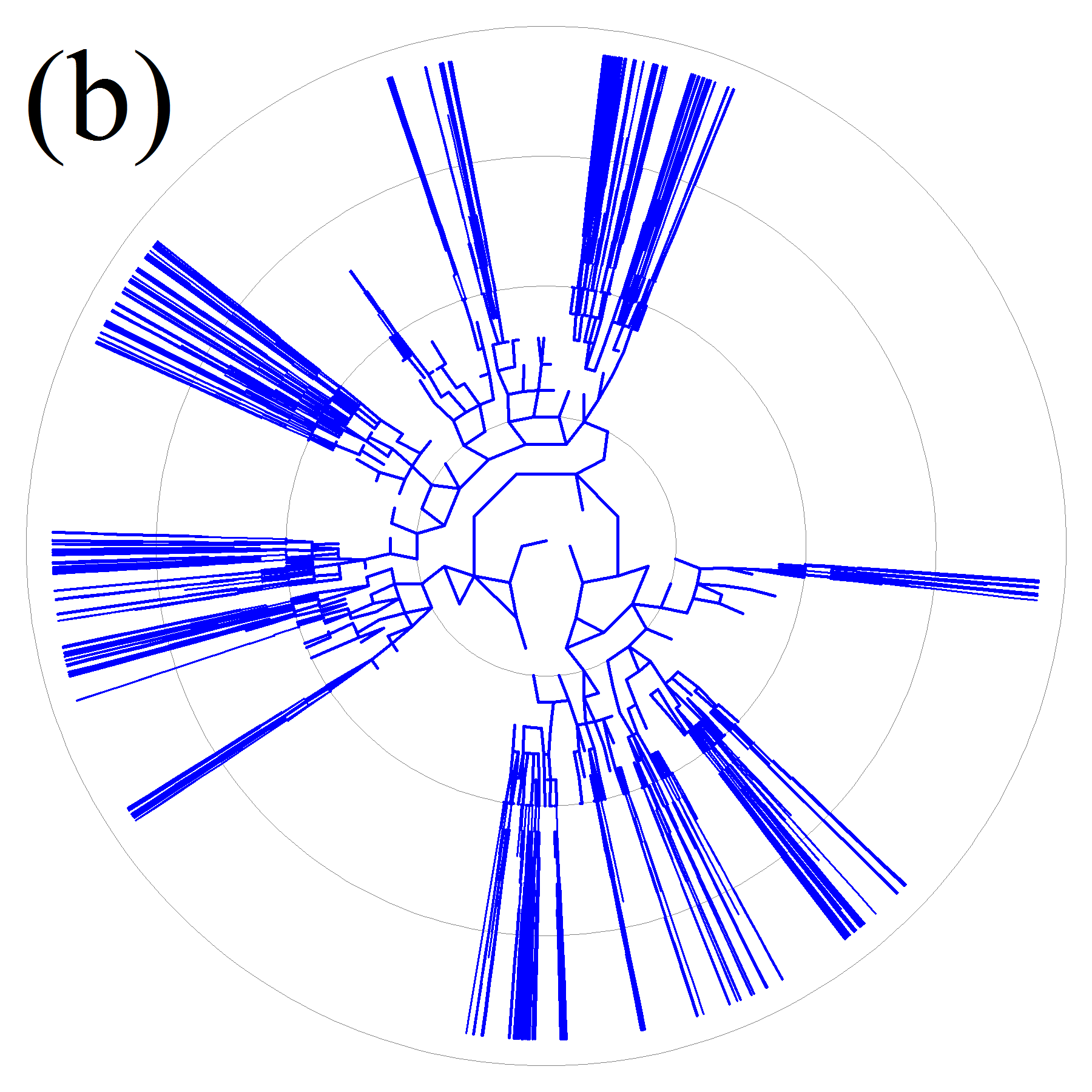}
\includegraphics[width=55mm,clip]{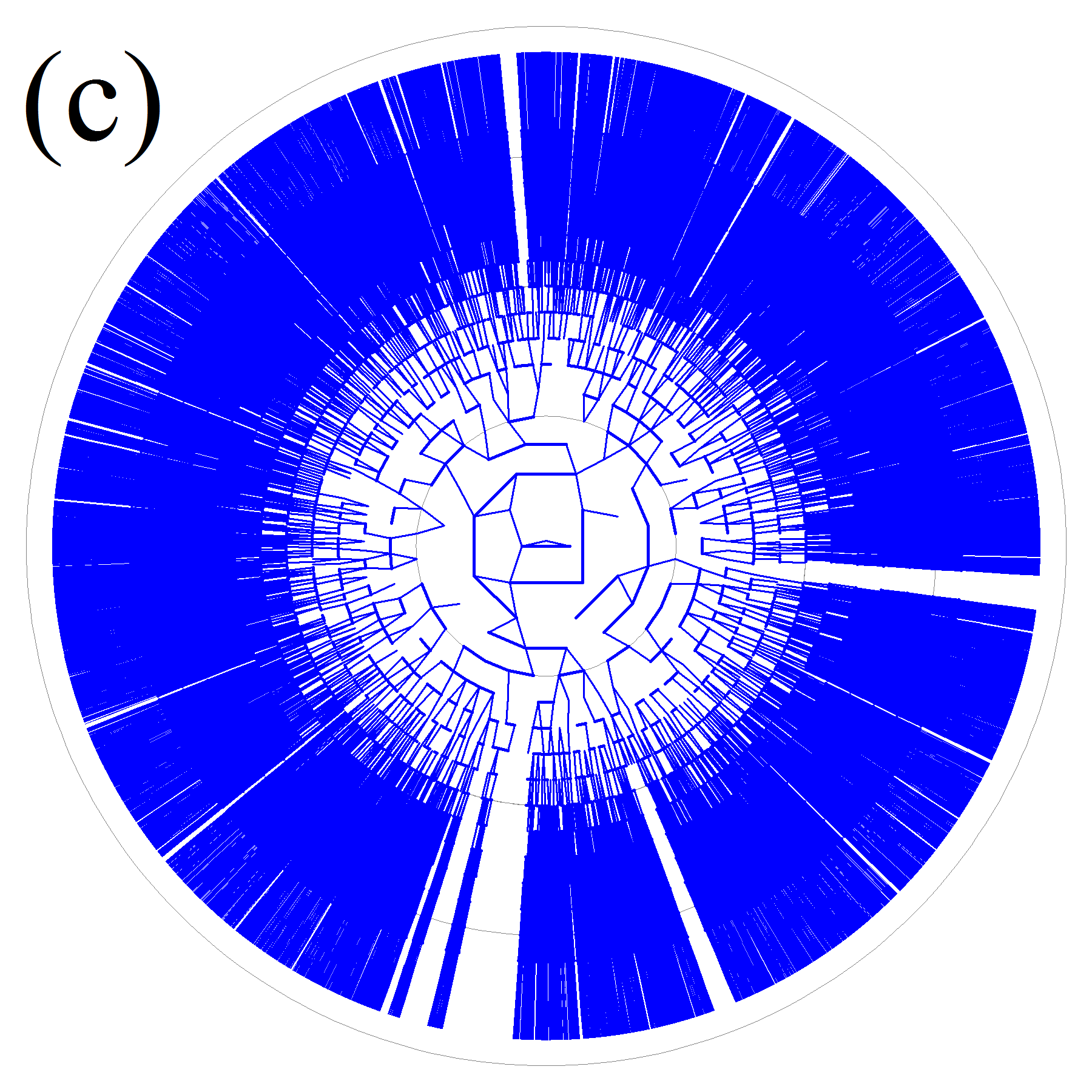}
\end{center}
\caption{Typical geometry of the root cluster of the EBT with $L=20$ for (a) $p=0.304$, (b) $p=0.400$, and (c) $p=0.564$.
}
\label{fig:NAG:snapshot}
\end{figure}

\begin{figure}\begin{center}
\includegraphics[width=120mm,clip]{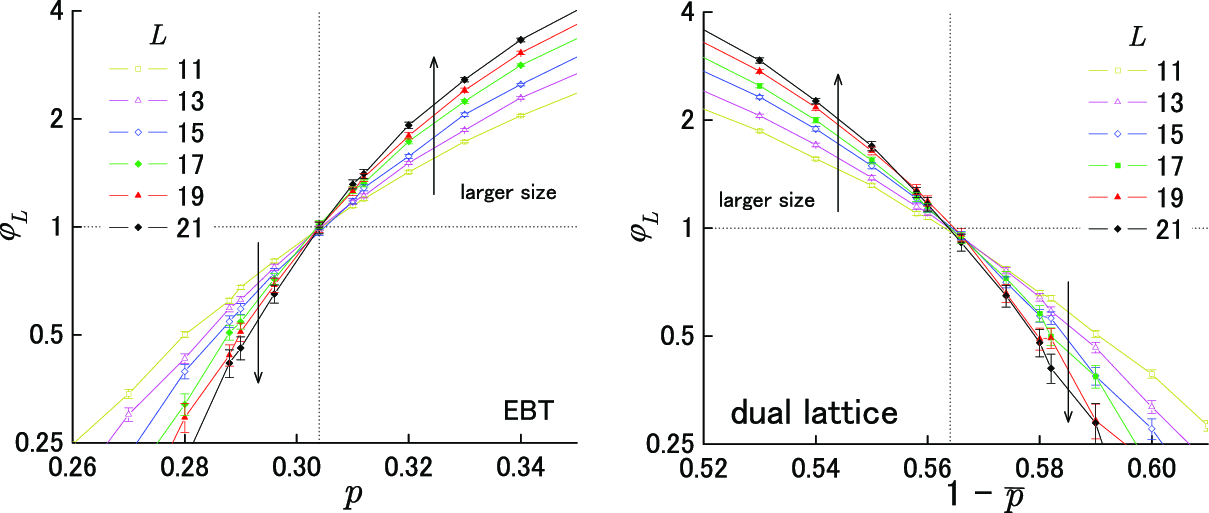}
\end{center}
\caption{
The $p$-dependence of $\vphi_L(p,N_L)$ for the EBT (left) and its dual lattice (right). 
The crossing points give the precise value of the first transition point $\pl$.
}
\label{fig:NAG:pc1}
\end{figure}

\subsection{Percolation on the enhanced binary tree}

\noindent
Next, we consider the EBT, on which percolation shows a two-stage transition at $\pl$ and $\pu$ 
(see our original articles \cite{nogawa2009monte,nogawa2009reply}).
We ``enhance'' the binary tree with generation $L$ by adding the intra-generation (circumferential) bonds between $v_{n,m}$ and $v_{n,m+1}$ ($1 \le n \le L-1$ and $0 \le m \le 2^n-2$) and $v_{n,2^n-1}$ and $v_{n,0}$ to create the EBT with generation $L$.
We performed Monte Carlo simulations for percolation on the EBT with $L=10-22$. 
Each data at a given value of $p$ is averaged over 480,000 samples. 

We also investigated percolation on the dual lattice of the EBT, which we call the dual EBT.
Each node of the dual EBT is put on the center of the triangular or rectangular cells of the EBT, 
and each bond crosses with the conjugate bond of the EBT (Fig.~\ref{fig:NAG:example}(c)). 
Let us denote the open bond probability of the dual lattice by $\pbar$.
Just like a well-known duality relation between a planar Euclidean lattice and its dual lattice, $p_c+\pbar_c=1$, 
we assume the following duality relations for our case,  
\begin{eqnarray}
\pl+\pbar_{c2} = 1 \quad \mrm{and} \quad \pu + \pbar_{c1} = 1,
\label{eq:duality}
\end{eqnarray}
which indeed hold for transitive planar NAGs \cite{benjamini2001percolation}.

Figure~\ref{fig:NAG:psi} shows the $p$-dependences of $\proot(p,N)$ for both of the EBT and the dual EBT. 
In Fig.~\ref{fig:NAG:psi}, we reverse the values of $\pbar$ of the dual EBT (as $1-\pbar$) for the check of the duality relations. 
As expected, we find the three phases: the nonpercolating phase below $\pl \approx 0.30$, 
the critical phase between $\pl$ and $\pu \approx 0.56$, and the percolating phase above $\pu$. 
The fractal exponent $\proot(p,N)$ of the EBT continuously increases from zero to one in the critical phase. 
Conversely, $\proot(p,N)$ of the dual EBT decreases from unity to zero between $1-\pbar_{c2} \approx 0.30$ and $1-\pbar_{c1} \approx 0.56$.
This suggests the duality relations indeed hold. 

As a reference, we show some snapshots of the root cluster in the critical phase (Fig.~\ref{fig:NAG:snapshot}). 
At $p=\pl$, the root cluster survives marginally. 
The root cluster percolates along the radial direction, but its branches do not show spreading behavior; 
therefore, the mass of this cluster is roughly proportional to the number of generations (Fig.~\ref{fig:NAG:snapshot}(a)). 
The size of this root cluster diverges in the limit $N_L \to \infty$, 
but it occupies a very small part of the whole system due to the fact that the spreading rate is slower than that of the EBT itself. 
Consequently, this root cluster does not produce any macroscopic order, 
and the system has space for other clusters to diverge in the limit $N_L \to \infty$. 
When $p$ increases, the root cluster grows; it has more branches, a larger spreading rate, and a larger size (Fig.~\ref{fig:NAG:snapshot}(b)). 
When $p$ reaches $\pu$, the connections of the ``enhanced'' circumferential connections are effective to form 
the giant component such that it occupies the finite fraction of the whole system in the large size limit (Fig.~\ref{fig:NAG:snapshot}(c)). 

The precise value of $\pl$ is given by measuring the $L$-dependence of the root cluster size, 
{\it i.e.}, by measuring whether the root cluster can survive or not. 
Approaching $p_{c1}$ from above, $\proot(p)$ goes to zero, but $\sroot(p,N_L)$ diverges as $\ln N_L$ ($\propto L$). 
In Fig.~\ref{fig:NAG:pc1}, we plot 
\begin{equation}
\vphi_L(p,N_L) = \frac{d \ln \sroot(p,N_L)}{d \ln L}, 
\end{equation}
as a function of $p$.
As $L$ goes to infinity, $\vphi_L(p,N_L)$ decreases to zero for $p<p_{c1}$, and diverges for $p>p_{c1}$. 
Only at $p=p_{c1}$ does $\vphi_L$ rapidly converge to unity, which is consistent with $\sroot(p,N_L) \propto L$. 
Our numerical result for the $L$-dependence of the root cluster size in systems of various size shows one cross point with $\vphi_L=1$, 
which gives the precise values of the first critical point of both the EBT and the dual EBT, 
$\pl=0.304(1)$ and $\bar{p}_{c1}=0.446(1)$. 
We can also determine $\pl$ from the point where $\xi(p)^{-1}=\ln 2$, as is done in \cite{nogawa2009reply}.

\begin{figure}\begin{center}
\includegraphics[width=90mm,clip]{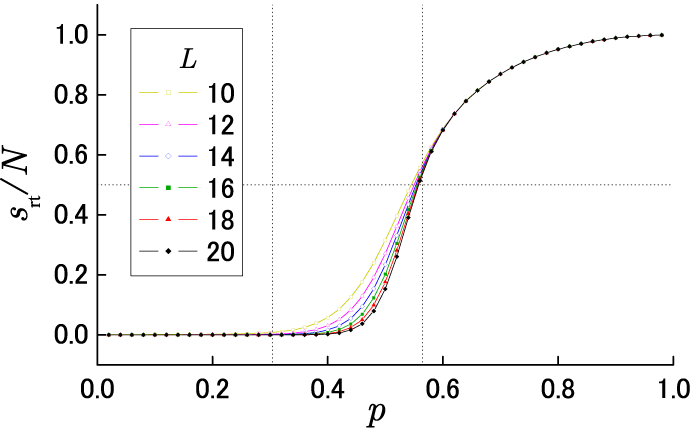}
\end{center}
\caption{
Order parameter $\sroot(p,N_L)/N_L$. The two vertical lines represent $\pl=0.304$ and $\pu=0.564$.
}
\label{fig:NAG:m}
\end{figure}

Compared to the first transition, it is more difficult to determine the precise value of $\pu$ directly from Monte Carlo simulations
(see arguments in \cite{nogawa2009monte,baek2009comment,nogawa2009reply,minnhagen2010analytic,gu2012crossing,baek2012upper}). 
In \cite{nogawa2009monte}, we only assumed $p_{c2}=0.564(1)$ and $\overline{p}_{c2}=0.696(1)$ from the duality relations (\ref{eq:duality}) 
\footnote{Our values of $\pl$ and $\pu$ of the EBT are consistent with the recent numerical result \cite{gu2012crossing}.}
to observe the $N$-dependence of the order parameter at $\pu$. 
The order parameter $\sroot(p,N_L)/N_L$ is well-fitted by $\sroot(p,N_L)/N_L = 0.49+0.58L^{-0.083}$ at $p_{c2}$ (Fig.~\ref{fig:NAG:m}). 
This means that $\sroot(p,N_L)/N_L$ has a finite limit value in $N_L \goto \infty$ at $\pu$.
We must be careful when determining whether or not this transition is truly discontinuous,
but $\beta$ of the order parameter would be very small even if it were continuous.

\begin{figure}[t]\begin{center}
\includegraphics[width=70mm,clip]{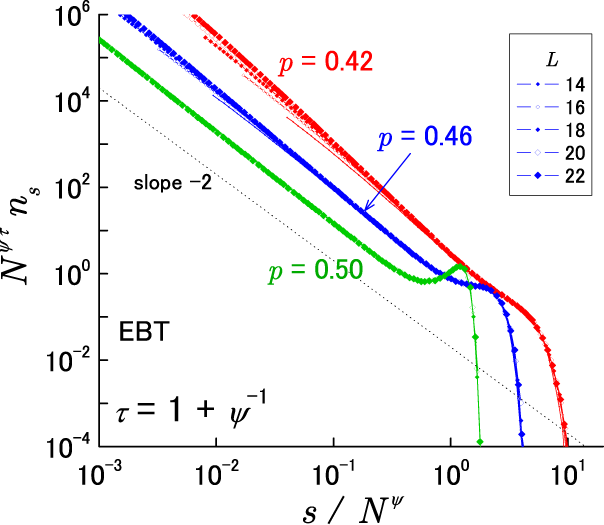}
\end{center}
\caption{
Finite-size scaling for $n_s(p)$ on the EBT.
We used the values of $\proot(p)$ shown in Fig.~\ref{fig:NAG:psi}.
Here we omitted the data for $s<16$ since the data in this range of $s$ does not obey to the scaling law. 
}
\label{fig:NAG:ns}
\end{figure}

Finally, we check the cluster size distribution $n_s$ in the critical phase. 
We assume a finite-size scaling law for $n_s$ as 
\begin{eqnarray}
n_s(p,N) = N^{-\psi(p) \tau(p)} \tilde{n}( s  N^{-\psi(p)} ),
\label{eq:ns-scaling}
\end{eqnarray}
at each $p$ in the critical phase. 
Here the scaling functions $\tilde{n}(x)$ and $\tilde{\tilde{n}}(x)$ behave as
\begin{equation}
\tilde{n}(x) 
\sim
\begin{cases}
\text{rapidly decaying func.} & \text{for $x\gg1$,}\\
x^{-\tau} & \text{for $x\ll 1$.}
\end{cases}
\end{equation}
In Fig.~\ref{fig:NAG:ns}, we show a finite-size scaling result for $n_s$ at several values of $p$ in the critical phase. 
Our results strongly support that our scaling holds in the critical phase, and therefore $n_s$ is indeed a power law for $N \to \infty$ 
at the all points in the critical phase. 

\subsection{Remarks on other NAGs}

\noindent
Other representative examples of NAGs are hyperbolic lattices.
In recent years, numerical approaches for percolation on the hyperbolic lattices have been investigated in \cite{baek2009percolation,gu2012crossing,lee2012bounds}. 
Their numerically-obtained results also showed the existence of the MPT.

To summarize percolation on NAGs, the critical phase where the system is in a critical state appears.
The critical phase is not characterized by the standard order parameter $m(p)$, which catches only macroscopic order, 
but by the fractal exponent $\psi(p)$, which catches a subextensive order. 
As we showed in this section, the uniqueness of the critical point, {\it i.e.}, $\pl=\pu$, is likely violated for non-Euclidean lattices.
Therefore, we should keep in mind the possibility of the critical phase when we study percolation on general graphs.
\section{Percolation on complex networks \label{sec:Nets}}

\noindent
One of the important issues in network science is the robustness of real networks against 
random failures, which remove a set of nodes randomly, and against 
intentional attacks, which preferentially remove nodes having large degrees 
\cite{albert2002statistical,newman2003structure,dorogovtsev2008critical,barrat2008dynamical}.
Albert {\it et al.} \cite{albert2000error} numerically examined the robustness 
of scale-free networks with small $\dgamma$ to show that 
they are highly robust against random failures,
{\it i.e.}, the network remains intact until almost all nodes have been removed.
On the other hand, such networks are very fragile against intentional attacks because 
the removal of a small fraction of hubs is enough to destroy the network. 
Random failures and intentional attacks can be interpreted as percolation problems 
when the removal of a node/bond 
is regarded as a node vacancy in site percolation/a closed link in bond percolation. 
Therefore, both site and bond percolation models on various networks have been extensively studied 
(see \cite{dorogovtsev2008critical} and references therein).

An elementary theoretical framework of percolation in complex networks is provided by 
the local tree approximation for uncorrelated networks~\cite{cohen2000resilience,cohen2002percolation}.
Uncorrelated networks with arbitrary degree distribution $P(k)$ are prepared by the configuration model \cite{molloy1995critical}.
The configuration model with $N$ nodes is given as follows: 
(i) generate a degree sequence $\{ k_1, k_2, \cdots, k_N \}$ of $N$ nodes, according to a desired distribution $P(k)$, 
(ii) attach $k_i$ stubs (half edges that are the ends of edges-to-be) to node $i$,
and (iii) make links by connecting randomly chosen pairs of stubs
\footnote{Since two stubs of a node may be paired together or a pair of nodes may be selected in many times, 
the configuration model allows a number of self-loops or multiple edges in principle. 
But these may be neglected in many cases.}.
The local tree approximation shows that the critical properties for uncorrelated networks 
with scale-free degree distribution $P(k) \propto k^{-\dgamma}$ are determined by the exponent $\dgamma$~\cite{cohen2000resilience,cohen2002percolation}: 
the critical point between the nonpercolating phase and the percolating phase is given 
by $p_c=\langle k \rangle/\langle k^2-k \rangle$, where $\langle x \rangle =\sum_k x(k) P(k)$,  
and the critical behavior of the order parameter $m(p)$ depends only on $\dgamma$ \cite{cohen2002percolation} 
as given by 
\begin{eqnarray}
m(p) \propto 
\left\{
\begin{array}{ll}
p-p_c, & \dgamma  \ge 4, \\
(p-p_c)^{1/(\dgamma-3)}, & 3< \dgamma < 4, \\
p \exp(-2/p \langle k \rangle), & \dgamma =3, \\
p^{1/(3-\dgamma)}, & 2< \dgamma < 3. 
\end{array} 
\right.  \label{betaConfig}
\end{eqnarray}
The local tree approximation for uncorrelated networks
confirms that scale-free networks with heavy-tailed degree distributions are robust against random failures \cite{callaway2000robustness,cohen2000resilience}, 
{\it i.e.}, $p_c$ is zero for both bond and site percolations when $\dgamma \le 3$, while $p_c>0$ for $\dgamma>3$.
This approximation can be applied to the case of intentional attacks 
to show that uncorrelated scale-free networks with small $\dgamma$ are fragile against such attacks \cite{callaway2000robustness,cohen2001breakdown}. 
This theory can be extended to treat clustered networks and correlated networks \cite{goltsev2008percolation,tanizawa2012robustness,newman2003properties,gleeson2009bond,gleeson2009analytical,gleeson2010how}.

As long as a network is uncorrelated, the percolation on the network 
shows a conventional second order phase transition between the nonpercolating phase and the percolating phase 
at the critical point $p_c$.
However, the systems on networks made with the growth mechanism show quite a different picture.
Here ``growth'' means the number of nodes in the graph increases with time. 
Previous analytical studies for several growing and hierarchical small-world (deterministically growing) networks have revealed that 
the system exhibits an unusual phase transition, 
termed the inverted Berezinskii-Kosterlitz-Thouless (BKT) transition~\cite{callaway2001randomly,dorogovtsev2001anomalous,lancaster2002cluster,kim2002infinite,coulomb2003asymmetric,
zalanyi2003properties,krapivsky2004universal,bollobas2004phase,bollobas2005phase,bollobas2005slow,riordan2005small,
pietsch2006derivation,zhang2008degree}: 
(i) The singularity of the transition at $\pu$ is infinitely weak.
When $p$ is nearly above the percolation threshold $\pu$, the order parameter follows 
\begin{equation}
m(p) \propto \exp [-\alpha/(\Delta p)^{\beta'}], \quad {\rm for} \quad \Delta p \ge 0, \label{eq:infiniteorder}
\end{equation}
where $\Delta p=p-p_{c}$.
(ii) In the whole region below $\pu$, $n_s(p)$ obeys a power law, {\it i.e.}, $\pl=0$. 
If this unusual phase on growing networks is regarded as the critical phase on NAGs, 
these systems have a new scenario of phase diagrams: $\pl=0$ and $\pu > 0$. 
In this section, we consider such networks by calculating the fractal exponent in order to substantiate this conjecture.

\subsection{Stochastically growing networks}

\noindent
Among several mathematical models of growing networks, 
the most famous one is the Barab{\'a}si-Albert (BA) model.
It has a preferential attachment mechanism, 
in which new nodes prefer to connect to pre-existing nodes with larger degrees.
The percolation on the BA model is known to indicate $\pu=0$ \cite{albert2000error}.
On the other hand, the percolation on a growing network without preferential attachment, 
which is called $m$-out graph, 
has been studied rigorously and is found to have a finite percolation threshold $p_{c2}>0$, 
at which an inverted BKT transition occurs.
Here, we consider percolation on a growing random network (GRN) proposed 
in \cite{dorogovtsev2000structure,krapivsky2000connectivity}, 
which interpolates between the BA model and the $m$-out graph by introducing the initial attractiveness.

\subsubsection{Model}

\noindent
The GRN stochastically generates a graph with $N$ nodes as follows: 
(i) At the initial time, we start with a complete graph of $m_0(\ge m)$ nodes, where $m$ is a given positive integer. 
We call these nodes the roots.
(ii) At each time step, a new node joins the network by attaching to $m$ pre-existing nodes. 
The probability that a new edge attaches to a node with degree $k$ is proportional to a quantity called the linear attachment kernel $A_k$, given by 
\begin{equation}
A_k=k+k_0, \label{eq:preferential}
\end{equation}
where $k_0(>-m)$ is a constant called the initial attractiveness. 
(iii) The process (ii) continues until the number of nodes reaches $N$.
For $N \gg 1$, the degree distribution $P(k)$ becomes stationary and is given by 
\begin{equation}P(k) \propto k^{-\dgamma}, 
\quad {\rm where } \quad 
\dgamma=3+\frac{k_0}{m},
\end{equation}
for $k_0 < \infty$ and $P(k) \propto [(m+1)/m]^{-k}$ for $k_0 = \infty$ \cite{dorogovtsev2000structure}.
Here $\dgamma$ is the degree exponent controlled by the initial attractiveness $k_0$ as $\dgamma =3+k_0/m$ \cite{krapivsky2000connectivity}; 
$\dgamma$ increases from $2$ to $\infty$ as $k_0$ increases from $-m$ to $\infty$.
The cases of $k_0=0$ and $k_0=\infty$ are reduced to the BA model and the $m$-out graph, respectively.
We also call the case of $m_0=m=1$ the growing random tree (GRT) because the resulting network is a tree. 

\subsubsection{Percolation on the GRT}

\noindent
In \cite{hasegawa2010critical}, the authors studied percolation on the GRT.
This network model remains a tree but grows with time.
Since we expect that growing networks have no nonpercolating phase (except at $p=0$) 
while any tree has no percolating phase (except at $p=1$), 
it is not surprising that the 
percolation on the GRT is always critical in the whole range of $p$ (except at $p=0,1$), 
regardless of the degree exponent $\dgamma$.

By the rate equation approach, 
we can obtain the approximate forms of $\sroot(p,N)$ and $n_s(p)$ for the GRT with arbitrary $\dgamma$ 
(see \cite{hasegawa2010critical}).
The fractal exponent $\proot(p)$ of $\sroot(p,N)$ and the exponent $\tau(p)$ of $n_s(p)$ are 
\begin{equation}
\proot(p)=\frac{1+(\dgamma-2)p}{\dgamma-1}, \quad \tau(p)=\frac{\dgamma+(\dgamma-2)p}{1+(\dgamma-2)p}, \label{eq:GRtree}
\end{equation}
respectively.
The relation (\ref{eq:psi-tau}) holds for this case, as expected.
Equation (\ref{eq:GRtree}) indicates that $\proot(p)$ increases from $1/(\dgamma-1)$ to 1 as $p$ increases from $0+$ to $1$.
Thus, the GRT is always in the critical phase (except at $p=0,1$), regardless of the value of $\dgamma$, {\it i.e.}, 
the network heterogeneity. 

\begin{figure}\begin{center}
\includegraphics[width=7.0cm]{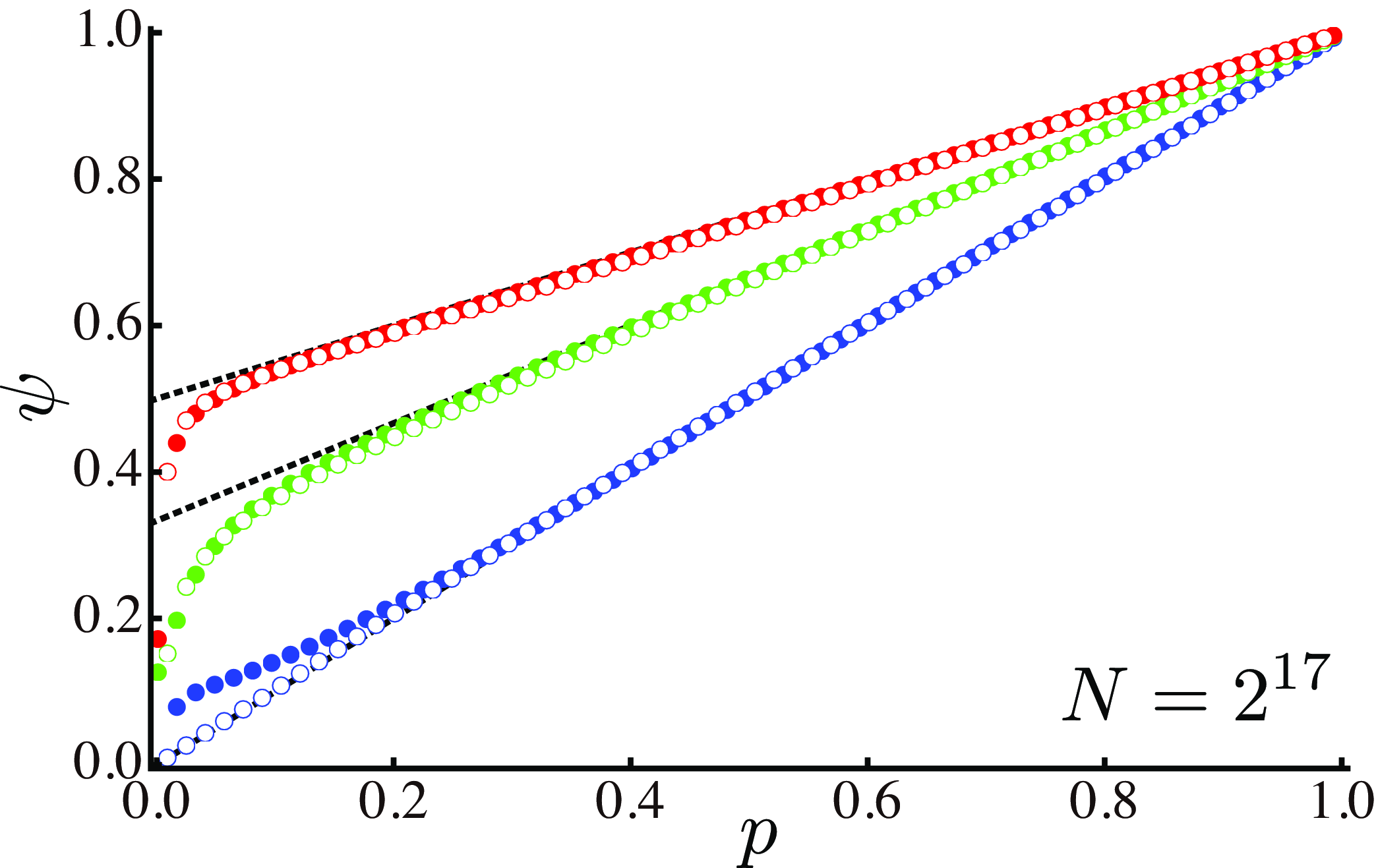}
\end{center}
\caption{
Numerical results of $\psi$ (full circles) and $\proot$ (open circles) for the GRT with $\dgamma=3, 4$ and $\infty$, from top to bottom. The dashed lines are drawn from Eq.(\ref{eq:GRtree}).
}
\label{fig:tree:psi}
\end{figure}

\begin{figure}\begin{center}
\includegraphics[width=5.5cm]{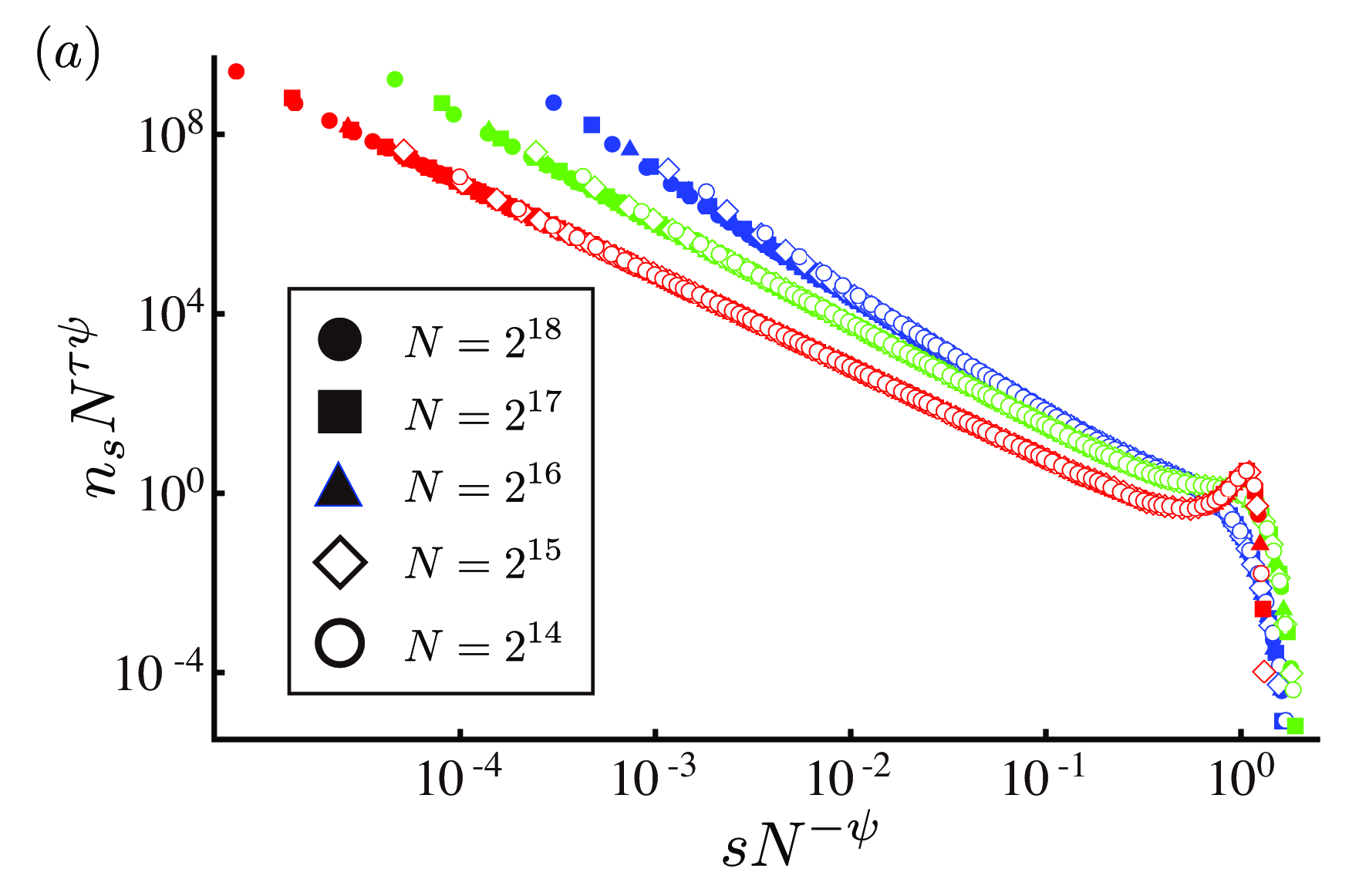}
\includegraphics[width=5.5cm]{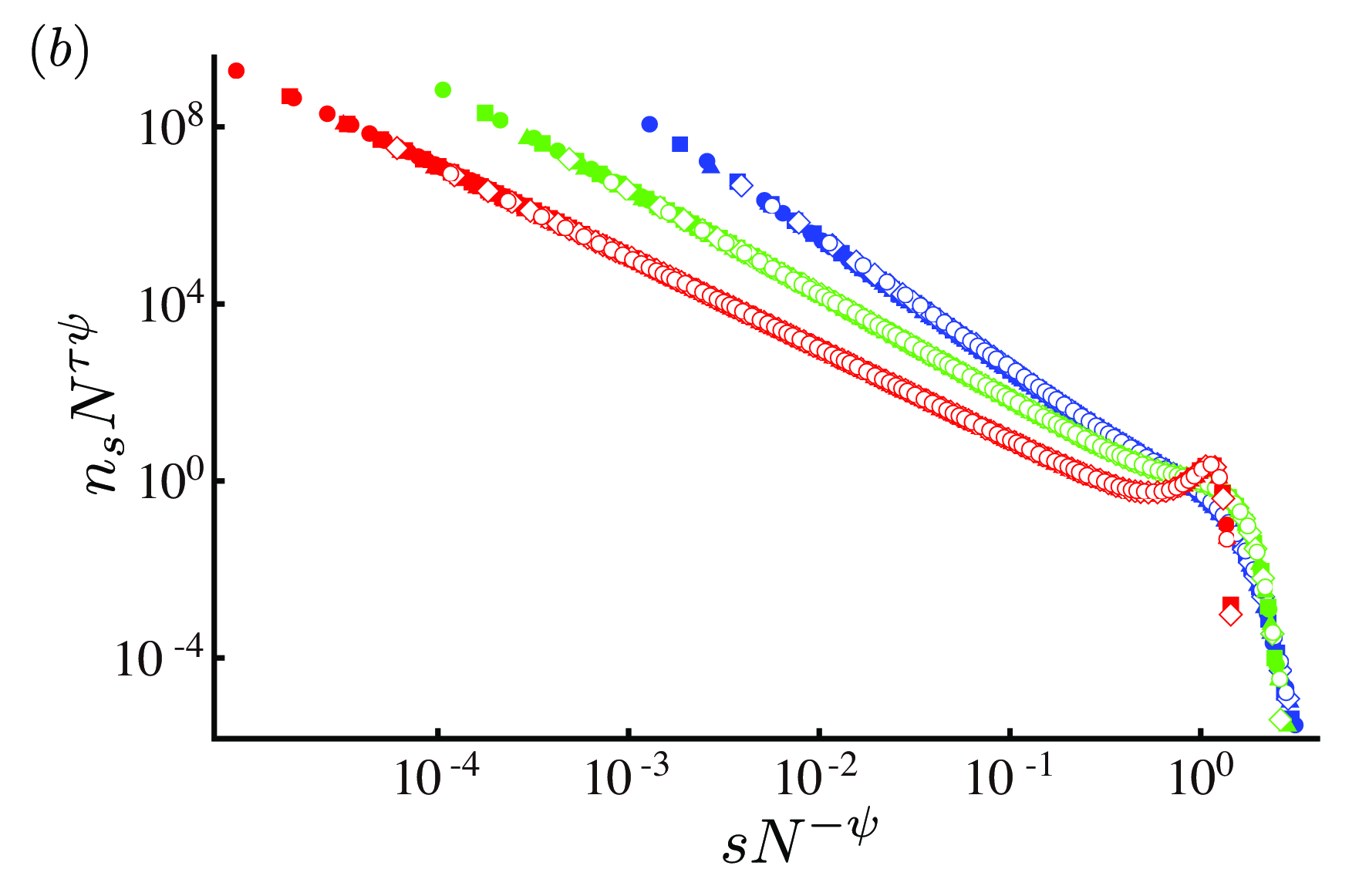}
\includegraphics[width=5.5cm]{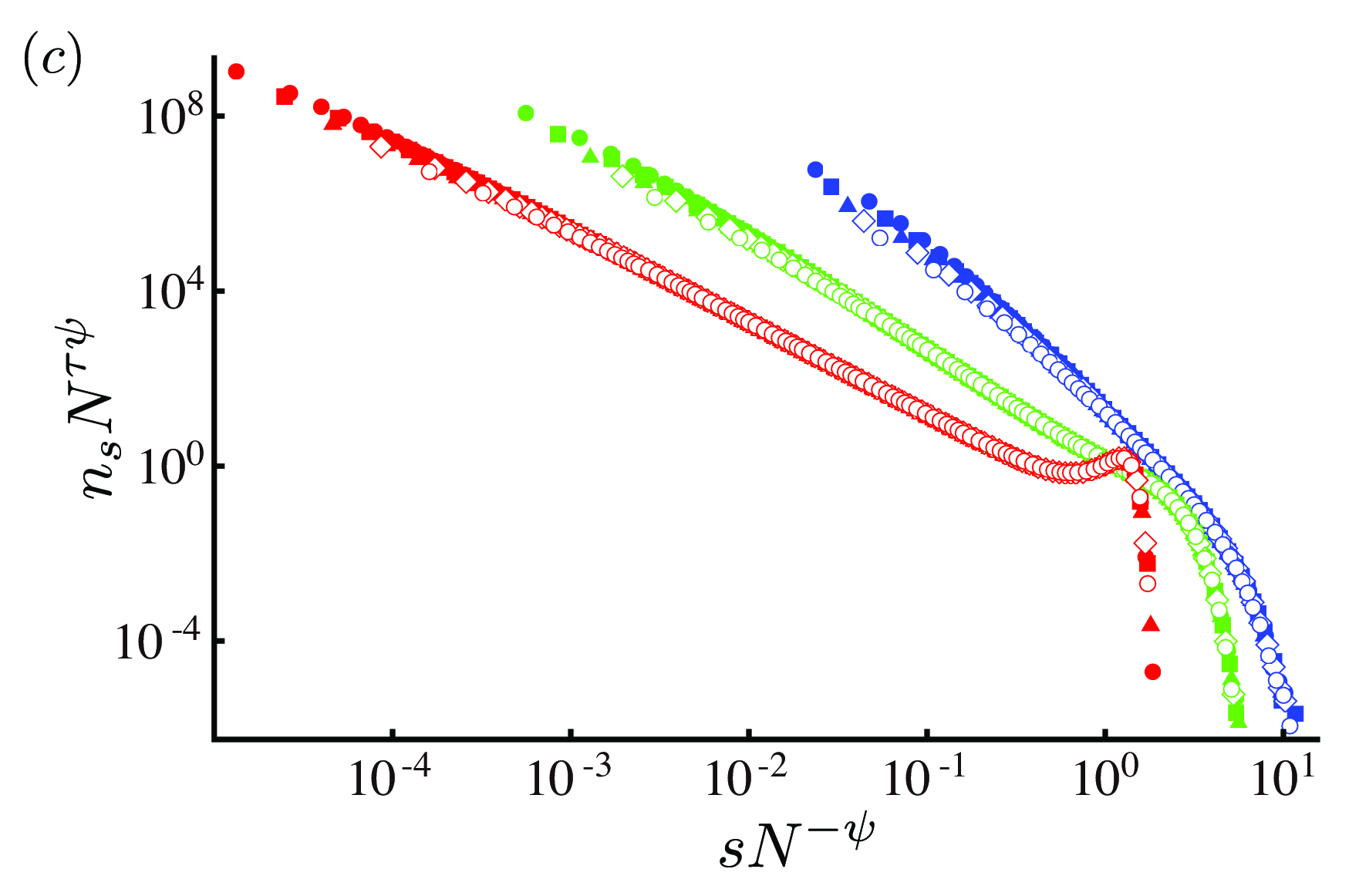}
\end{center}
\caption{
Finite size scaling of $n_s$ for the GRT with (a) $\dgamma=3$, (b) $\dgamma=4$, and (c) $\dgamma=\infty$.
In each panel, we set $p=0.3$ (blue symbols), $p=0.6$ (green symbols), and $p=0.9$ (red symbols).
}
\label{fig:tree:ns}
\end{figure}

To check this analytical prediction, we performed the Monte Carlo simulations.
The number of nodes is taken from $2^{14}$ to $2^{18}$.
To average each data at a given value of $p$, 
we generated 1,000 graph samples and generated the bond percolation 1,000 times on each sample.
Figure \ref{fig:tree:psi} plots the numerical result of $\proot(p,N)$ 
and its analytical prediction (\ref{eq:GRtree}). 
We see that the numerically-obtained $\proot(p)$ corresponds well with the line generated by (\ref{eq:GRtree}) 
for $p \gtrsim 0.2$.
The deviation seen for $p\lesssim 0.2$ tends to diminish with increasing $N$ (not shown).
We also measure the fractal exponent $\psi(p)$ of the largest cluster, as shown in Fig.~\ref{fig:tree:psi}.
The estimated value of $\psi(p)$ shows good agreement with $\proot(p)$, 
which indicates that the root cluster is one of the infinite clusters (for $N\to \infty$ ) with finite probability, 
as well as being the largest cluster.
Figure \ref{fig:tree:ns}(a) shows our finite-size scaling (\ref{eq:ns-scaling}) for $n_s(p)$ on the GRT with $\dgamma=3$.
The scaling is quite good over a wide range of $p$.
We also find similar results for other values of $\dgamma$, 
as shown in Figs.~\ref{fig:tree:ns}(b) and (c).

\subsubsection{Percolation on the GRN}

\begin{figure}[!h]\begin{center}
\includegraphics[width=5.5cm]{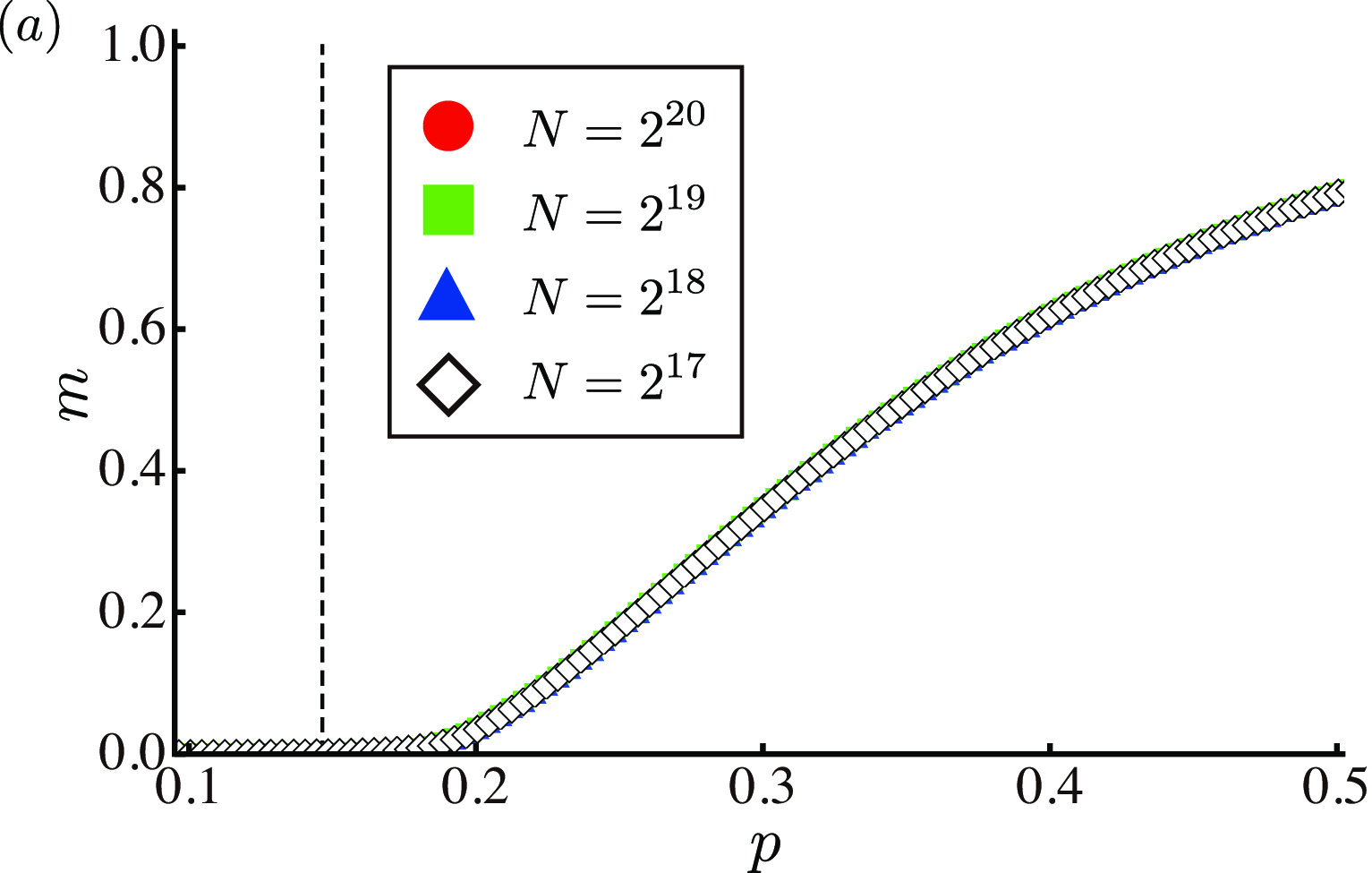}
\includegraphics[width=5.5cm]{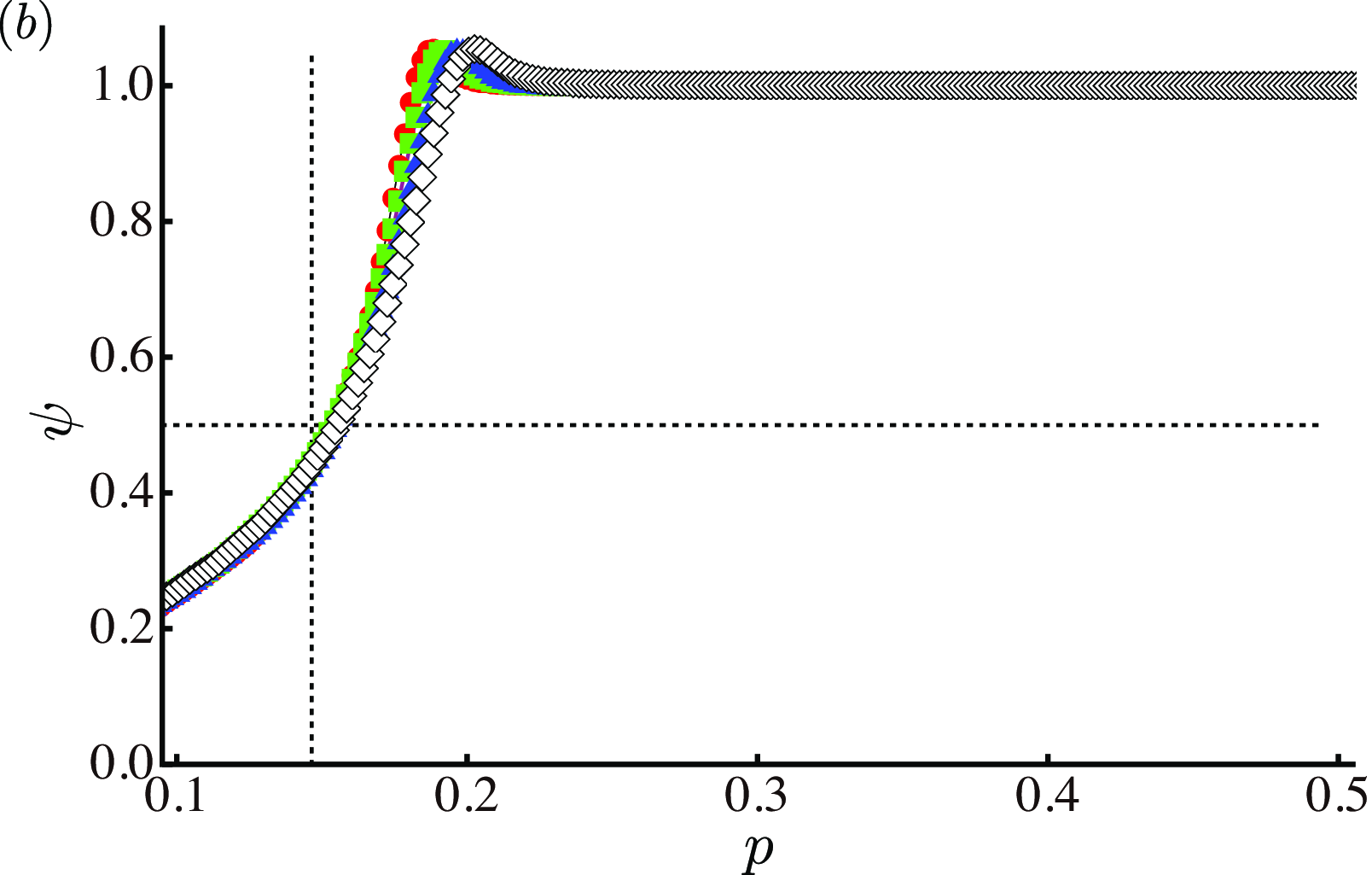}
\includegraphics[width=5.5cm]{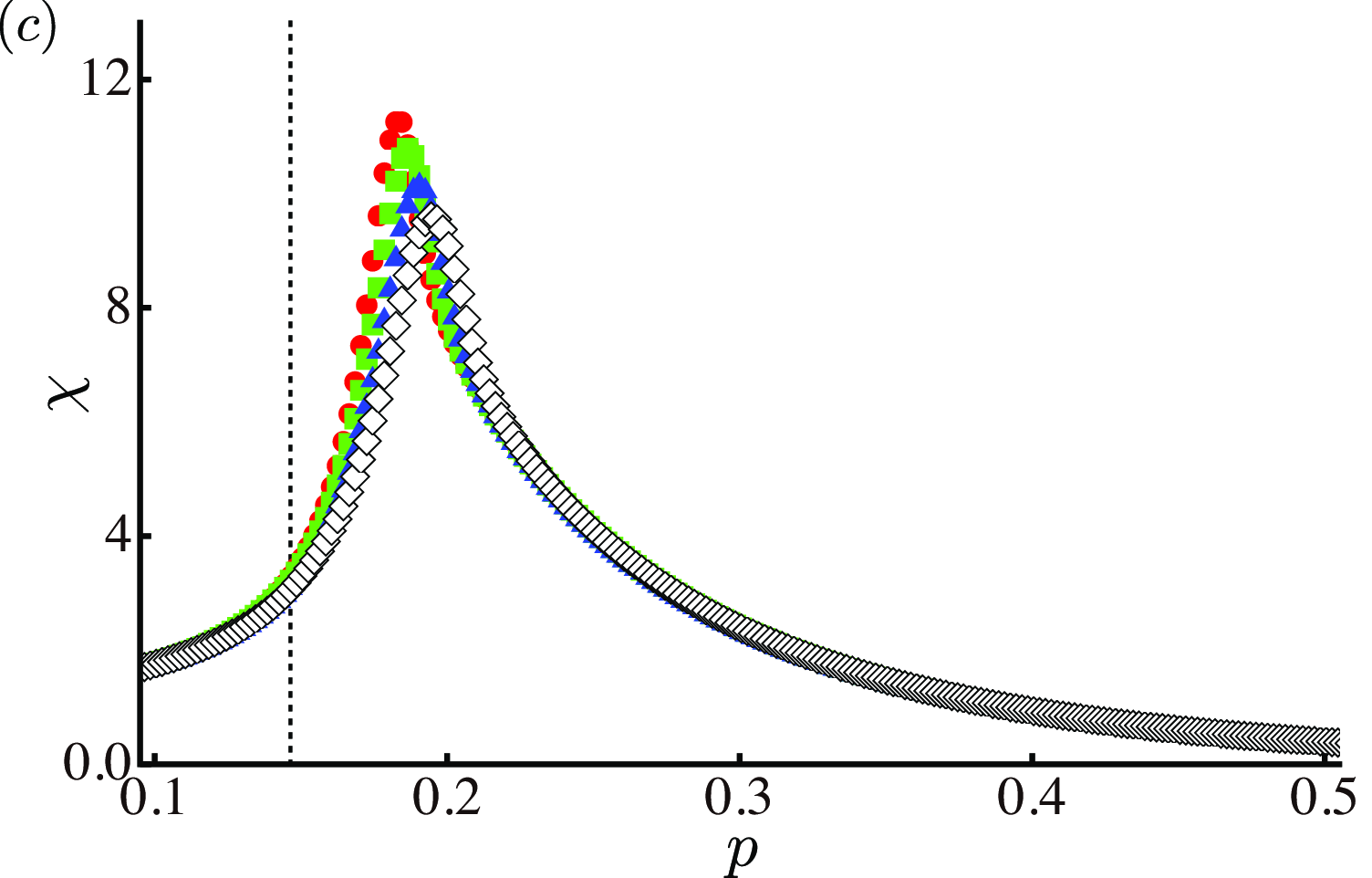}
\end{center}
\caption{ 
(a) The order parameter $m(p, N)$, (b) the fractal exponent $\psi(p, N)$, and (c) the susceptibility $\chi(p,N)$ of the $m$-out graph. 
The vertical dashed line represents $p_{c2}=(1-1/\sqrt{2})/2$, and the horizontal dashed line represents $\psi_c = 1/2$.
}
\label{fig:GRN:m}
\end{figure}

\noindent
Next, we consider the case of $m>1$. 
In \cite{hasegawa2013profile}, we studied percolation on the $m$-out graph (the GRN with $k_0 \to \infty$).
As given in \cite{riordan2005small,bollobas2005slow} after \cite{zalanyi2003properties}, 
the percolation threshold $\pu$ of the $m$-out graph is given as
\begin{equation}
\pu =\frac{1}{m}\Big(1-\sqrt{\frac{m-1}{m}}\Big),
\end{equation}
and $m(p)$ follows Eq.(\ref{eq:infiniteorder}) with $\alpha=\pi/2[m(m-1)]^{1/4}$ and $\beta^\prime=1/2$.
In \cite{krapivsky2004universal}, Krapivsky and Derrida also derived a similar result on a generalized model 
and showed that 
the cluster size distribution $n_s$ is a power-law in the whole region below $\pu$, meaning $\pl=0$.
Moreover, we have $\psi(\pu)= 1/2$ from the power-law behavior of $n_s$ at $p_{c2}$ (see Eq.(6) in \cite{krapivsky2004universal}).
For Monte Carlo simulations, we set $m_0=3$ and $m=2$.
The number of graph realizations is 1,000 and the number of percolation trials on each realization is 100.
The order parameter $m(p, N)$ and the fractal exponent $\psi(p, N)$
on the $m$-out graph are shown in Fig.~\ref{fig:GRN:m}.
For each value of $p$ below $p_{c2}$, $\psi(p, N)$ almost converges to a certain value, 
while $\psi(p, N)$ for $p>p_{c2}$ varies very slowly but approaches unity with increasing $N$.
In spite of our extensive simulations, 
$\psi(p, N)$ at $p_{c2}$ looks slightly smaller than $\psi_c$ 
due to a logarithmic correction in the power law of $n_s$ \cite{krapivsky2004universal}.
However, $\psi(p)$ grows continuously with $p$ for $0(=p_{c1})<p<p_{c2}$ up to $\psi_c \simeq 1/2$ at $p_{c2}$, 
and then jumps to $\psi=1$.

\begin{figure}[!h]\begin{center}
\includegraphics[width=5.5cm]{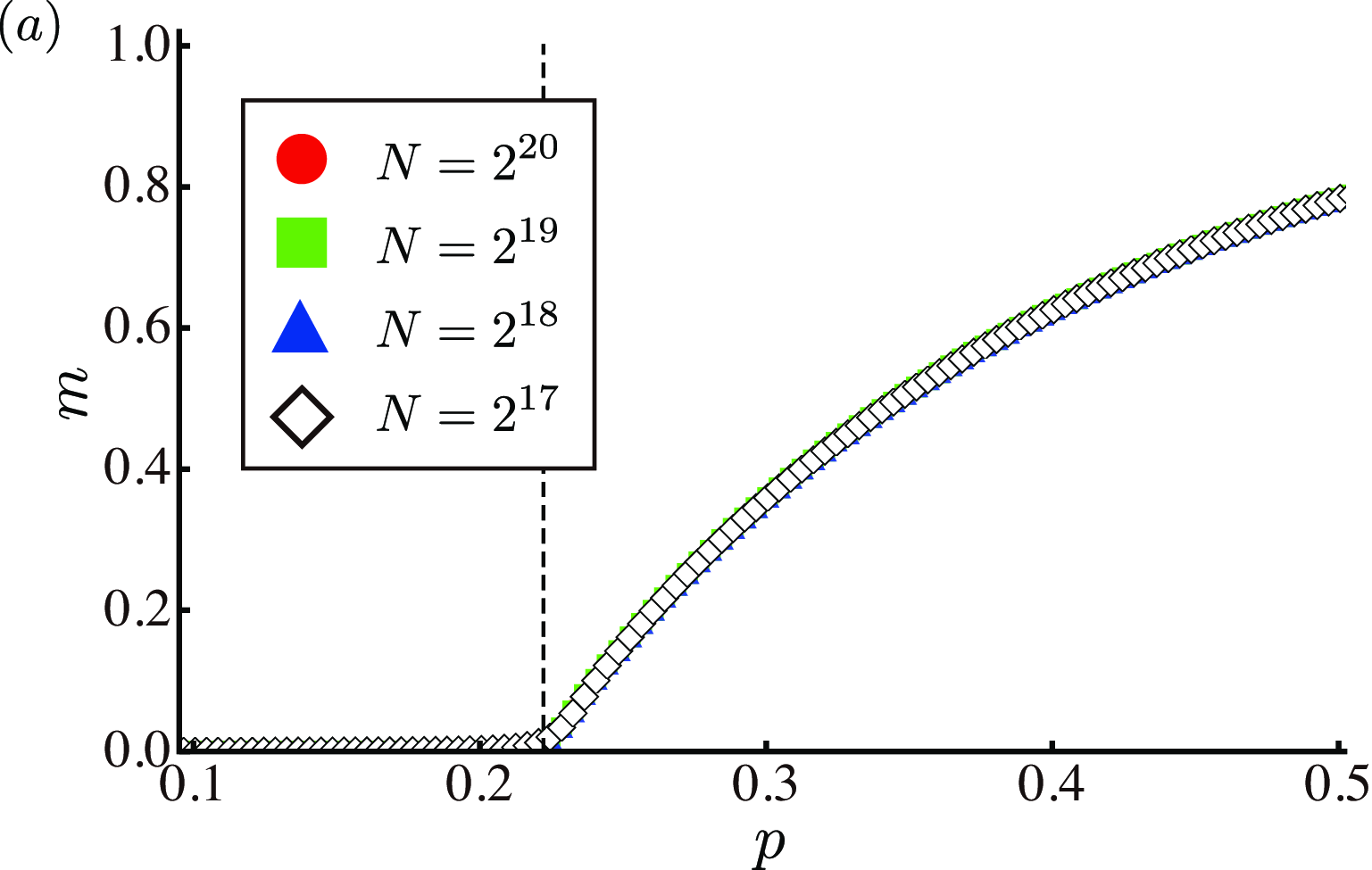}
\includegraphics[width=5.5cm]{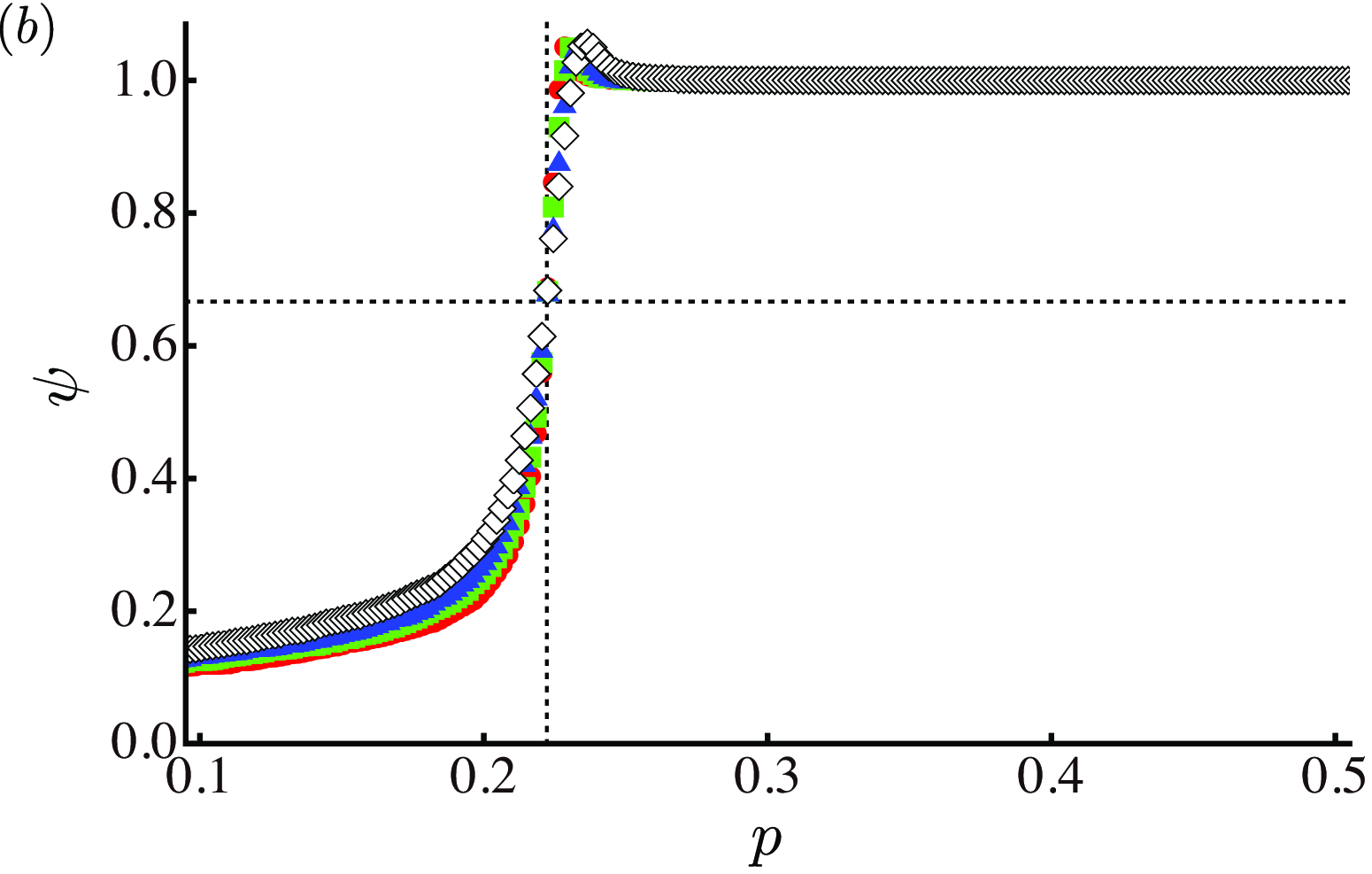}
\includegraphics[width=5.5cm]{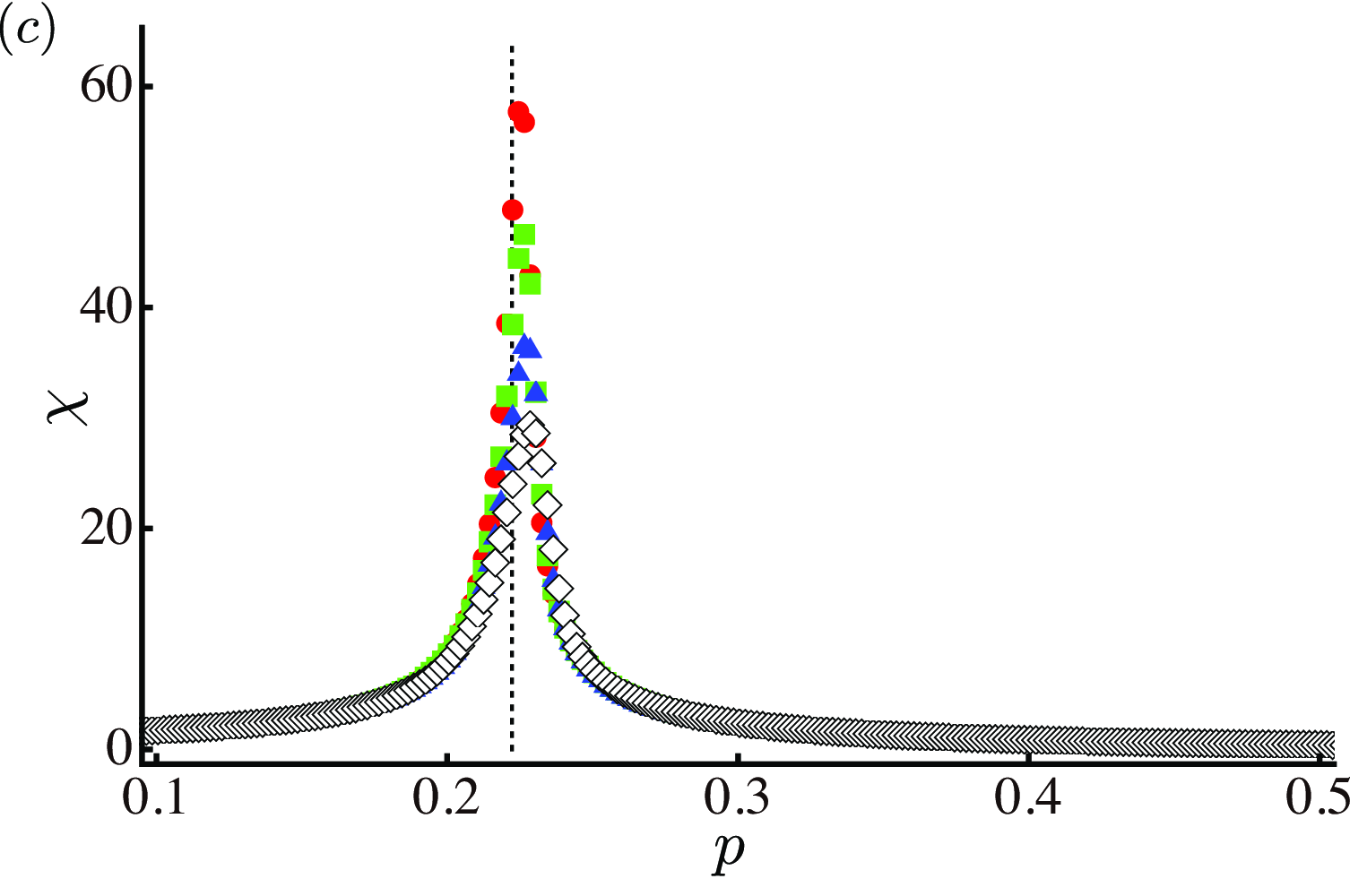}
\end{center}
\caption{
(a) The order parameter $m(p, N)$, (b) the fractal exponent $\psi(p, N)$, 
and (c) the susceptibility $\chi(p,N)$ of the configuration model.
The vertical dashed line represents $p_{c2}=2/9$, and the horizontal dashed line represents $\psi_c=2/3$.
}
\label{fig:GRN:m_config}
\end{figure}

\begin{figure}[!h]\begin{center}
\includegraphics[width=5.5cm]{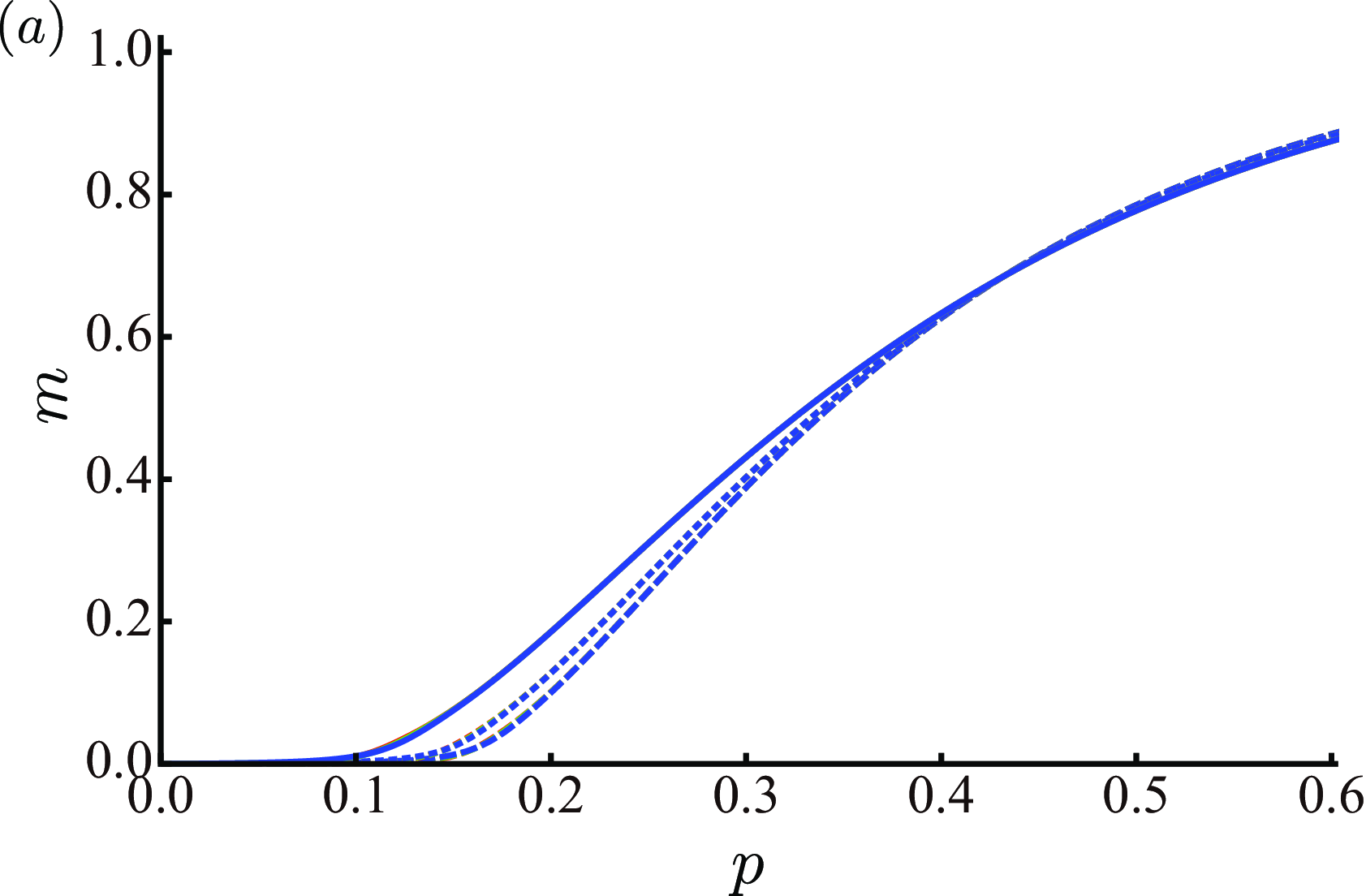}
\includegraphics[width=5.5cm]{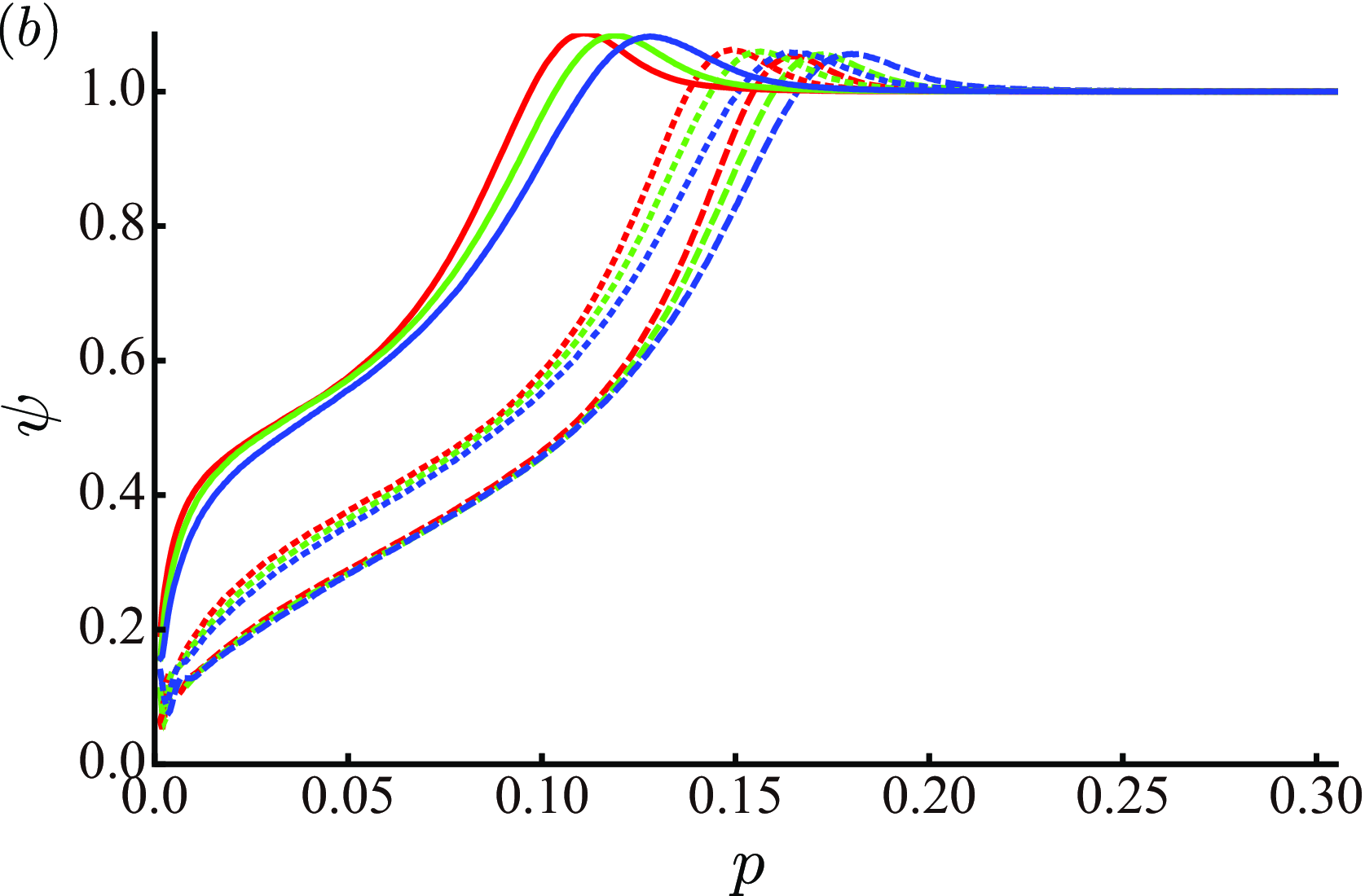}
\includegraphics[width=5.5cm]{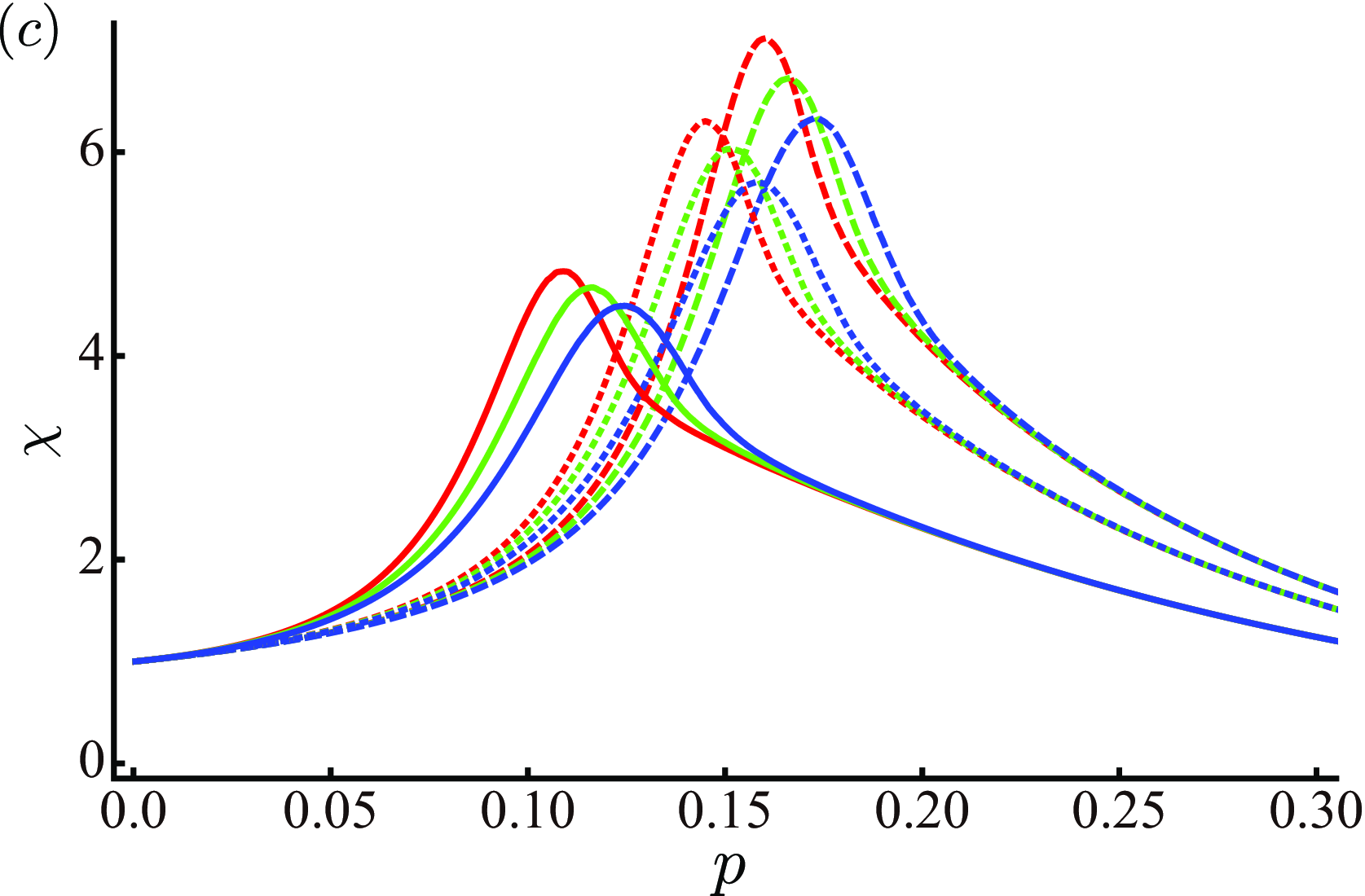}
\end{center}
\caption{
(a) The order parameter $m(p, N)$ and (b) the fractal exponent $\psi(p, N)$, 
and (c) the susceptibility $\chi(p,N)$ of the GRN with 
$\dgamma=3$ (solid lines), $\dgamma=4$ (dotted lines), and $\dgamma=5$ (dashed lines). 
The number of nodes is $N=2^{17}$ (red), $2^{16}$ (green), and $2^{15}$ (blue).
}
\label{fig:GRN:general}
\end{figure}

\noindent
Apparently, the susceptibility apparently has a larger peak when the network is larger (Fig.~\ref{fig:GRN:m}(c)), 
but it does not diverge in the limit $N \to \infty$. 
Analytical results \cite{zalanyi2003properties} show that the mean cluster size has a finite jump at $\pu$.

For comparison, we also performed simulations for the configuration model that has the same degree distribution as the $m$-out graph.
In this case, the critical phase shrinks to a unique critical point $p_c$. 
By using the local tree approximation, we have $p_c=2/9$, and the critical exponents in the mean field universality class, $\beta=1$.
In Fig.~\ref{fig:GRN:m_config}, $m(p,N)$, $\psi(p, N)$, and $\chi(p, N)$ on the configuration model are plotted as a function of $p$.
In this model, $\psi(p, N)$ of various size cross at $(p_c, \psi_c) = (2/9, 2/3)$.
Here the largest cluster size, just at the transition point, is of $O(N^{2/3})$, which is also observed on the Erd\"os-R\'enyi model~\cite{janson1993birth}.
In the limit $N \to \infty$, $\psi=0(=1)$ for $p<p_c (>p_c)$, 
which indicates that the transition at $p_c$ is between the nonpercolating phase and the percolating phase.
The difference between the $m$-out graph and its randomized version leads us to an interesting question. 
What is the geometrical origin of the critical phase in complex networks?
The present result means that the standard network properties, {\it i.e.}, 
the degree distribution, the mean shortest path length, and the clustering coefficient 
are not the answer because these are essentially the same between the $m$-out graph and the corresponding configuration model. 
The answer will be proposed in future studies.

In Fig.~\ref{fig:GRN:general}, we show the fractal exponent for GRN with finite $\dgamma$. 
The behavior is same as that of the case without a preferential attachment, except for the value of $\pu$. 
As $\dgamma$ decreases, the fractal exponent approaches one at smaller $p$. 
As for the case of $\dgamma=3$ (the BA model), $\psi(p,N)$ seems not to converge and to increase with $N$ in the whole region of $p$.
Similar to the configuration model, $\pu$ may be zero when $\dgamma \le 3$, 
although a numerical estimation would be difficult to obtain since $\pu$ is very small (if exists).

\subsection{Deterministically growing networks} 

\noindent
The critical phase is understood as a set of fixed points of a renormalization group (RG).
Hierarchical networks have a great advantage in analytical treatments because 
the structural properties of, and cooperative behaviors on, networks 
can be analyzed from real space RGs.
As mentioned before, hierarchical small-world networks (deterministically growing networks) 
are expected to have only the critical phase and the percolating phase.
This is not the case for hierarchical ``large-world'' networks 
and also some hierarchical small-world networks, 
such as the Apollonian network \cite{auto2008finite} and the Dorogovtsev-Goltsev-Mendes network \cite{dorogovtsev2003renormalization}, 
which have only the percolating phase due to their strong network heterogeneity.
Here, we consider the percolation on the (2,2) flower, 
also known as the diamond lattice or Migdal-Kadanoff lattice in the field of statistical physics, 
and the decorated flower, which is generated by adding the shortcuts to the (2,2) flower.
The former is a large-world network, whereas the latter a small-world network.
In \cite{hasegawa2010generating}, the authors analyzed these models by using generating functions 
to obtain the root cluster size and the cluster size distribution. 
Our results show that the (2,2) flower has nonpercolating and percolating phases,
 while the decorated flower has critical and percolating phases.

\begin{figure}[t]\begin{center}
\includegraphics[width=7.5cm]{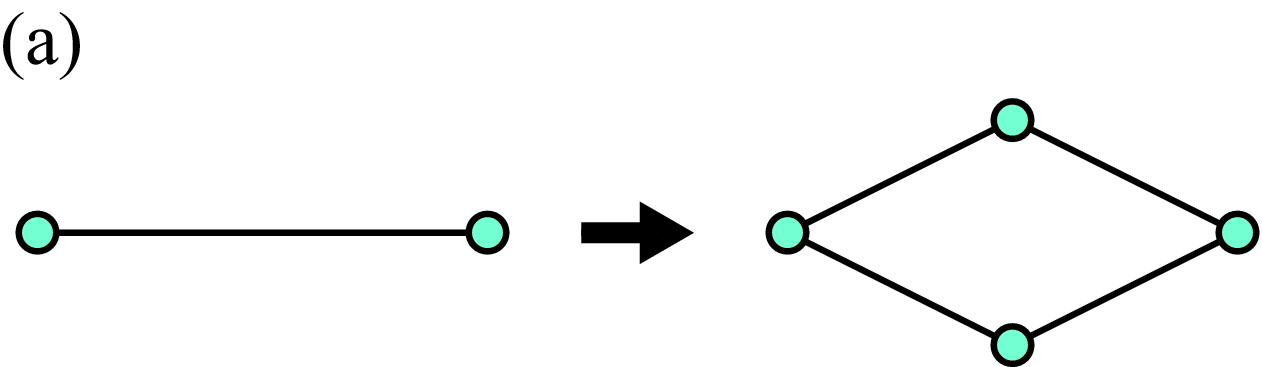}
\hspace{0.5cm}
\includegraphics[width=7.5cm]{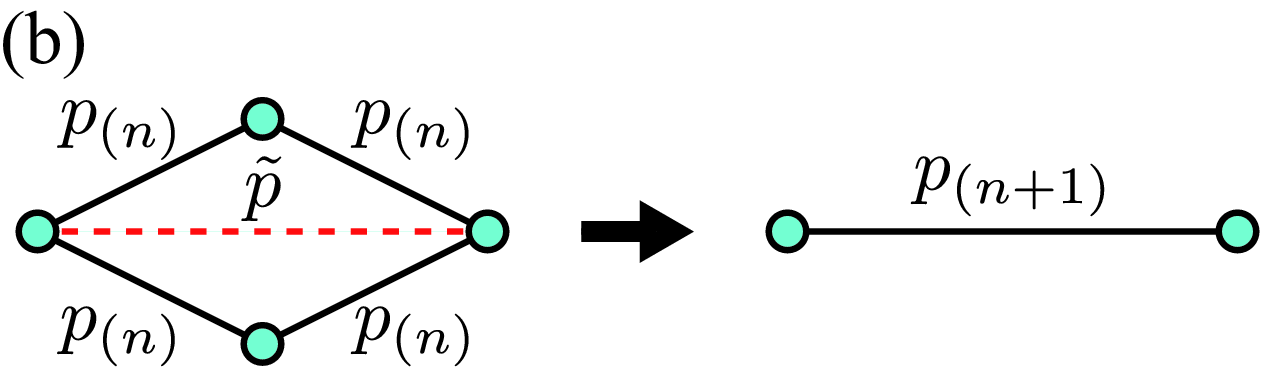}
\end{center}
\caption{
(a) Bond replacement rule of the (2,2) flower and (b) schematic for the decimation of the decorated flower.
}
\label{fig:flower:rule}
\end{figure}

\begin{figure}[t]\begin{center}
\includegraphics[width=120mm]{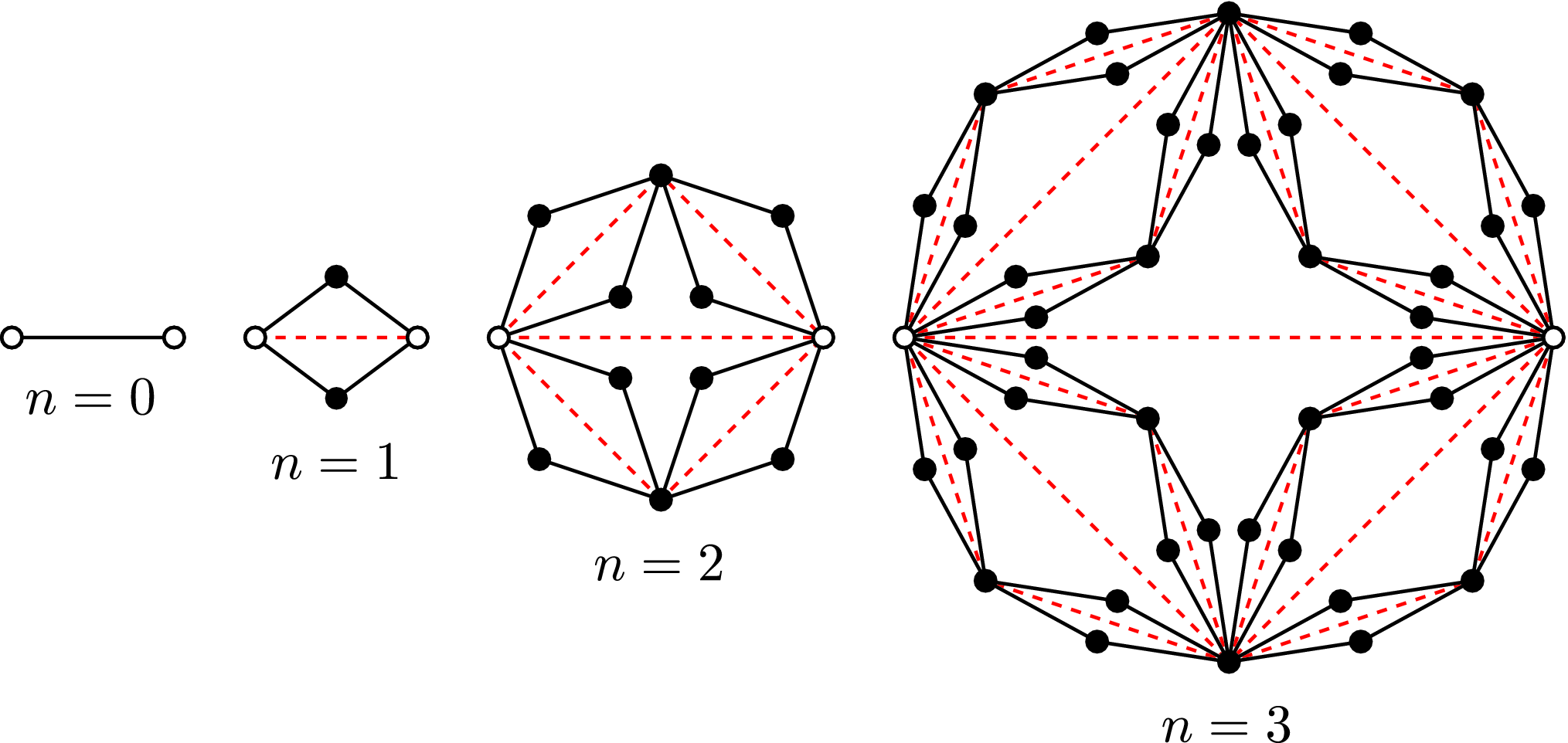}
\end{center}
\caption{
Realizations of the (2,2) flower and the decorated flower with $n=0$--$3$. 
The decorated flower is obtained by adding the shortcuts (dotted lines) to the (2,2) flower. 
Note that in each iteration, the shortcuts are not replaced.
The open circles are called the roots.
}
\label{fig:flower:example}
\end{figure}

\subsubsection{Construction of flowers}

\noindent
The (2,2) flower with generation $n$, denoted by F$_n$, is recursively constructed as follows \cite{rozenfeld2007fractal,rozenfeld2007percolation}:
At generation $n=0$, the flower F$_0$ consists of two nodes connected by a bond. 
We call these nodes the {\it roots}.
For $n \ge 1$, F$_n$ is obtained from F$_{n-1}$, such that each existing bond in F$_{n-1}$ is replaced by 
two parallel paths consisting of two edges and one node each (Fig.~\ref{fig:flower:rule}(a)).
In Fig.~\ref{fig:flower:example}, we show realizations of F$_n$ with $n=$ 1, 2, and 3.
The network properties of the (2,2) flower are given in \cite{rozenfeld2007fractal,rozenfeld2007percolation}: 
(i) the number of nodes of F$_n$, $N_n$, is $N_n=4^n (2/3)+4/3$, 
(ii) the number of edges is $4^n$,
(iii) the degree distribution is $P(k) \propto k^{-3}$, 
(iv) the clustering coefficient $C$ is zero, 
and (v) the network is {\it not} small-world because the diameter of F$_n$ is as $2^n \propto \sqrt{N}$,
which increases as a power of $N$ like the finite-dimensional Euclidean lattice.
A hierarchical small-world network, which we call the decorated flower $\tilde{\rm F}_n$, 
is achieved by adding some long-range bonds to F$_n$, as shown in Fig.~\ref{fig:flower:example}. 
The decorated flower is also regarded as a deterministically growing network: 
the network starts from single bond between two nodes at time (= generations) $n=0$, 
and grows with time such that every edge in $\tilde{\rm F}_n$ at time $n$ adds two new nodes, 
which link to both end nodes of the edge, to create $\tilde{\rm F}_{n+1}$.
The decorated flower has both the small-world property $\bar{\ell} \propto \log N$ and a high clustering coefficient $C \simeq 0.82$, 
while also keeping a power-law degree distribution $P(k) \propto k^{-3}$ 
\cite{rozenfeld2007fractal,hinczewski2006inverted}.

\begin{figure}\begin{center}
\includegraphics[width=6.5cm]{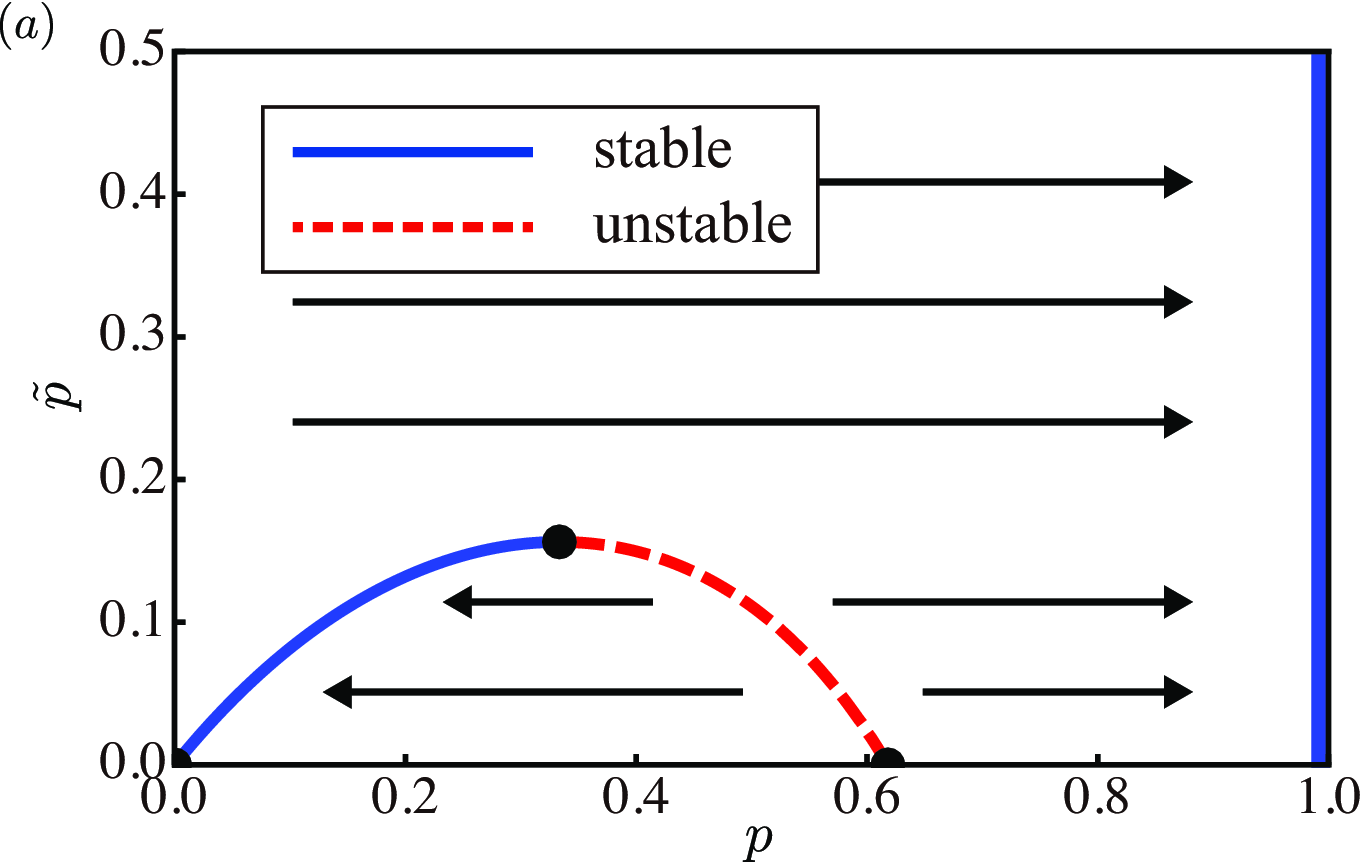}
\includegraphics[width=6.5cm]{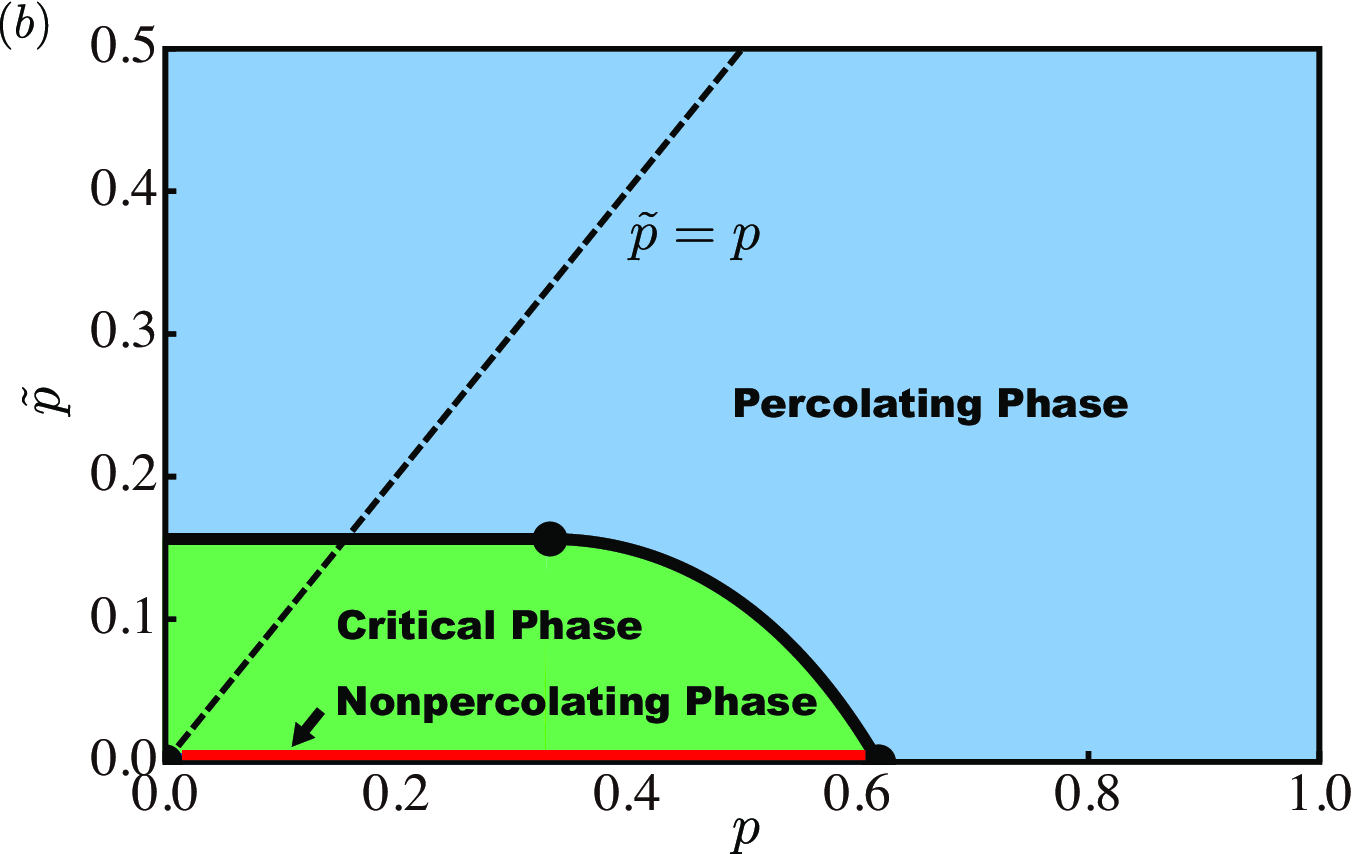}
\end{center}
\caption{
(a) RG flow diagram and (b) the phase diagram. 
}
\label{fig:flower:diagram}
\end{figure}

\subsubsection{RG analysis}

\noindent
We consider bond percolation on $\tilde{\rm F}_n$, 
with open bond probability $p$ of the ``short-range'' bonds constituting F$_n$ 
and that of the ``long-range'' bonds $\tilp$, being given independently. 
The cases of $\tilde{p}=0$ and $\tilp=p$ correspond 
to the uniform bond percolation on F$_n$ and $\tilde{\rm F}_n$, respectively.

The phase diagram is given by the RG technique \cite{rozenfeld2007percolation,berker2009critical,hasegawa2010generating}.
We denote by $p_{(n)}$ the open bond probability of the short-range bonds after $n$ renormalizations.
The recursion relation for $p_{(n)}$ is obtained by replacing each unit of the flower 
by a renormalized short-range bond (Fig.~\ref{fig:flower:rule}(b)) as 
\begin{equation}
p_{(n+1)}= 1 - (1-\tilp)(1-p_{(n)}^2)^2. \label{eq:flower:RG}
\end{equation}
Here, the initial condition is $p_{(0)}=p$. 
Note that probability $\tilp$ for long-range bonds is not renormalized.
This recursion equation has trivial stable fixed points at $p=1$ for arbitrary $\tilp$ and $(p,\tilp)=(0,0)$. 
The region where flow converges onto a fixed point at $p=1$ and $p=0$ 
corresponds to the percolating phase and the nonpercolating phase, respectively. 
Other nontrivial fixed points, $p_*=p_{(n)}=p_{(n+1)}$, are given from the solution of 
\begin{eqnarray}
p_*= 1 - (1-\tilp)(1-{p_*}^2)^2.
\label{eq:flower:FP}
\end{eqnarray}
For a fixed $\tilp$ in $0 < \tilp<\tilp_c=5/32$, 
there is one stable point $p=\fps(\tilp)$ and one unstable fixed point $p=\fpu(\tilp) >\fps(\tilp)$.
RG flow starting at $0 \le p < \fpu(\tilp)$ goes to $\fps(\tilp)$, and thus this region is regarded as the critical phase. 
The region for $p > \fpu(\tilp)$, where flow goes to $p=1$, corresponds to the percolating phase. 
Thus the curve $(\tilp, \fpu(\tilp))$ gives the phase boundary.
Two fixed lines of $\fps$ and $\fpu$ terminate at $\tilp=\tilp_c$. 
Then, the line from $p=0$ to $p=1/3$ at $\tilp=\tilp_c$ also gives the phase boundary.
For $\tilp>\tilp_c$ there is only one stable fixed point at $p=1$, so that the system is always percolating.
In Fig.~\ref{fig:flower:diagram}, we show the RG flow diagram and the phase diagram.
The case of $\tilp=0$ indicates that bond percolation on ${\rm F}_n$ has 
the nonpercolating phase for $p<p_c=(\sqrt{5}-1)/2$ and the percolating phase for $p>p_c$, 
while the case of $\tilp=p$ indicates that bond percolation on $\tilde{\rm F}_n$ has 
the critical phase for $p<\pu=\tilp_c$ and the percolating phase for $p>\pu$, but no nonpercolating phase, {\it i.e.}, $\pl=0$.

\begin{figure}\begin{center}
\includegraphics[width=6.5cm]{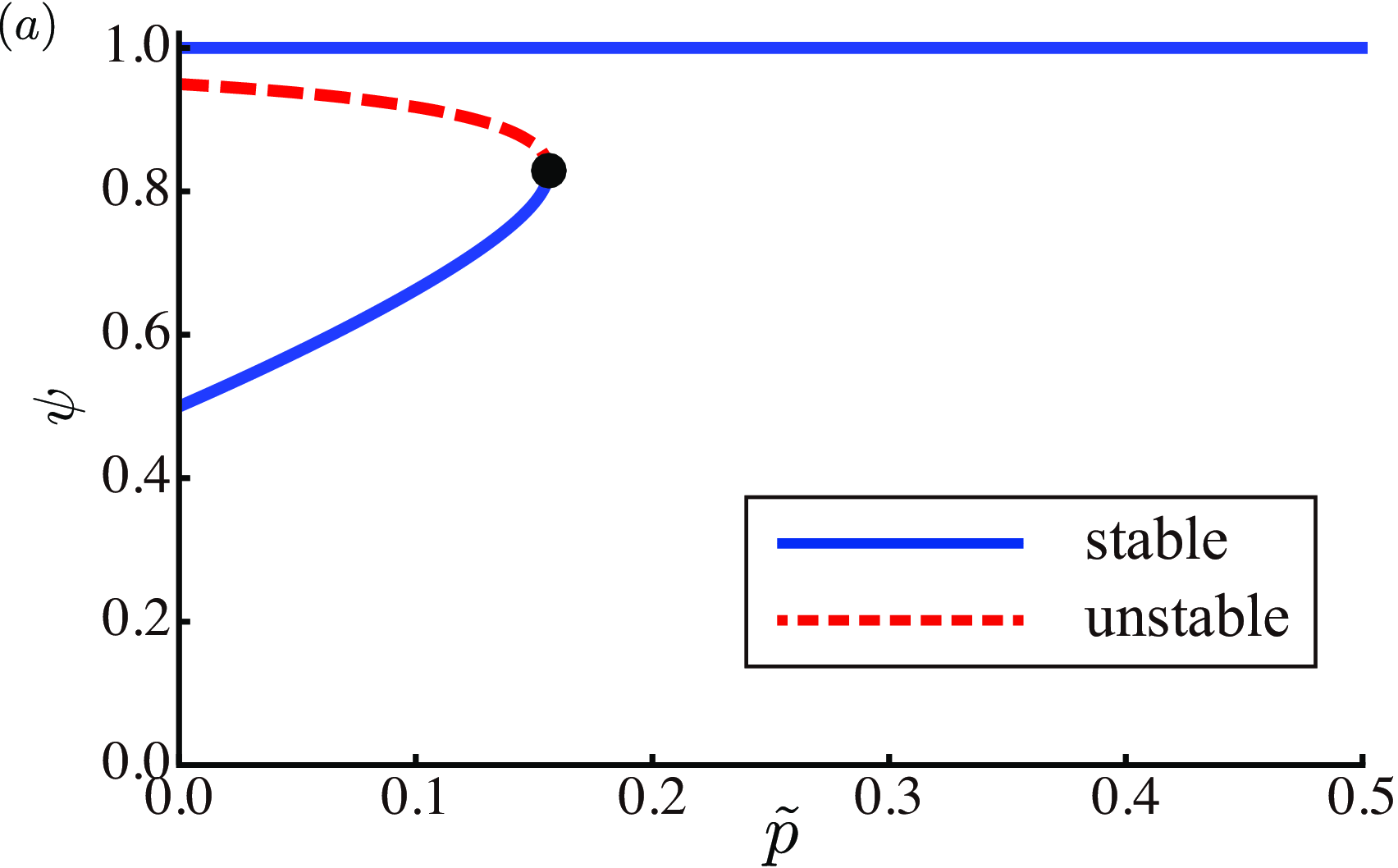}
\includegraphics[width=6.5cm]{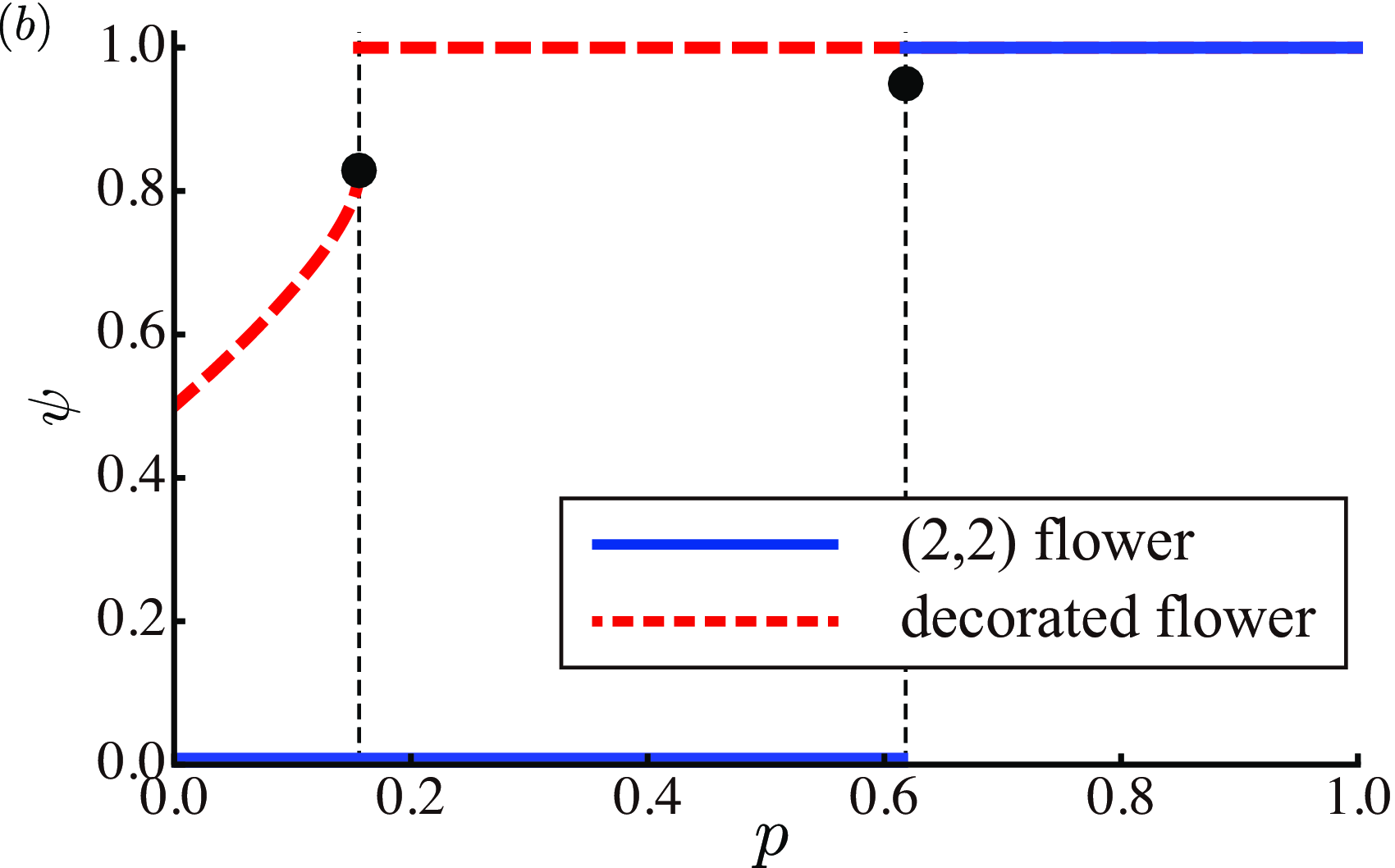}
\end{center}
\caption{
(a) $\tilde{p}$-dependence of $\proot$ and 
(b) $p$-dependence of $\proot$ on F$_n$ ($\tilp=0$) and $\tilde{\rm F}_n$ ($\tilp=p$). 
The vertical lines in (b) indicate $\pu=5/32$ of $\tilde{\rm F}_n$ and $p_c=(\sqrt{5}-1)/2$ of F$_n$.
}
\label{fig:flower:psi}
\end{figure}

\begin{figure}\begin{center}
\includegraphics[width=6.5cm]{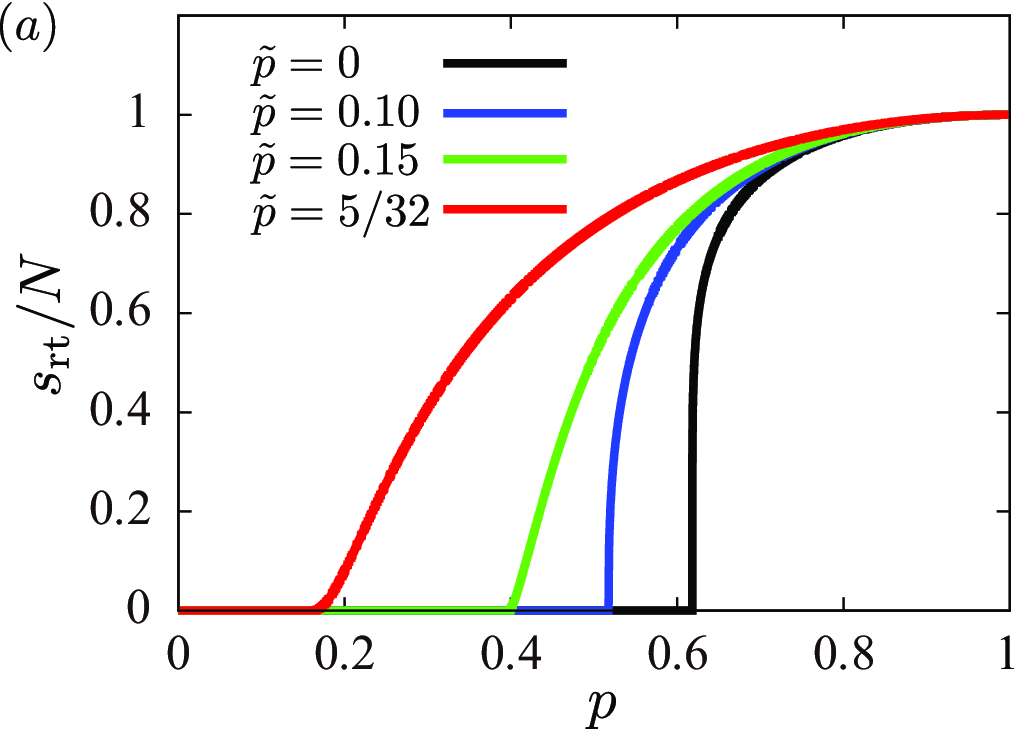}
\includegraphics[width=6.5cm]{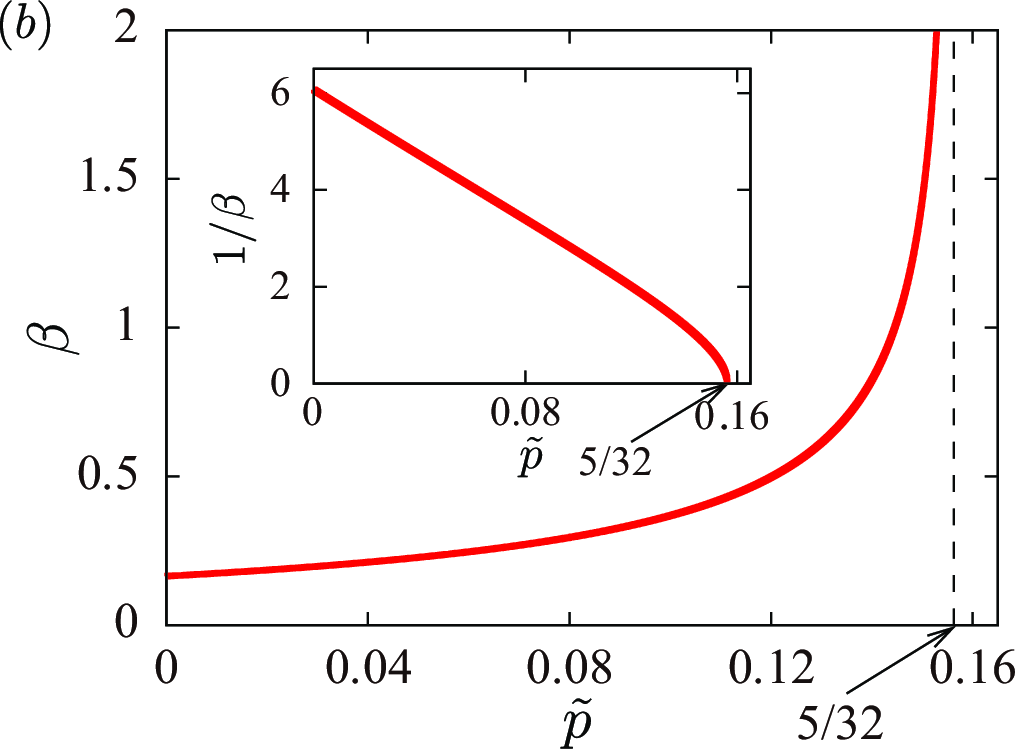}
\end{center}
\caption{
(a) Order parameter $\sroot(p,N_n)/N_n$ on $\tilde{\rm F}_n$ with several values of $\tilp$ 
and (b) $\tilde{p}$-dependence of the critical exponent $\beta$.
}
\label{fig:flower:beta}
\end{figure}

Next, we utilize generating functions to calculate the fractal exponent $\proot(p)$ of the root cluster.
The details of the technique are given in \cite{hasegawa2010generating} and we can obtain
the recursion relations of the mean fraction of the clusters, 
including two roots, $\tau_n$,
and including either of the roots, $\sigma_n$, as
 \begin{eqnarray}
\begin{pmatrix}
    \sigma_{n+1} \\
    \tau_{n+1} 
\end{pmatrix}
  &\simeq&
\begin{pmatrix}
    \frac{1}{2}(1-\tilp)(1-p_*)(1+p_*)^2 & (1-\tilde{p})(1-p_*)^2(1+p_*) \\
    \frac{1}{2}(1+p_*)\left[p_*^2+\tilp (1-p_*)(1+p_*)\right] & 1-(1-\tilp)(1-p_*)^2(1+p_*) 
   \end{pmatrix}  
   \begin{pmatrix}
    \sigma_{n} \\
    \tau_{n} 
   \end{pmatrix}  
  \quad \mbox{for $n \gg 1$}.
  \label{Matrix_inf}
 \end{eqnarray}
Since the fixed points $p_*(<1)$ satisfy Eq.~(\ref{eq:flower:FP}), the recursion relation (\ref{Matrix_inf}) is then reduced to 
\begin{equation}
   \begin{pmatrix}
    \sigma_{n+1} \\
    \tau_{n+1} 
   \end{pmatrix}  
=
\begin{pmatrix}
   \frac{1}{2} & \alpha\\
   \frac{1}{2}p_* & 1-\alpha
  \end{pmatrix}  
   \begin{pmatrix}
    \sigma_{n} \\
    \tau_{n} 
   \end{pmatrix},
\end{equation}
where $\alpha=(1-p_*)/(1+p_*)$.
By using the largest eigenvalue $\lambda(p_*)$ of this matrix,
\begin{eqnarray}
 \lambda(p_*)=\frac{1}{4}\left[(3-2\alpha)+\sqrt{1-4\alpha(1-2p_*)+4\alpha^2}\right],
  \label{eigen_value}
\end{eqnarray}
we can calculate the fractal exponent $\proot$ on the fixed points:
\begin{eqnarray}
 \proot=1+\frac{\ln{\lambda(p_*)}}{\ln{4}}.
  \label{eq:flower:psi}
\end{eqnarray}

Equation (\ref{eq:flower:psi}) tells us the $\tilde{p}$-dependence of $\proot$ (Fig.~\ref{fig:flower:psi}(a)):
(i) for $\tilde{p}<\tilde p_c=5/32$, $\proot$ on the (un)stable fixed points increases (decreases) with increasing $\tilde{p}$, 
and (ii) for $\tilde{p}>\tilde p_c$, $\proot$ is equal to one regardless of $p$ or $\tilde{p}$. 
The open bond probability $\tilde{p}$ of the long-range bonds essentially determines 
the degree of the criticality of the system in the critical phase.
In Fig.~\ref{fig:flower:psi}(b), we show the fractal exponent on F$_n$ ($\tilde{p}=0$) and $\tilde{\rm F}_n$ ($\tilde{p}=p$). 
For the (2,2) flower, $\proot$ is a step function: $\proot=0$ for $p<p_c=(\sqrt{5}-1)/2$, $\proot \approx 0.949644$ at $p_c$, 
and $\proot=1$ for $p>p_c$.
For the decorated flower, $\proot$ increases continuously from $\proot=1/2$ at $p=0+$ 
to $\proot \approx 0.828752$ at $\pu$, and then $\proot$ jumps to unity.

We numerically iterate the generating functions defined in \cite{hasegawa2010generating} 
to obtain the order parameter $\sroot(p,N_n)/N_n$ on $\tilde{\rm F}_n$ with arbitrary combination of $p$ and $\tilp$.
The result for $n=10^6$ is shown in Fig.~\ref{fig:flower:beta}(a).
The initial growth of the order parameter becomes moderate with increasing $\tilp$.
To examine the critical exponent $\beta$ of the order parameter on the phase boundary $p=p_c(\tilp)$, 
we follow the scaling argument in \cite{rozenfeld2007percolation} to obtain 
\begin{equation}
\beta(\tilp)=-\frac{\ln{\lambda(p_c(\tilp))}}{\ln{\Lambda(p_c(\tilp))}}, \label{beta}
\end{equation}
where
\begin{equation}
\Lambda(p_*)=\left.\frac{\partial p_{(n+1)}}{\partial p_{(n)}}\right|_{p_*}=4(1-\tilp)p_*(1-p_*^2)=\frac{4 p_*}{1+p_*}. \label{Lambda}
\end{equation}
Figure \ref{fig:flower:beta}(b) shows the $\tilp$-dependence of $\beta$.
We find that $\beta$ increases continuously with $\tilde{p}$, 
from $\beta =0.164694$ at $\tilde{p}=0$ to $\beta=\infty$ at $\tilde{p}=\tilde p_c=5/32$. 
A plausible discussion gives us that the order parameter at $\tilp=\tilp_c$ indeed follows Eq.(\ref{eq:infiniteorder}), 
{\it i.e.}, an essential singularity in the order parameter \cite{hasegawa2010generating}.

In \cite{hasegawa2010generating}, the authors also numerically evaluate the generating functions 
to obtain the cluster size distribution $n_s$ on $\tilde{F}_n$. 
The result shows that our finite-size scaling for $n_s$ is indeed well fitted on both stable and unstable fixed points. 
The scaling also works at any $p$ in the critical phase, but the convergence is not as rapid as that on the fixed points.

\begin{figure}[!t]\begin{center}
\includegraphics[width=5.2cm]{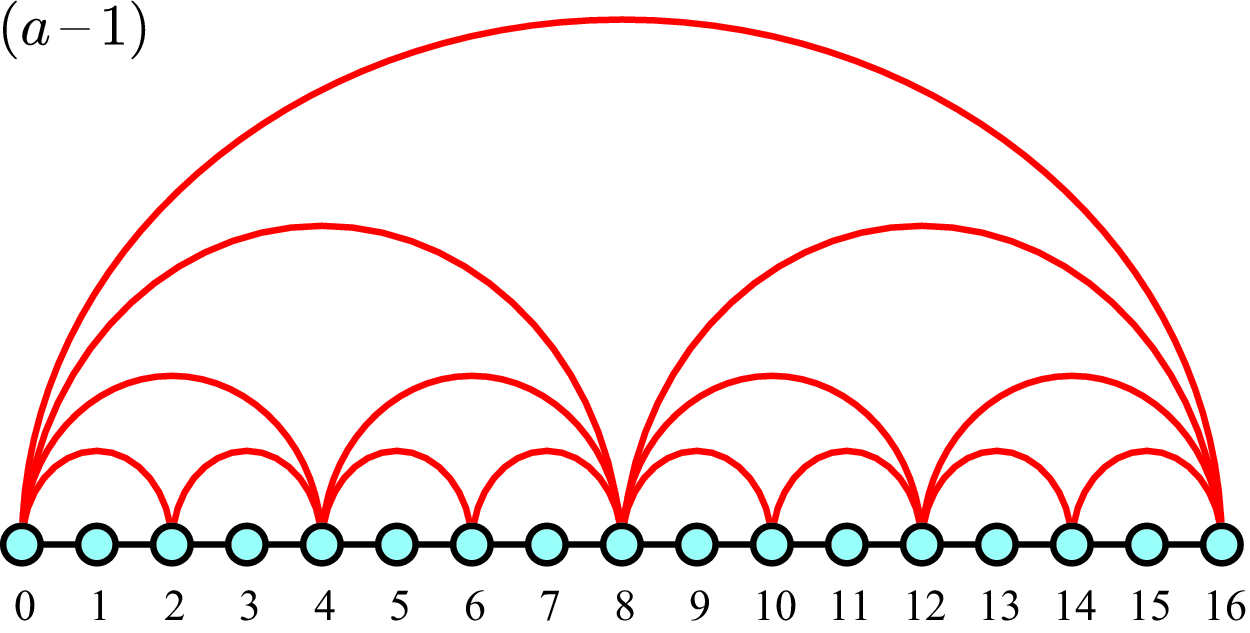}
\hspace{0.1cm}
\includegraphics[width=5.2cm]{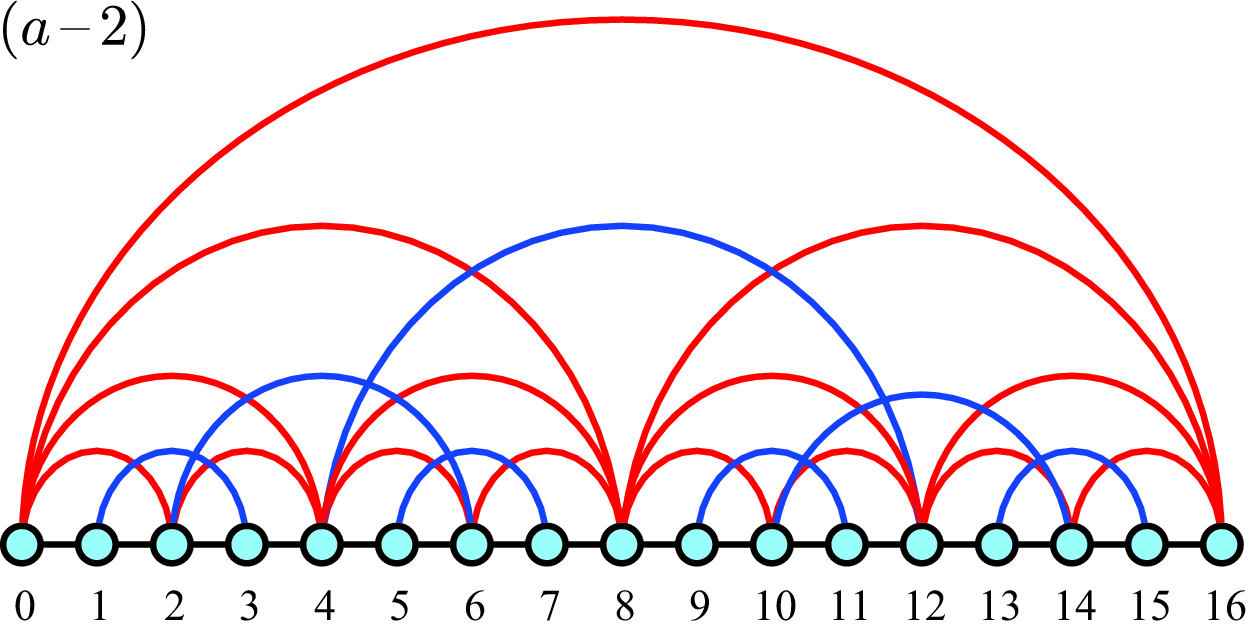}
\hspace{0.1cm}
\includegraphics[width=5.2cm]{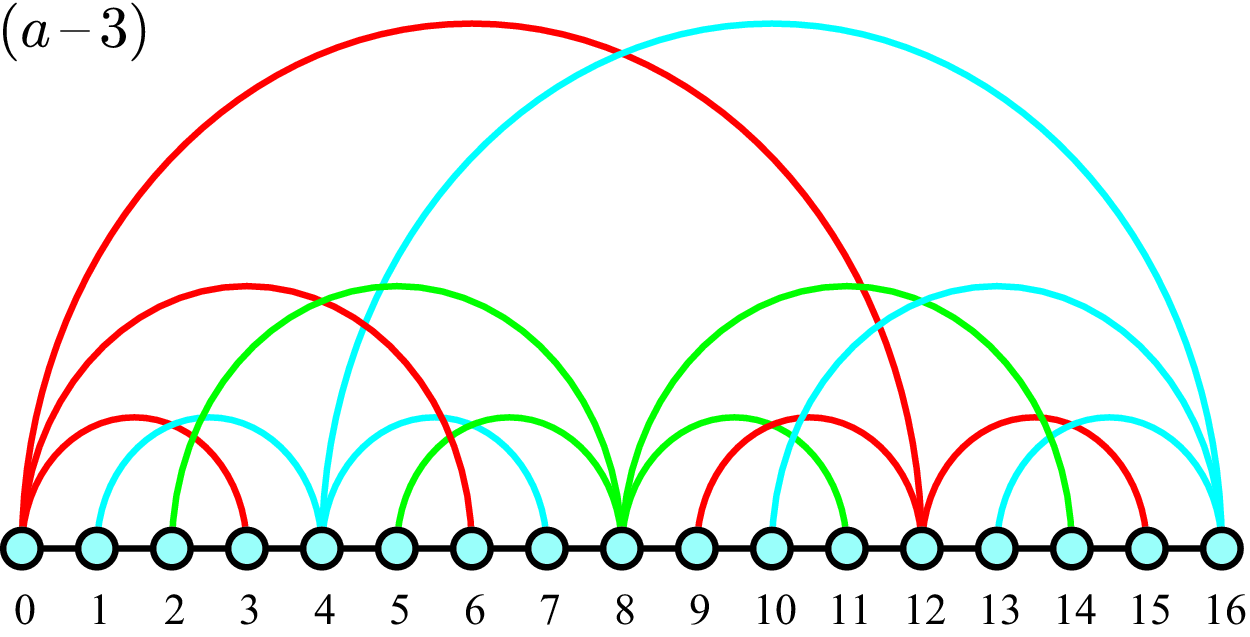}

\includegraphics[width=5.2cm]{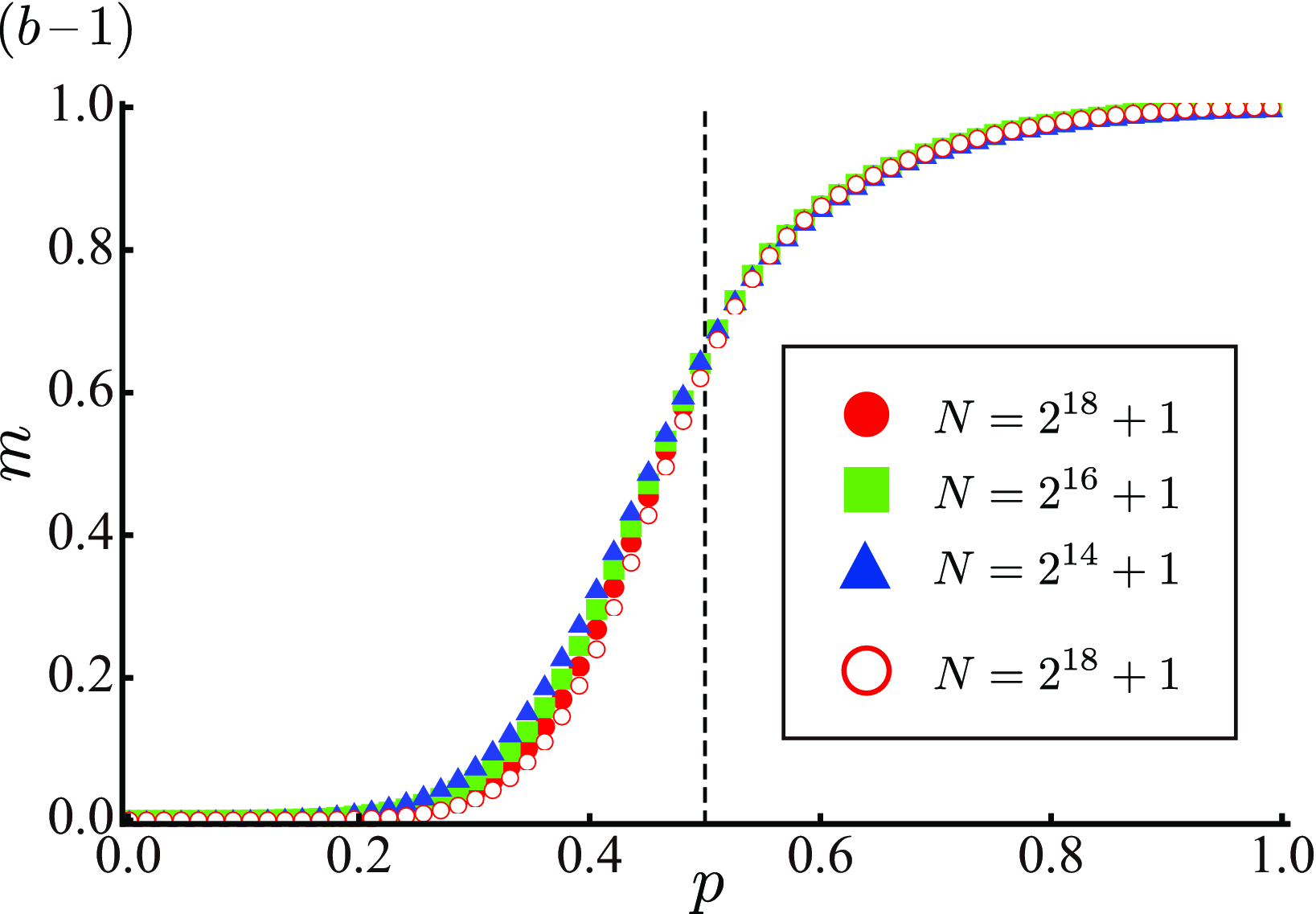}
\hspace{0.1cm}
\includegraphics[width=5.2cm]{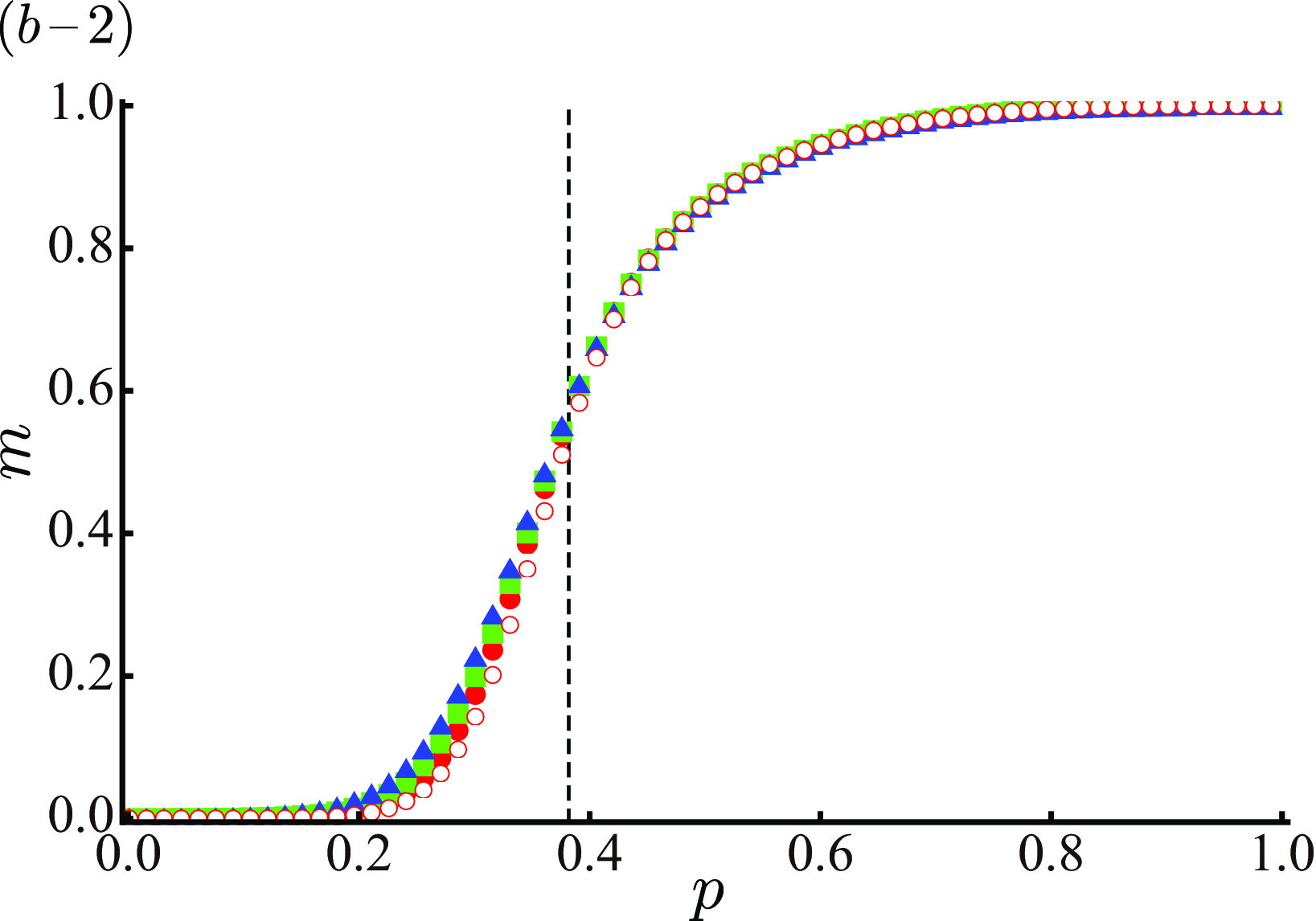}
\hspace{0.1cm}
\includegraphics[width=5.2cm]{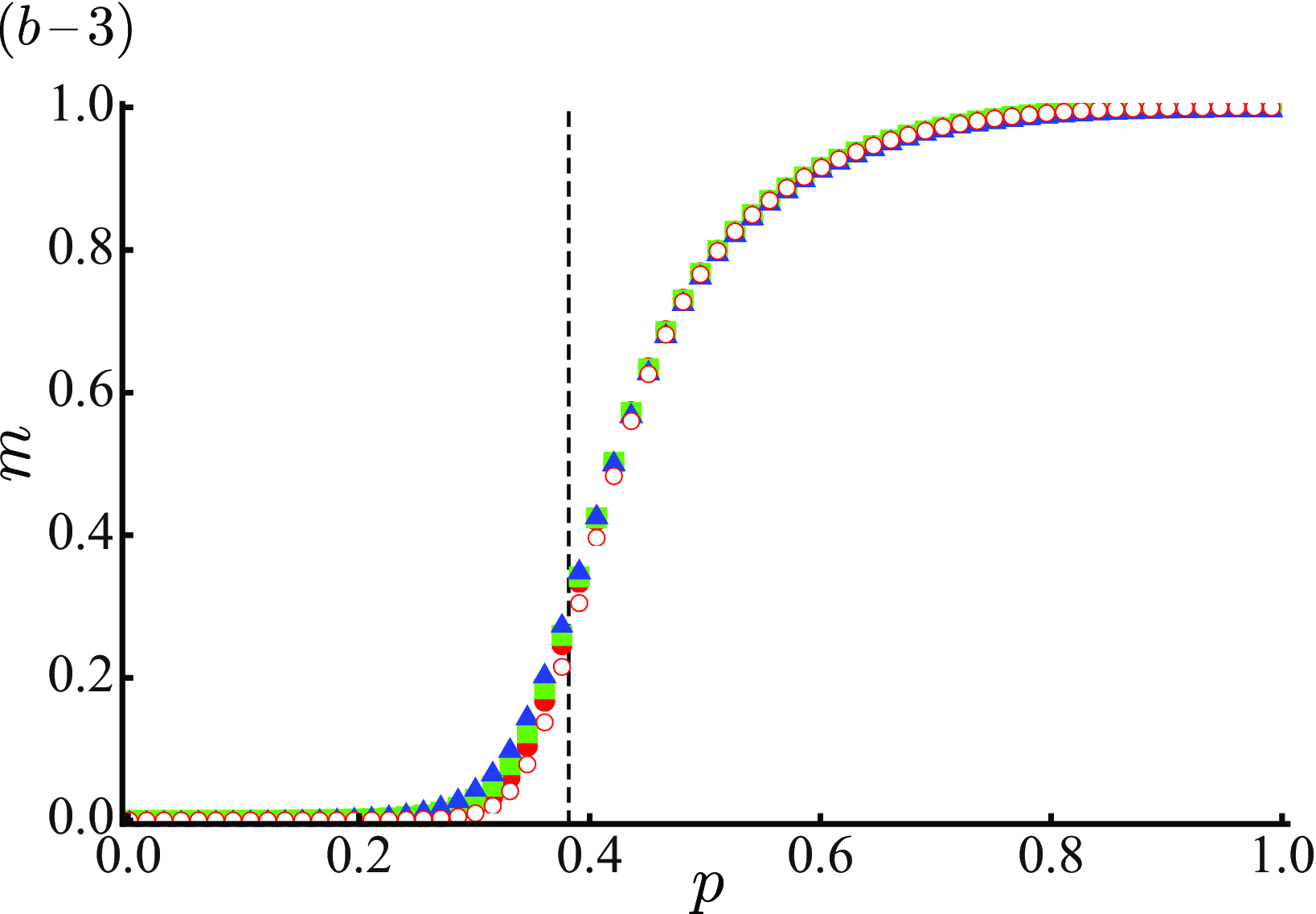}

\includegraphics[width=5.2cm]{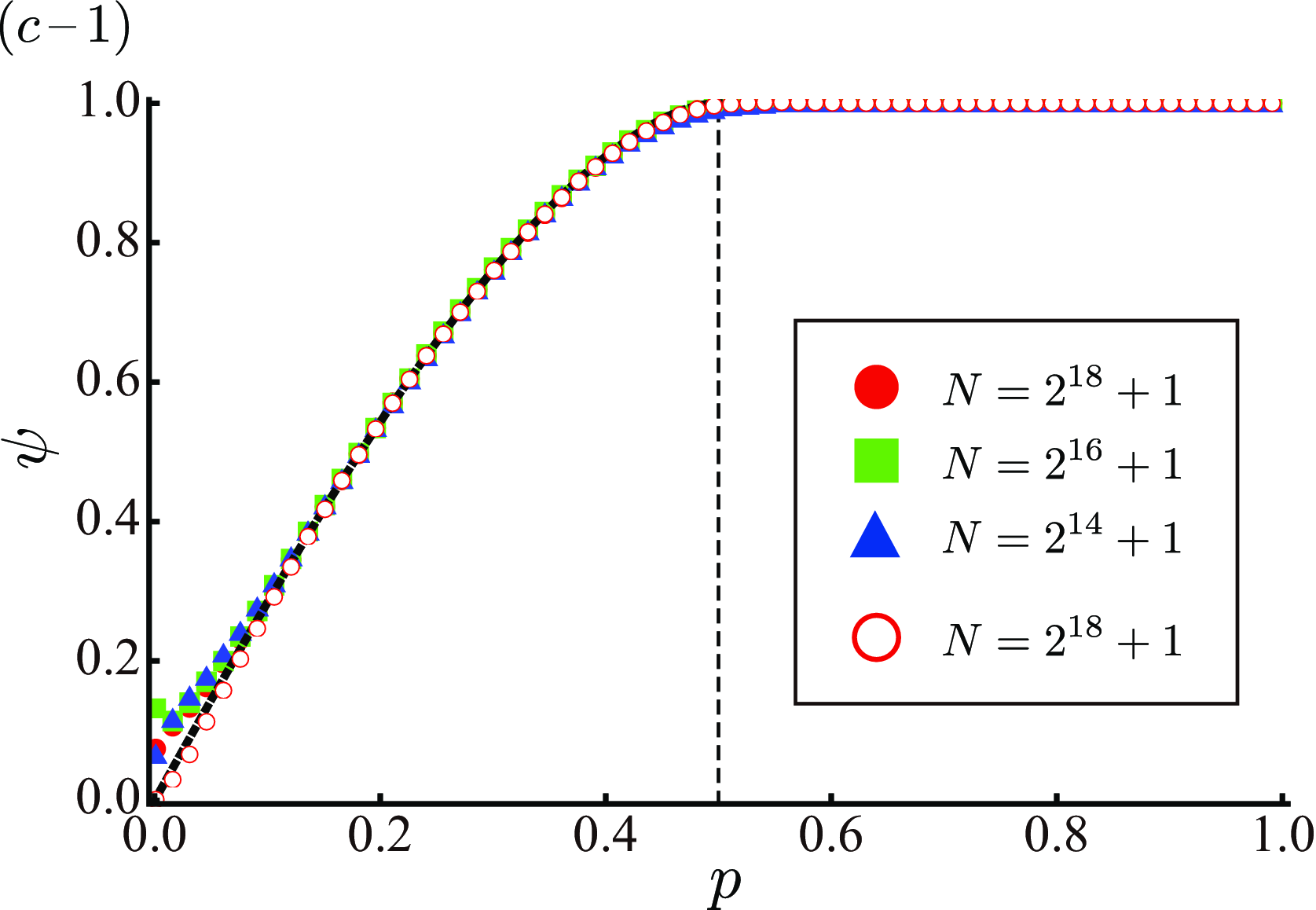}
\hspace{0.1cm}
\includegraphics[width=5.2cm]{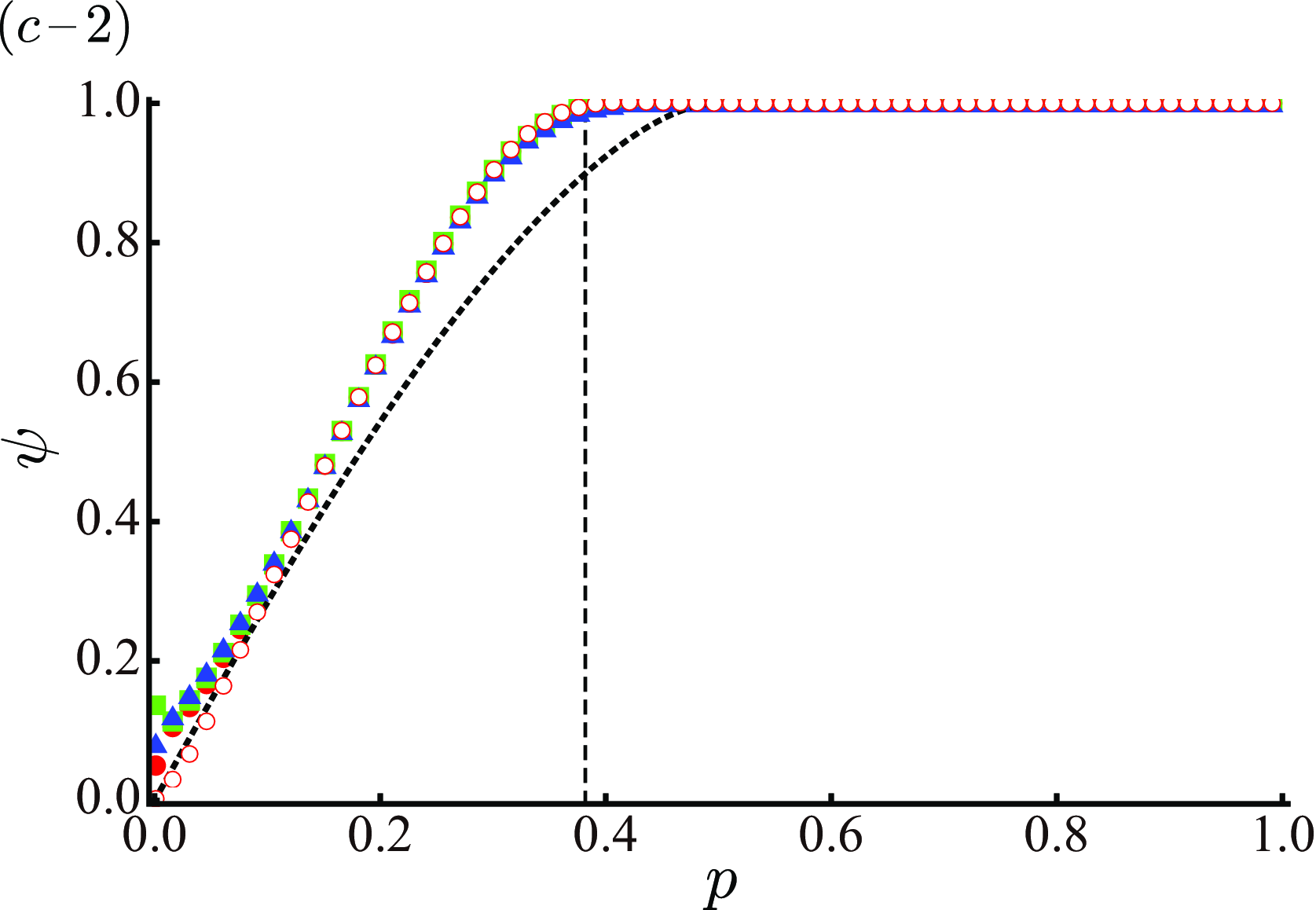}
\hspace{0.1cm}
\includegraphics[width=5.2cm]{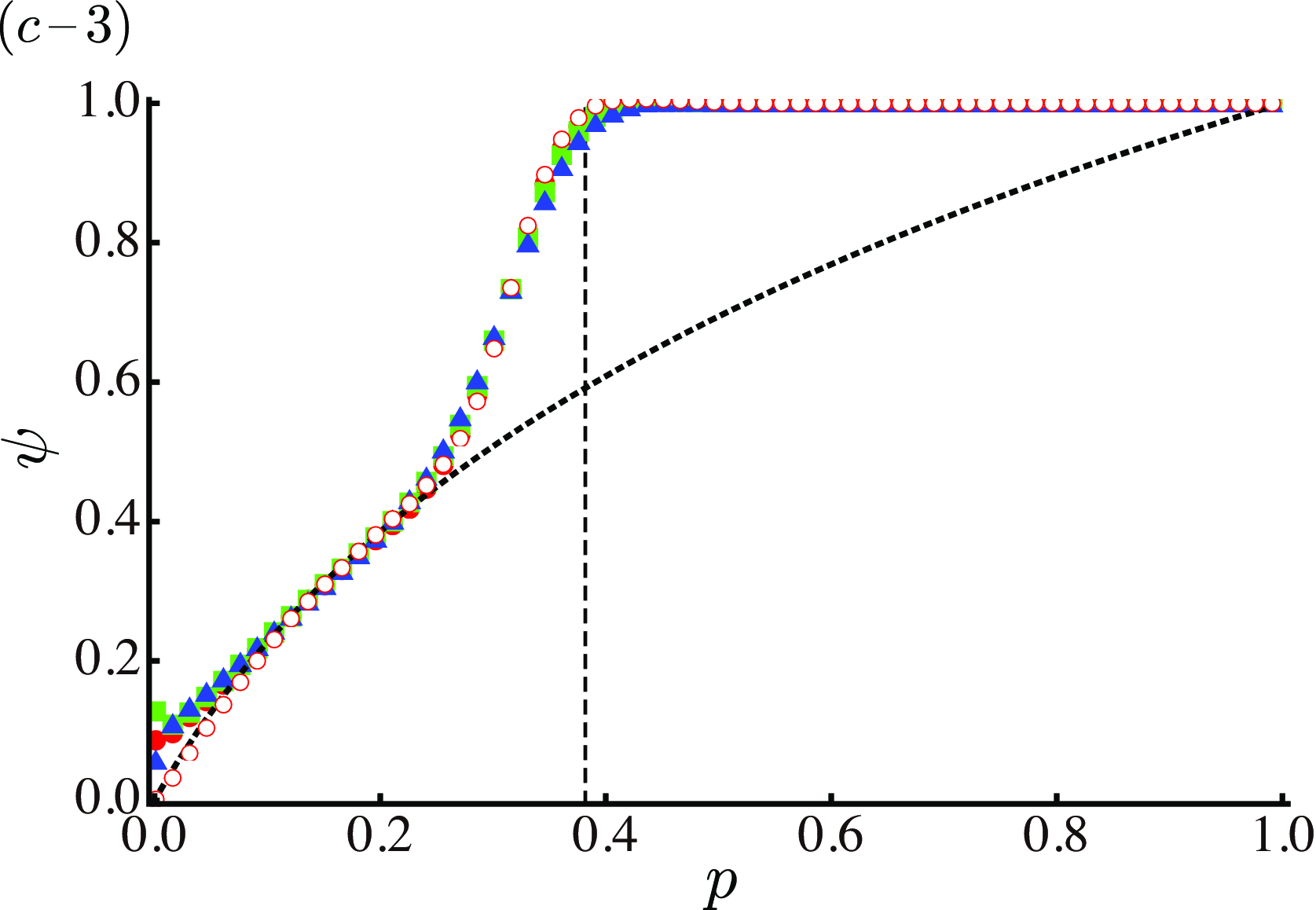}
\end{center}
\caption{
(a) Depiction of graph, (b) the order parameter, and (c) the fractal exponent for (1) HSWN, (2) HN5, and (3) HN-NP. 
A HSWN is composed of a one-dimensional backbone (black lines) and long-range bonds (red lines). 
In the HSWN that has $n$ generations, each node $2^i j$ ($i=0,\cdots,n$ and $j=0,\cdots,2^{n-i}-1$) is connected to node $2^i (j+1)$, 
and the number of nodes is $N_n=2^n+1$.
A HN5 is HSWN with additional bonds (blue lines), 
such that for each combination of $i(=0, \cdots, n-2)$ and $j(=0,\cdots, \lfloor (2^{n-i}-3)/2 \rfloor)$, 
node $2^i(2j+1)$ is connected to node $2^i(2j+3)$.
Here, HN5 is a planar graph.
To create HN-NP of $n$ generations, we add long-range bonds to one-dimensional chain of $N_n$ nodes, 
such that for each combination of $i (=0,1,2,\cdots, n-2)$ and $j (=0,1,2,\cdots, 2^{n-i-2}-1)$, 
nodes $2^i(4j)$ and $2^i(4j+1)$ are connected to $2^i(4j+3)$ and $2^i(4j+4)$, respectively. 
Each data is averaged over 100,000 percolation trials.
The full and open symbols in (b) represent $m(p,N_n)$ and $\sroot(p,N_n)/N_n$, respectively. 
The full and open symbols in (c) represent $\psi(p,N_n)$ and $\proot(p,N_n)$, respectively.
The vertical lines represent $\pu$. 
The dotted lines in (c-1) and (c-2) are generated by Eq.~(\ref{eq:psi:MK1}). 
The dotted line in (c-3) drawn from Eq.~(\ref{eq:psi:HNNP}) is a lower bound of $\proot(p)$.
}
\label{fig:Hanoi}
\end{figure}

\subsubsection{Remarks on other hierarchical small-world networks}

\noindent
The critical phase appears in other hierarchical small-world networks. 
Boettcher and his collaborators studied bond percolation on 
some hierarchical small-world networks, {\it i.e.}, a hierarchical small-world network with one-dimensional backbone (HSWN) \cite{boettcher2012ordinary} 
(also known as the Farey graph \cite{zhang2011farey}), HN5 \cite{boettcher2009patchy}, and HN-NP \cite{boettcher2009patchy},
by RG analysis to discover a critical phase (called ``patchy'' phase in \cite{boettcher2009patchy}). 
In Fig.~\ref{fig:Hanoi}, we show our Monte Carlo results of the order parameter and the fractal exponent for these models.
In \cite{boettcher2012ordinary}, Boettcher {\it et al.} analytically obtained $\proot(p)$ of HSWN as 
\begin{equation}
\proot(p)=\frac{\ln \lambda(p)}{\ln 2}, \quad {\rm where} \quad \lambda(p)=\frac{1+3p-4p^2}{2(1-p)}+\sqrt{\frac{1-p(1-4p)^2}{4(1-p)}}, \label{eq:psi:MK1}
\end{equation}
implying that $\pl=0$ and $\pu=1/2$.
This analytical result corresponds to the numerical results of both $\psi(p,N_n)$ and $\proot(p,N_n)$. 
Boettcher {\it et al.} also showed from the RG flow that HN5 has only the critical phase and the percolating phase, $\pl=0$ and $\pu \approx 0.381966$ \cite{boettcher2009patchy}. 
This result is reasonable because HN5 contains HSWN, as a subgraph. 
Our numerical result in Fig.~\ref{fig:Hanoi} shows that the fractal exponents of HN5 is always equal or larger than that of HSWN, 
and approaches continuously unity as $p$ approaches $\pu$. 
The RG scheme of HN-NP is more complicated than that of HSWN and HN5. 
In \cite{hasegawa2013absence}, the authors derived the lower bound of the fractal exponent by considering a subtree of the HN-NP as, 
\begin{equation}
\proot(p) \ge \frac{\ln (1+\sqrt{1+8p})}{\ln 2}-1, \label{eq:psi:HNNP}
\end{equation}
meaning $\pl=0$.
Our numerical result indicates $\pl=0$ and $\pu \approx 0.381966$, where the latter value was obtained from the RG analysis \cite{boettcher2009patchy}.
The phase diagram ( $0=\pl<\pu<1$ ) of these networks is similar as that of other growing networks discussed in this sections.
However, we should note that the type of the transition at $\pu$ is quite different: 
for these networks, the transition at $\pu$ is not inverted BKT one, but probably is discontinuous one.
At least, this is confirmed carefully for HSWN \cite{boettcher2012ordinary}, 
although it is tedious to get an evidence from finite size simulations. 
The relation between the continuity/discontinuity of the order parameter the and 
discontinuity/continuity of the fractal exponent was argued in our recent manuscript \cite{nogawa2013meta}, 
although its theory is still limited.

\subsection{Finite-size scaling for complex networks \label{Sec:FSS}}

\noindent
Finite-size scaling is a powerful tool for extracting transition points, 
as well as their critical exponents, from numerical data obtained from simulations of various sizes.
In fact, it is very successful in analyzing phase transitions of some static networks
such as the configuration model \cite{hong2007finite}.
In the case of exhibiting critical phase such as in growing networks, 
however, we are faced with the following two difficulties 
in the estimation of the transition point $p_{c2}$ from their critical behaviors. 
First, as aforementioned in Sec.~\ref{sec:finite_NAG}, only the data for $p \ge p_{c2}$
should be used to perform scaling analysis, and the data for $p < p_{c2}$ are useless.
Secondly, the singularity at $p_{c2}$ is often infinitely weak (Eq.~(\ref{eq:infiniteorder})), 
for which the standard scaling theory based on power-law behaviors does not work.
To overcome this difficulty, the authors proposed 
a novel finite-size scaling method described in terms of the network size $N$ and the fractal exponent $\psi$ \cite{hasegawa2013profile}.
Our finite-size scaling analysis works well for both inverted BKT and second order transitions in network systems.

Let us derive a scaling form with $\psi$ and $N$ for $p>p_{c2}$ with a heuristic method.
We consider the $N$-dependence of $\smax(p, N)$ at $p$ slightly larger than $p_{c2}$
by assuming the existence of a crossover size $N^*(p)$, which diverges as $p$ approaches $p_c$.
For $N \ll N^*(p)$, the system behaves as if it were critical even for $p > p_{c2}$, 
so that $\smax(p, N) \propto N^{\psi_c}$, where $\psi_c$ is the fractal exponent at $p=p_{c2}$.
For $N \gg N^*(p)$, 
on the other hand, we observe that the behavior is essentially same as that in the thermodynamic limit, 
{\it i.e.}, $\smax(p, N)=Nm(p)$.
By connecting these two limits at $N=N^*$, we expect $N^*(p)$ to satisfy 
\begin{equation}
N^*(p) \propto m(p)^{1/(\psi_c-1)} \,, \label{crossover}
\end{equation}
and the finite-size scaling form for $\smax(p, N)$ to satisfy 
\begin{equation}
\smax(p, N) =N^{\psi_c} f_1 \Big[ \frac{N}{N^*(p)} \Big] \,, 
\label{SmaxScalingForm}
\end{equation}
where $f_1(x)$ is a scaling function satisfying 
\begin{equation}
f_1(x) \propto 
{\Biggl\{}
\begin{array}{ccl}
{\rm const} & {\rm for} & x \ll 1 \\
x^{1-\psi_c} & {\rm for} & x \gg 1 \,. 
\end{array} \label{SmaxScaling1}
\end{equation}
Or equivalently, 
\begin{equation}
\smax(p, N) =N^*(p)^{\psi_c} f_2 \Big[ \frac{N}{N^*(p)} \Big] \,, 
\label{SmaxScalingForm2}
\end{equation}
where
\begin{equation}
f_2(x) = x^{\psi_c} f_1(x) \propto 
{\Biggl\{}
\begin{array}{ccl}
x^{\psi_c} & {\rm for} & x \ll 1 \\
x & {\rm for} & x \gg 1 \,.
\end{array} \label{SmaxScaling2}
\end{equation}
The derivative of  Eq.~(\ref{SmaxScalingForm2}) with respect to $\ln N$ gives us the scaling form for $\psi(p, N)$:
\begin{eqnarray}
\psi(p, N) = g \Big[ \frac{N}{N^*(p)} \Big] \,, 
\label{psiScalingForm}
\end{eqnarray}
where
\begin{equation}
g(x) =\frac{{\rm d} \ln f_2(x)}{{\rm d} \ln x}=
{\Biggl\{}
\begin{array}{ccl}
\psi_c & {\rm for} & x \ll 1 \\
1 & {\rm for} & x \gg 1 \,.
\end{array} \label{psiScaling}
\end{equation}
Similarly, the finite-size scaling form for the susceptibility $\chi(p,N)$ (as well as the mean cluster size $\sav(p,N)$) 
can be assumed to be 
\begin{equation}
\chi (p, N) = N^{\pavc} h \Big[ \frac{N}{N^*(p)} \Big] \,, 
\label{chiScalingForm}
\end{equation}
where 
\begin{equation}
h(x) =
{\Biggl\{}
\begin{array}{ccl}
{\rm const} & {\rm for} & x \ll 1 \\
x^{-\pavc} & {\rm for} & x \gg 1 \,,
\end{array} \label{chiScaling}
\end{equation}
$\pavc$ is the fractal exponent of the mean cluster size at $p=p_{c2}$, defined as $\sav(p_{c2}, N) \propto N^{\pavc}$ 
and is related to $\psi_c$ as $\pavc = 2 \psi_c-1$ (see Sec.~\ref{SecBT}).

In the case of a second order transition, the order parameter behaves as $m(p) \propto (\Delta p)^\beta$ 
for $p \ge p_{c2}$ ($=p_{c1}$).
Then, the finite-size scaling forms for $\smax(p, N)$, $\psi (p, N)$, and $\chi(p, N)$ are 
\begin{equation}
\smax(p, N) = N^{\psi_c} f_1 [ N (\Delta p)^{\beta/(1-\psi_c)}] \,, 
\label{SmaxScalingForm-recon}
\end{equation}
\begin{equation}
\psi(p, N) = g [ N (\Delta p)^{\beta/(1-\psi_c)}] \, , 
\label{PsiScalingForm-recon}
\end{equation}
and
\begin{equation}
\chi(p, N) = N^{\pavc} h [ N (\Delta p)^{\beta/(1-\psi_c)}] \,, 
\label{SavScalingForm-recon}
\end{equation}
respectively.

In the case of an inverted BKT transition, $m(p)$ follows Eq.~(\ref{eq:infiniteorder}). 
Then, we obtain the scaling form for $\smax(p, N)$ and $\psi(p, N)$ 
by substituting Eq.~(\ref{crossover}) into Eqs.~(\ref{SmaxScalingForm}) and (\ref{psiScalingForm}) as 
\begin{equation}
\smax(p, N) =N^{\psi_c} f_1 ( N \exp [-\alpha/(1-\psi_c)(\Delta p)^{\beta^\prime}]) \,, 
\label{SmaxScalingFormInfiniteOrder}
\end{equation}
and  
\begin{equation}
\psi(p, N)= g ( N \exp [-\alpha/(1-\psi_c)(\Delta p)^{\beta^\prime}]) \,, 
\label{PsiScalingFormInfiniteOrder}
\end{equation}
respectively.

Note that the present scaling form includes the conventional finite-size scaling. 
For $d$-dimensional lattice systems, the fractal exponent satisfies $\psi_c=1-\beta /d \nu$, 
so that Eq.~(\ref{SmaxScalingForm}) reduces to the conventional scaling for $N m$ 
provided that $N/N^*(p) =(L/\xi)^d$, 
where $L$ is the linear dimension, and $\nu$ is the critical exponent of the correlation \textit{length} 
$\xi \propto (\Delta p)^{-\nu}$.

\begin{figure*}[ttttt]\begin{center}
\includegraphics[width=150mm,clip]{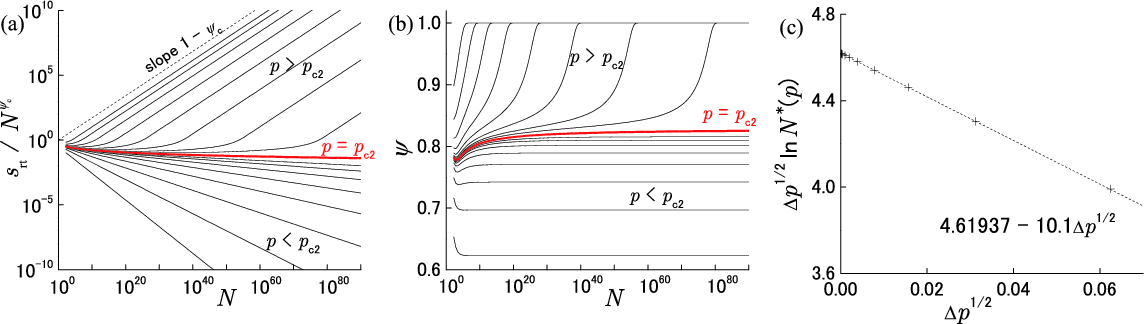}
\end{center}
\caption{ 
Size dependence of (a) $\sroot$ and (b) $\proot$ of the decorated flower. 
(c) The dependence of the crossover scale $N^*$ at which $\proot(p, N)=0.95$ on $\Delta p = p - p_{c2}$ for $p > p_{c2}$. 
}
\label{fig:FSS:flower1}
\end{figure*}

\begin{figure}[t]\begin{center}
\includegraphics[width=55mm,clip]{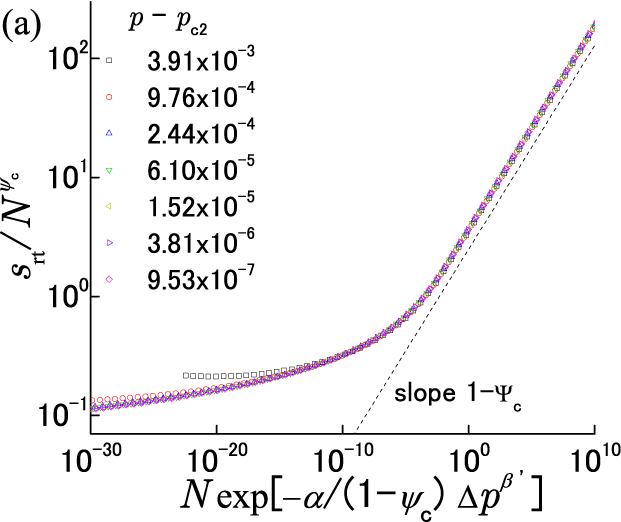}
\includegraphics[width=55mm,clip]{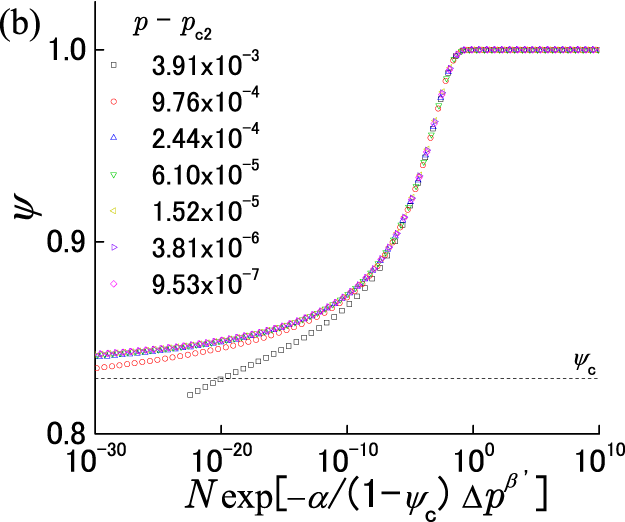}
\end{center}
\caption{ 
Scaling plot of (a) $\sroot(p, N)$ and (b) $\proot(p, N)$ of the decorated flower, for $p>p_c$. 
Here we used $p_{c2}=5/32$, $\psi_c = 1/2 + \log_4 ( 1 + 1/\sqrt{3} )$, $\beta'$=1/2, and $\alpha$ = 0.791049. 
}
\label{fig:FSS:flower2}
\end{figure}

\subsubsection{Example: the decorated flower}

\noindent
As a demonstration, we apply our method to the decorated flower.
We plot $\sroot (p, N) /N^{\psi_c}$ and $\psi(p, N)$ as a function of $N$ 
at several values of $p$ in Figs.~\ref{fig:FSS:flower1}(a) and (b), respectively.
For $p<p_{c2}$, $\sroot (p, N)$ is proportional to ${N}^{\psi(p)}$ with $\psi(p)<\psi_c$ for large $N$. 
For $p>p_{c2}$, on the other hand, $\sroot (p, N)$ shows a crossover from $N^{\psi_c}$ to $N^1$
around certain $N^*(p)$, as mentioned in Eqs.~(\ref{SmaxScalingForm})-(\ref{SmaxScaling2}).
Similarly, $\psi(p, N)$ converges to $\psi(p)$ for $p<p_{c2}$, and shows a stepwise change from $\psi_c$ to 1 for $p>p_{c2}$ 
around $N\sim N^*(p)$.
Specifically, we estimate the value of $N^*(p)$ so as to satisfy $\psi(p,N^*(p)) = 0.95$, 
and we show its $p$-dependence in Fig.~\ref{fig:FSS:flower1}(c).
It is consistent with our hypothesis 
$N^*(p) \propto e^{\alpha/(1-\psi_c) \Delta p^{\beta'} }$, {\it i.e.}, $\Delta p^{\beta'} \ln N^*(p) = \alpha/(1-\psi_c) + {\rm const.} \times \Delta p^{\beta'}$ by assuming $\beta' = 1/2$. 
From this plot, we obtain $\alpha = 0.791049$. 
We then perform the scaling plot of $\sroot (p, N)$ and $\psi (p, N)$ in Figs.~\ref{fig:FSS:flower2}(a) and (b), respectively, 
by using the exponents mentioned above. 
The collapsing of data to a universal scaling curve is very nice for data with large $N$ 
(as seen in Fig.~\ref{fig:FSS:flower1}(b), some correction to the scaling cannot be neglected for $N < 10^{20}$). 

\subsubsection{Example: the $m$-out graph and the configuration model}

\begin{figure}[!t]\begin{center}
\includegraphics[width=5.5cm]{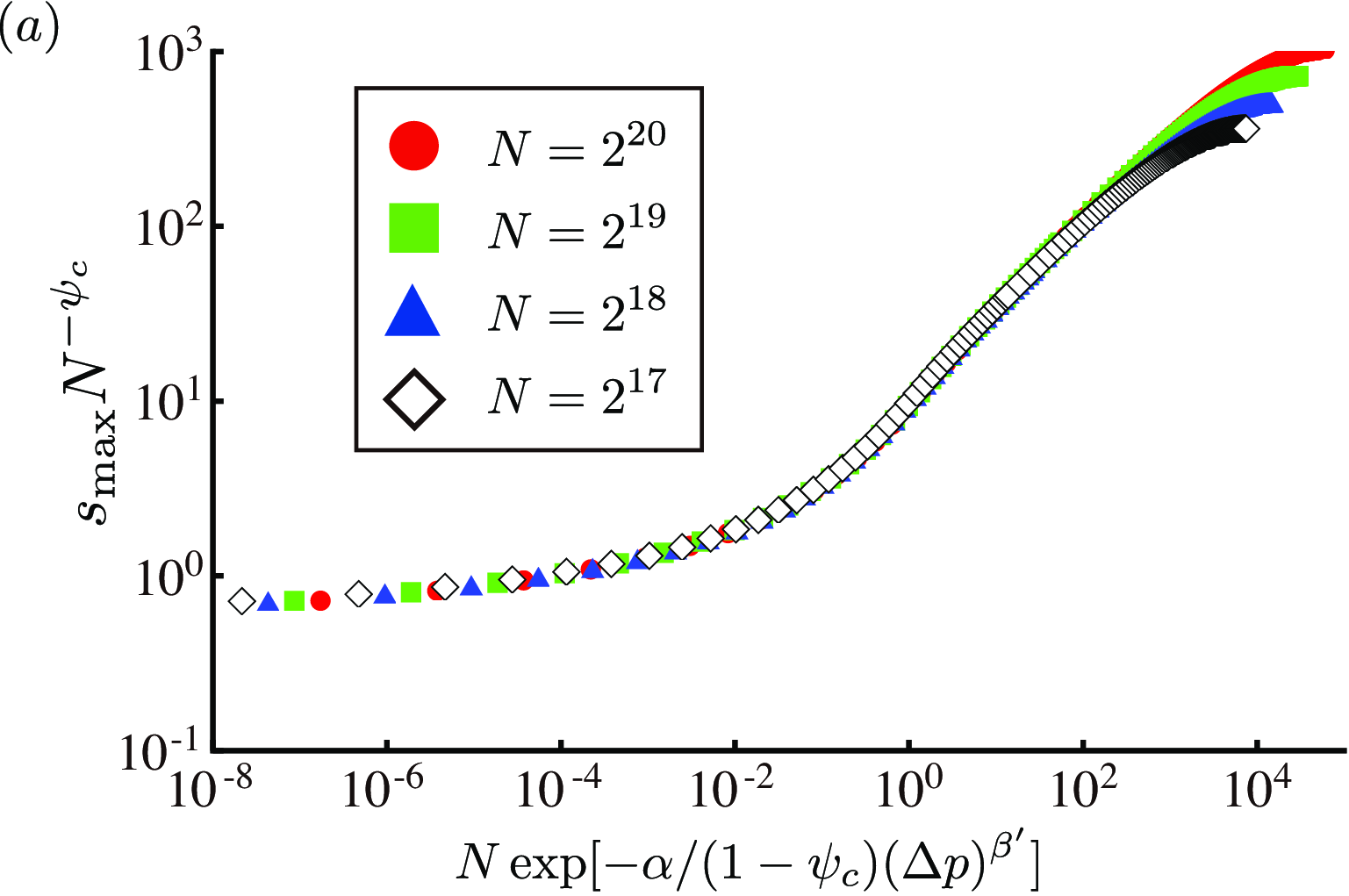}
\includegraphics[width=5.5cm]{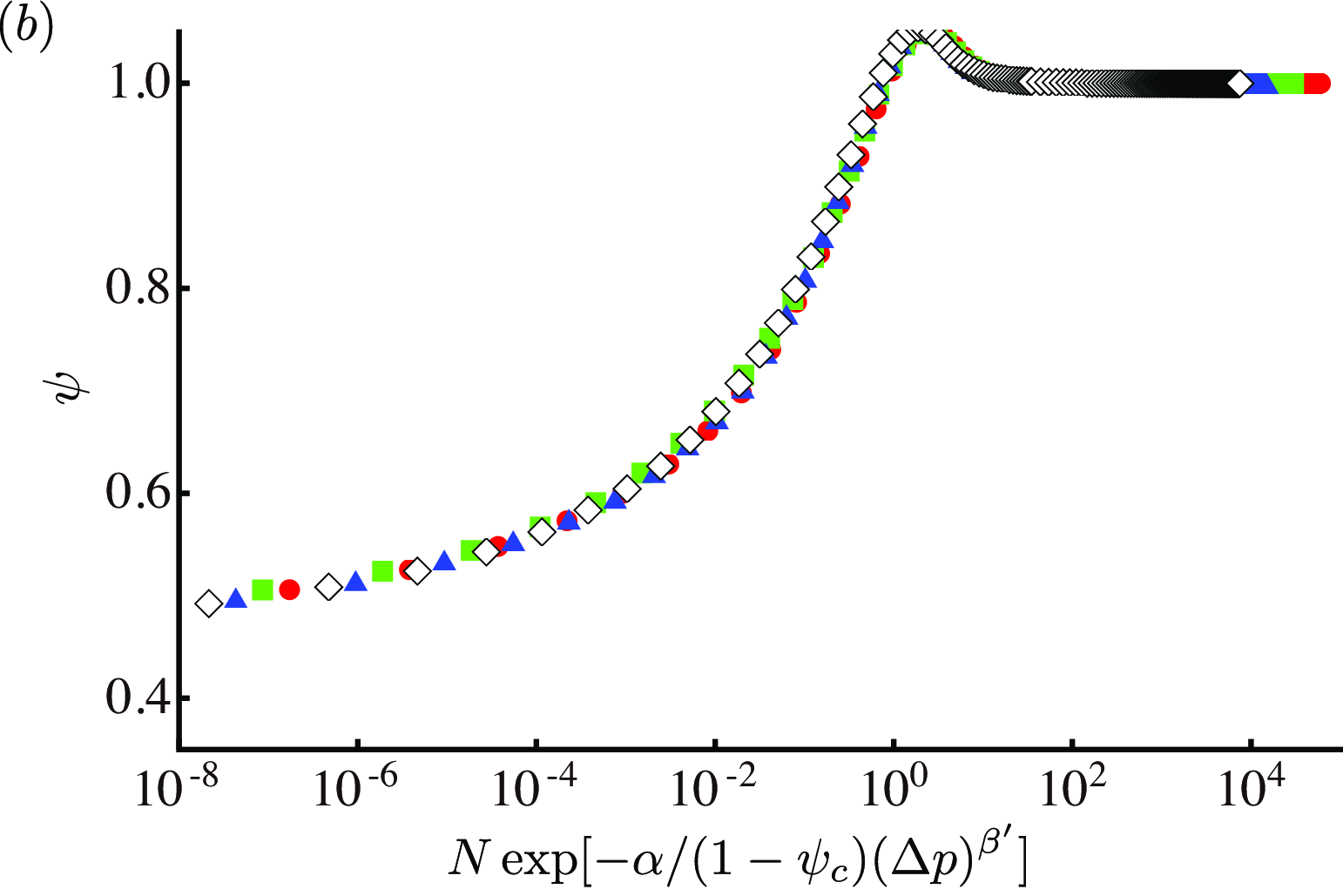}
\end{center}
\caption{
Finite-size scaling for (a) $\smax (p, N)$ by Eq.~(\ref{SmaxScalingFormInfiniteOrder}) 
and (b) $\psi (p, N)$ by Eq.~(\ref{PsiScalingFormInfiniteOrder}) of the $m$-out graph.
}
\label{fig:GRN:scaling}
\end{figure}

\noindent
Our second example is the $m$-out graph and the configuration model.
We obtain good data collapses for both $\smax(p, N)$ and $\psi(p, N)$ by the finite-size scaling, as shown in Fig.~\ref{fig:GRN:scaling}.
We did not perform the finite-size scaling of $\chi(p,N)$ for the $m$-out graph 
since our numerical result for the $m$-out graph shows $\psi_c=1/2$ and $\pavc=0$ at the transition point $p=p_{c2}$, 
meaning that the susceptibility of the $m$-out graph does not diverge. 
Here, parameters $\alpha$ and $\beta^\prime$ are set to the analytically obtained values.

\begin{figure*}[!t]\begin{center}
\includegraphics[width=5.5cm]{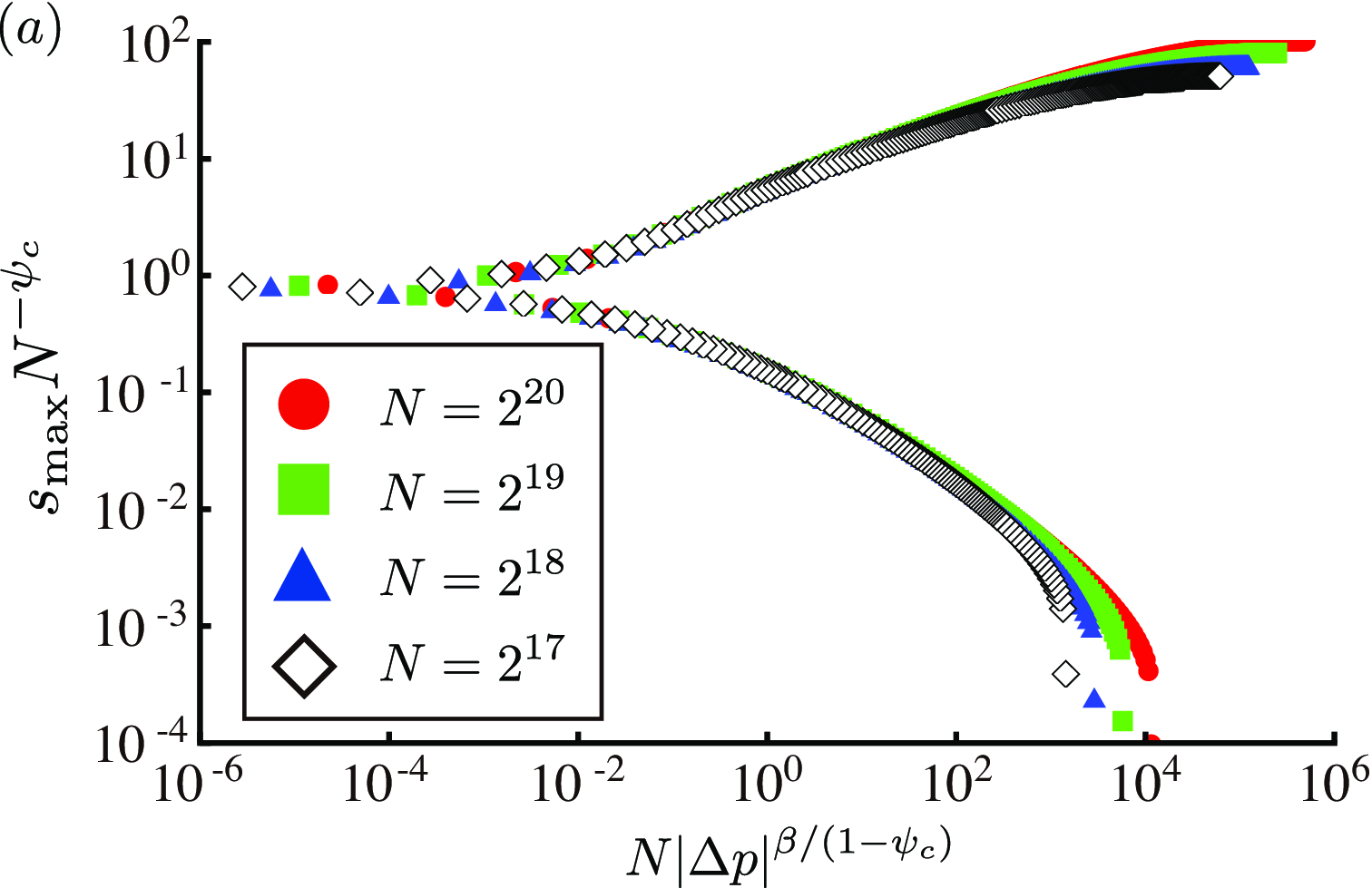}
\includegraphics[width=5.5cm]{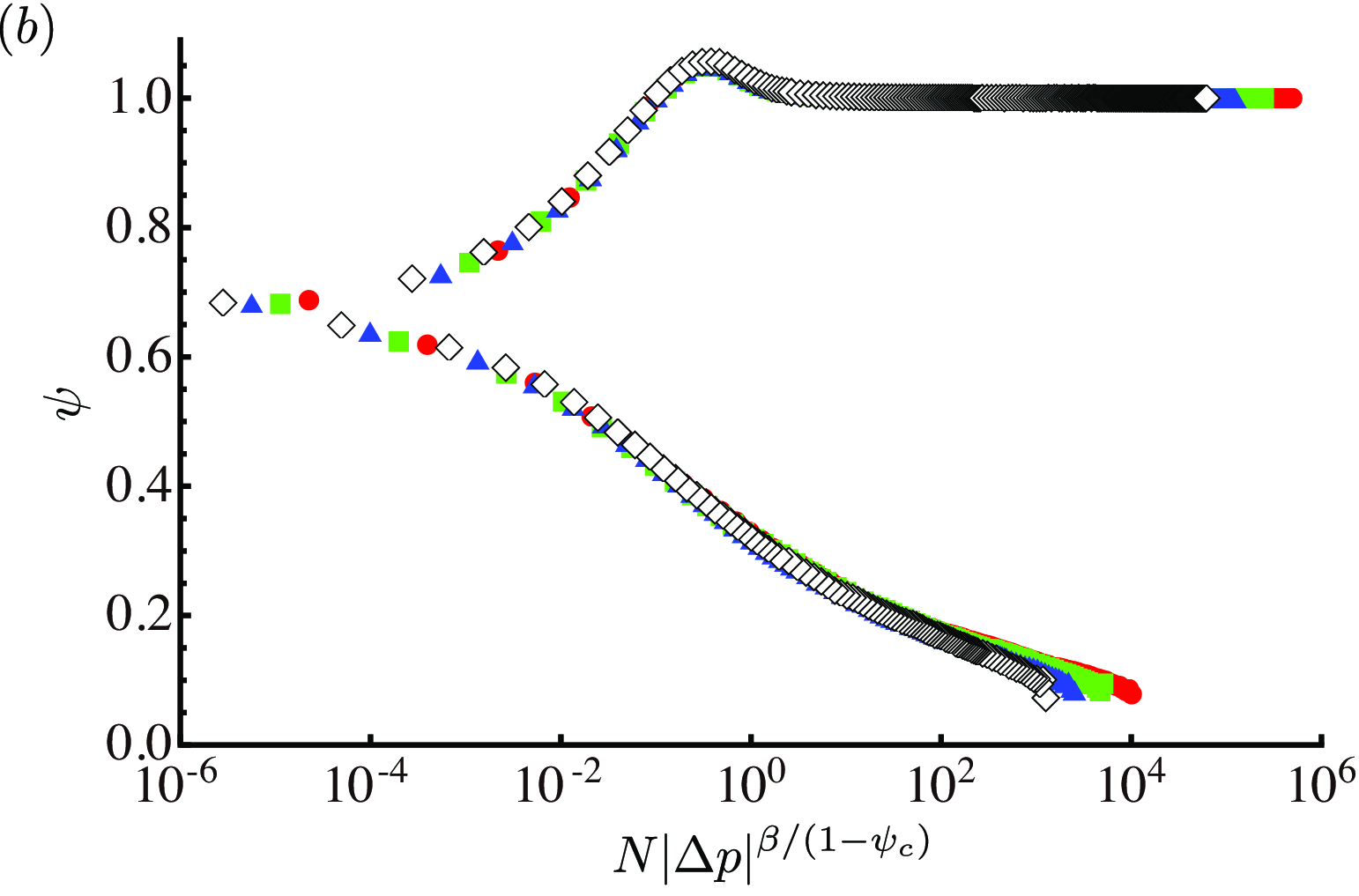}
\includegraphics[width=5.5cm]{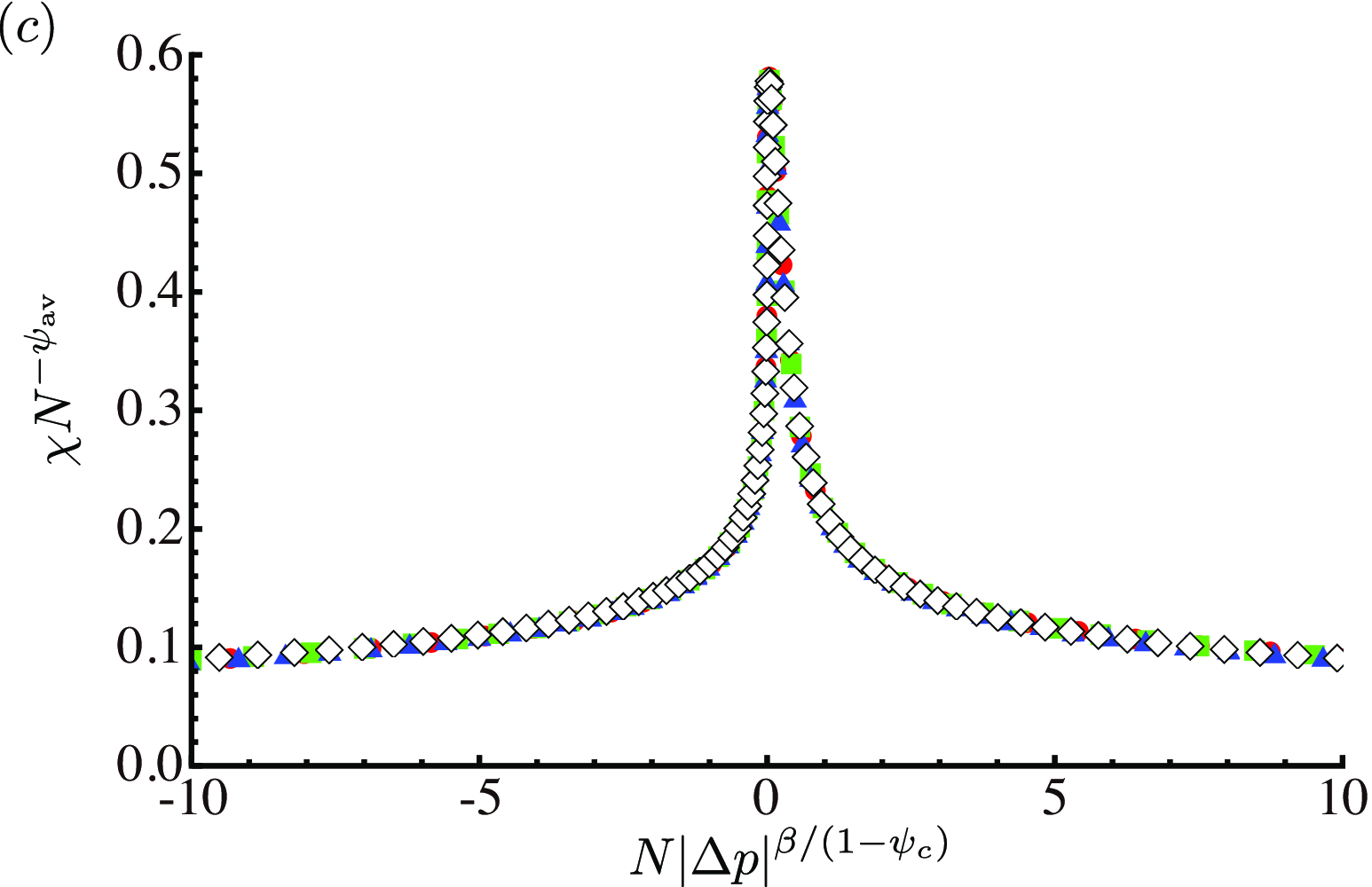}
\end{center}
\caption{
Finite-size scaling for (a) $\smax(p, N)$ by Eq.~(\ref{SmaxScalingForm-recon}), 
(b) $\psi(p, N)$ by Eq.~(\ref{PsiScalingForm-recon}), and (c) $\chi(p, N)$ by Eq.~(\ref{chiScalingForm}) of the configuration model.
}
\label{fig:GRN:scaling_config}
\end{figure*}

In Fig.~\ref{fig:GRN:scaling_config}, we show the results for finite-size scalings of $\smax(p, N)$, $\psi(p, N)$, and $\chi(p, N)$ with $\beta=1$ on the configuration model that has the same $P(k)$ as that of the $m$-out graph.
Again, we observe good data collapses for these values.

\begin{table*}[!t]
\caption{Two critical points of bond percolations on various graphs.} 
\label{relation}
\begin{tabular}{llc}
\hline graph type & example & scenario \\
\hline \hline amenable graph with two ends 
& chain 
& $0<p_{c1}=p_{c2}=1$ \\ \hline 
amenable graph with one end & $d(\ge 2)$-dimensional Euclidean lattice & $0<p_{c1}=p_{c2}<1$\\ \hline 
NAG with infinitely many ends & BT & $0<p_{c1}< p_{c2}=1$ \\ \hline 
NAG with one end & EBT \cite{nogawa2009monte,baek2009comment,nogawa2009reply,baek2012upper,gu2012crossing}, 
hyperbolic lattice~\cite{baek2009percolation,lee2012bounds,gu2012crossing} & $0<p_{c1}<p_{c2}<1$ \\ \hline 
stochastic growing tree & GRT~\cite{hasegawa2010critical,zhang2008degree,lancaster2002cluster,zalanyi2003properties,pietsch2006derivation} & $0=p_{c1}<p_{c2}=1$ \\ \hline 
stochastically growing network & GRN~\cite{zalanyi2003properties,riordan2005small,bollobas2005slow}, 
CHKNS model~\cite{dorogovtsev2001anomalous,callaway2001randomly} & $0=p_{c1}<p_{c2}<1$ \\ \hline 
deterministically growing network & decorated flower~\cite{rozenfeld2007percolation,berker2009critical,hasegawa2010generating}, 
HN5~\cite{boettcher2009patchy} & $0=p_{c1}<p_{c2}<1$ \\
& 
HN-NP~\cite{boettcher2009patchy,hasegawa2013absence}, HSWN~\cite{boettcher2012ordinary} & 
\\
\hline \end{tabular} 
\end{table*}

\section{Discussion \label{sec:Discussion}}

\noindent
In this paper, we have considered percolation on various types of graphs. 
Their phase diagrams are summarized in Table~\ref{relation}.
The phase boundaries of amenable graphs including the Euclidean lattices and some static uncorrelated networks are 
$0 < p_{c1}=p_{c2} \le 1$ and for NAGs, they are $0 <p_{c1}<p_{c2} \le 1$. 
Networks with growth mechanisms lead us to a new scenario: $p_{c1}=0$ and $p_{c2} > 0$ 
(see the last three rows in table~\ref{relation}).
As for transitive graphs, 
we already have the condition for the existence of the critical phase (whether a graph is amenable or nonamenable). 
However, as for non-transitive graphs, {\it i.e.}, complex networks, the answer is still missing, 
although the small-world property is presumably a necessary condition 
because the critical phase can appear when the correlation length and correlation volume diverge at different points.
Discovery of the condition for the critical phase in complex networks is the next challenging work.

So far, we have focused on bond percolation. 
Here, we mention some remarks corresponding to other dynamics.
It already has been proven by mathematicians that a critical phase also exists in 
equilibrium spin systems on infinite NAGs \cite{wu1996ising,wu2000ising,lyons2000phase}.
In addition, 
the Ising model on hyperbolic lattices has been investigated by means of statistical physics, {\it i.e.}, Monte Carlo simulations \cite{shima2006geometric,shima2006dynamic,sakaniwa2009survival}
and the transfer-matrix method \cite{ueda2007corner,krcmar2008ising,iharagi2010phase,gendiar2012weak}.  
A critical phase and the inverted BKT transition has also been found in the Ising model on 
an inhomogeneous annealed network \cite{bauer2005phase}, the decorated flower \cite{hinczewski2006inverted}, 
Hanoi networks \cite{boettcher2011fixed,boettcher2011renormalization,boettcher2012classification}, 
and the Potts model on the HSWN \cite{nogawa2012generalized}.
Although a critical phase in spin systems has not been well understood,
a discussion parallel to that for percolation models may be also possible for
these systems and, if so, the fractal exponent will be useful to characterize the critical phase.
In spin systems, we have a local disconnected susceptibility $\tilde{\chi}(N)$, 
which is directly related to the cluster size in the percolation model, 
and its fractal exponent $\psi$ defined as $\tilde{\chi}(N) \propto N^{\psi}$. 
In \cite{nogawa2012generalized,nogawa2012criticality}, the average fractal exponent $\pav (= 2\psi - 1)$, 
rather than $\psi$, was investigated in the Ising and Potts models on hierarchical small-world network. 
Renormalization group studies lead us to a generalized scaling theory for equilibrium systems 
that is similar to the finite-size scaling theory shown in this article, 
which is based on the scale invariance of the free energy (or distribution function) \cite{nogawa2012generalized}. 
This theory includes the renormalization of the external field, 
and therefore the scalings of the order parameter and the susceptibility are derived straightforwardly 
since these quantities are given by the derivatives of the free energy with respect to the field.

As for the contact process (the susceptible-infected-susceptible model), 
the model on NAGs may have an intermediate phase between the absorbing phase, where the process dies out, 
and the active phase, where a finite fraction of nodes is active for all time. 
In the intermediate phase, the probability that process survives for all time is positive, 
but an arbitrary node is never infected after a long time, implying that the order parameter remains zero.
This is at least the case for tree \cite{pemantle1992contact,liggett1996multiple,stacey1996existence}. 
However, it is an open problem whether 
there exists a resemblance of this intermediate phase in finite NAGs with the large size limit or in complex networks presented in this paper. 

\section*{Acknowledgements}
This work was partially supported by the Grant-in-Aid for Young Scientists (B) of Japan Society for the Promotion of Science (Grant No.~24740054 to T.H.) and JST, ERATO, Kawarabayashi Large Graph Project.

\end{document}